\documentclass[apj]{emulateapj}
\usepackage{textcomp}
\usepackage{epsfig, epstopdf}
\usepackage{amssymb,amsmath}

\newcommand{\simgt}{\lower 2pt \hbox{$\, \buildrel {\scriptstyle >}\over {\scriptstyle\sim}\,$}}
\newcommand{\simlt}{\lower 2pt \hbox{$\, \buildrel {\scriptstyle <}\over {\scriptstyle\sim}\,$}}

\newcommand{\rxj}{RX~J1131$-$1231}

\newcommand{\qj}{QJ~0158$-$4325}
\newcommand{\sdss}{SDSS~1004$+$4112}

\slugcomment{Received 2016 Sept 20, Accepted 2017 January 29}
\shorttitle{Measuring the ISCO's of Black Holes}
\shortauthors{CHARTAS ET AL.}

\begin{document}

\def\sarc{$^{\prime\prime}\!\!.$}
\def\arcsec{$^{\prime\prime}$}
\def\beginrefer{\section*{References}%
\begin{quotation}\mbox{}\par}
\def\refer#1\par{{\setlength{\parindent}{-\leftmargin}\indent#1\par}}
\def\endrefer{\end{quotation}}

\title{Measuring the Innermost Stable Circular Orbits of Supermassive Black Holes }

\author{G. Chartas\altaffilmark{1,2},  H. Krawczynski\altaffilmark{3}, L. Zalesky\altaffilmark{1}, C. S. Kochanek\altaffilmark{4,5},  X. Dai\altaffilmark{6}, C. W. Morgan\altaffilmark{7}, and A. Mosquera\altaffilmark{7}}

\altaffiltext{1}{Department of Physics and Astronomy, College of Charleston, Charleston, SC, 29424, USA, chartasg@cofc.edu}
\altaffiltext{2}{Department of Physics and Astronomy, University of South Carolina, Columbia, SC, 29208}
\altaffiltext{3}{Physics Department and McDonnell Center for the Space Sciences, Washington University in St. Louis, 1 Brookings Drive, CB 1105, St. Louis, MO 63130, USA}
\altaffiltext{4}{Department of Astronomy, The Ohio State University, 140 West 18th Avenue, Columbus, OH 43210, USA}
\altaffiltext{5}{Center for Cosmology and AstroParticle Physics, The Ohio State University, 191 W. Woodruff Avenue, Columbus, OH 43210, USA}
\altaffiltext{6}{Homer L. Dodge Department of Physics and Astronomy, University of Oklahoma, Norman, OK 73019, USA}
\altaffiltext{7}{Physics Department, United States Naval Academy, Annapolis, MD 21403, USA}

\begin{abstract}
We present a promising new technique, the $g$-distribution method, for measuring the inclination angle ($i$), the innermost stable circular orbit (ISCO), and the spin of a supermassive black hole. The $g$-distribution method uses measurements of the energy shifts in the relativistic iron line emitted by the accretion disk of a supermassive black hole due to microlensing by stars in a foreground galaxy relative to the $g$-distribution shifts predicted from microlensing caustic calculations.  We apply the method to the gravitationally lensed quasars \rxj\ ($z_{s}$ = 0.658, $z_l$ = 0.295), \qj\ ($z_{s}$ = 1.294, $z_l$ = 0.317), and  \sdss\ ($z_{s}$ = 1.734, $z_l$ = 0.68).  For RX~J1131$-$1231 our initial results indicate that $r_{\rm ISCO}$~$\simlt$~8.5~gravitational radii ($r_{\rm g}$) and $i$ $\simgt$ 76$^{\circ}$. 
We detect two shifted Fe lines, in several observations, as predicted in our numerical simulations of caustic crossings.  
The current ${\Delta}E$-distribution of \rxj\ is sparsely sampled but further X-ray monitoring of RX~J1131$-$1231 and other lensed quasars will provide improved constraints on the inclination angles, ISCO radii and spins of the black holes of distant quasars. 

\noindent 
\\
\\

\end{abstract}


\section{INTRODUCTION}

One technique for measuring the innermost stable circular orbit (ISCO) and spin parameter $a$ ($a = Jc/GM_{BH}^{2}$, where $J$ is the angular momentum) of AGN relies on modeling the relativistically blurred Fe K$\alpha$ fluorescence lines originating from the inner parts of the disk (e.g., Fabian et al. 1989; Laor 1991; Reynolds \& Nowak 2003). 
This relativistic iron line method has be applied to about 20 relatively bright nearby Seyferts where the line is detectable with a high signal-to-noise ratio (Reynolds 2014 and Vasudevan et al. 2016).
The sample sizes are starting to become large enough  where the distribution of the spin parameter can be calculated and compared to simulated ones such as those presented in Volonteri et al.	(2013).
Even then, the Fe K${\alpha}$ line in most Seyferts is typically very weak, and constraining the spin and accretion disk parameters of Seyferts requires considerable observing time on {\sl XMM-Newton} and {\sl Chandra}. Moreover, the accuracy of the relativistic Fe line method for constraining the spin of a black hole from the broadened red wing of the Fe line profile is also questioned (e.g., Miller et al 2009; Sim et al. 2012). We note, however, that independent measurements of the size of the corona from microlensing and reverberation mapping indicate that the X-ray source is compact, consistent with the lampost model assumed in the relativistic Fe iron line method. Additional support of the relativistic Fe line method is provided by 3--50 keV observations with {\sl NuSTAR}, such as the recent observations of Mrk~335 that indicate a spin parameter of $>$~0.9 at 3 $\sigma$ confidence. The high-energy {\sl NuSTAR} spectra  can help to constrain the reflection component and better distinguish between models (Parker et al. 2014). 
Most of the measured spin parameters in Seyfert galaxies are found to be $\simgt$ 0.9 (e.g., Reynolds 2014 and references therein).  This may be the result of a selection bias in flux-limited samples (Vasudevan et al. 2016).  Specifically, high spin black holes are more luminous and hence brighter for a given accretion rate, and hence will simply be more highly represented in flux limited surveys. Recent observations and simulations (Fabian 2014; Keck et al. 2015; Vasudevan et al. 2016) also suggest that rapidly spinning black holes will tend to have stronger reflected relative to direct X-ray emission, making it easier to measure the spin parameter in these objects.

The relativistic iron line method has been applied to the gravitationally lensed quasars \rxj\ (Reis et al. 2014) and Q2237$+$0305 (Reynolds et al. 2014) and well as to a stacked spectrum of 27 lensed 1.0 $<$ $z$ $<$ 4.5 quasars observed with {\sl Chandra} (Walton et al. 2015). Specifically, the 
relativistic disk reflection features were fit with standard relativistic Fe~K$\alpha$ models to infer inclination angles and spin parameters of $i = {{15^{\circ}}^{+9^{\circ}}_{-15^{\circ}}}$ and $a = 0.87^{+0.08}_{-0.15}$ for \rxj\ (Reis et al. 2014) and $i  \simlt 11.5^{\circ}$ and $a =  0.74^{+0.06}_{-0.03}$ for Q2237$+$0305 (Reynolds et al. 2014). 

These studies have not, however, correctly accounted for the effects of gravitational microlensing. Gravitational microlensing is a well studied phenomenon in lensed quasars (e.g., see the review by Wambsganss 2006 and references therein) where stars near the lensed images produce time variable magnification of source components whose amplitude depends on the location and size of the emission region. In particular, in our analysis  of the X-ray spectra of lensed quasars (Chartas et al. 2016), we have frequently observed structural changes in the Fe K$\alpha$ emission indicating that the line emission is being differentially microlensed. Thus, applying the relativistic Fe~K$\alpha$ line method to stacked spectra of lensed quasars, without accurately accounting for microlensing, is likely to lead to unreliable and unrealistic results.

%Detections of relativistic disk reflection features have been claimed in the gravitationally lensed quasars \rxj\ (Reis et al. 2014) and Q2237$+$0305 (Reynolds et al. 2014) and in a stacked spectrum of 27 lensed 1.0 $<$ $z$ $<$ 4.5 quasars observed with {\sl Chandra} (Walton et al. 2015). These reflection features were fit with standard relativistic Fe~K$\alpha$ models to infer inclination angles and spin parameters of $i = {{15^{\circ}}^{+9^{\circ}}_{-15^{\circ}}}$ and $a = 0.87^{+0.08}_{-0.15}$ for \rxj\ (Reis et al. 2014) and $i  \simlt 11.5^{\circ}$ and $a =  0.74^{+0.06}_{-0.03}$ for Q2237$+$0305 (Reynolds et al. 2014). These studies have not correctly accounted for the effects of gravitational microlensing.  Gravitational microlensing is a well studied phenomenon in lensed quasars (e.g., see the review by Kochanek 2006 and references therein) where stars near the lensed images produce time variable magnification of source components whose amplitude depends on the location and size of the emission region. In particular, in our analysis  of the X-ray spectra of lensed quasars we have frequently observed structural changes in the Fe K$\alpha$ emission indicating that the line emission is being differentially microlensed. Thus, applying the relativistic Fe~K$\alpha$ line method to stacked spectra of lensed quasars is likely to lead to unreliable and unrealistic results.  

In Section~2 we present the X-ray observations and analyses of the {\sl Chandra} observations of \rxj, \qj, and \sdss.
In Section~3 we discuss our recently developed technique based on microlensing to provide a robust constraint on the inclination angle, the location of the ISCO, and the spin parameter, and present an analytic estimate
of the fractional energy shifts $g$~=~$E_{\rm obs}$/$E_{\rm rest}$  and numerical simulations of microlensing events.
In Section~4 we present our results from modeling the observed distribution of $g$ and the distribution of the measured energy separations of shifted Fe~K$\alpha$ lines in cases where two shifted lines are detected in an individual spectrum.
%In Section~4 we employ numerical simulations to evaluate if the microlensing of the X-ray emission from an accreting supermassive black hole can indeed produce energy spectra similar to those observed. 
Finally, in Section~5 we rule out several alternative scenarios to explain the shifted iron lines and present a summary of our conclusions.
Throughout this paper we adopt a flat $\Lambda$ cosmology with 
$H_{0}$ = 67~km~s$^{-1}$~Mpc$^{-1}$ $\Omega_{\rm \Lambda}$ = 0.69, and  $\Omega_{\rm M}$ = 0.31 (Planck Collaboration et al. 2013).

\section{X-RAY OBSERVATION AND DATA ANALYSIS}
We  have performed multiwavelength monitoring of several gravitationally lensed quasars 
%(Chartas et al. 2009, 2012, 2016; Chen et al. 2011, 2012; Morgan et al. 2006, 2008, 2010, 2012; Dai et al. 2010; Mosquera et al. 2009, 20011, 2013; Blackburne et al. 2006, 2011, 2014, 2015;  MacLeod et al. 2015)
(e.g., see Table~1 of Chartas et al. 2016 and references within) with the main scientific goal of measuring the emission structure near the black holes in the optical, UV, and X-ray bands in order to test accretion disk models.  The X-ray monitoring observations were performed with the {\sl Chandra X-ray Observatory}  (hereafter {\sl Chandra}). The Optical (B, R and I band) observations were made with the SMARTS Consortium 1.3m telescope in Chile. The UV observations were performed with the {\it Hubble Space Telescope}.
In this paper we focus on constraining the inclination angles, ISCO radii, and spin parameters  of quasars \rxj, \qj, and \sdss\ using the {\sl Chandra} observations of these  objects.

\rxj\ (hereafter RXJ1131) was observed with the Advanced CCD Imaging Spectrometer (ACIS; Garmire 2003) on board 
the {\sl Chandra X-ray Observatory}
38 times between 2004 April 12 and 2014 July 12. 
Results from the analysis of observations 1--6 of RXJ1131 were presented in 
Chartas et al. (2009), and  Dai et al. (2010) and those for observations 7--29 are in Chartas et al. (2012). 
Results from the analysis of a subset of these observations have also been presented in Blackburne et al. (2006), Kochanek et al. (2007), and Pooley et al. (2012).
Here we describe the data analysis for the remaining 9 observations, 
but we will use the results from all 38 observations in our microlensing analysis.
\qj\ (hereafter QJ0158) was observed with ACIS 12 times between 2010 November 6 and 2015 June 10, and  \sdss\ (hereafter SDSS1004)
was observed with ACIS 10 times between 2005 January 1 and 2014 June 2.
Results from the analysis of the first 6 observations of QJ0158, and the first 5 observations of SDSS1004
were presented in Chen et al. (2012).

We re-analyzed all the {\sl Chandra} observations of RXJ1131, QJ0158, and SDSS1004  using the software CIAO 4.8 with CALDB version 4.7.2,
provided by the {\sl Chandra X-ray Center} (CXC).  
A log of the observations for each object in our study that includes observation dates, observation identification numbers, exposure times, ACIS frame times 
and the observed 0.2--10~keV counts are presented in Tables 1, 2, and 3.
We used standard CXC threads to screen the data for status, grade, and time intervals of acceptable aspect solution and background levels. 

In Figures 1, 2, and 3 we show the 0.2--10 keV light-curves of the images of RXJ1131, QJ0158, and SDSS1004. The light-curves have been shifted by the time-delays estimated by Tewes et al. (2012), Faure et al. (2009), and Fohlmeister et al. (2008), respectively.
Microlensing will affect the images differently resulting in uncorrelated variability between images.  
Large uncorrelated events are noticeably present in the images of 
RXJ1131, QJ0158, and SDSS1004, indicating that the X-ray emission regions are significantly smaller than the projected Einstein radius of the stars and thus affected by microlensing (e.g., Chartas et al. 1995, 2009; Dai et al. 2003, 2010; Blackburne et al. 2006; Pooley 2007; Mosquera et al. 2013)

%suggesting that emission regions with sizes significantly smaller than the projected Einstein radius of the stars will be strongly affected by microlensing whereas emission regions with sizes significantly larger will not be affected.

%\subsection{\sl Determining the Significance of the Fe~K$\alpha$ Lines in the Spectra of the Individual Images in Single Epochs}

For the spectral analyses, we followed the approach described in Chartas et al. (2012). 
We extracted events from circular regions with radii of 1.5~arcsec
slightly off-center from the images to reduce contamination from nearby images. 
The backgrounds were determined by extracting events within an
annulus centered on the mean location of the images
with inner and outer radii of 7.5~arcsec and 50~arcsec, respectively. 
Spectral fits were restricted to events with energies between 0.4--10~keV.
Spectra with fewer than $\sim$~200~counts were fit using the $C$-statistic (Cash 1979)\footnote{The spectra were binned to have at least one count per bin.}, 
as appropriate for fitting spectra with low S/N. Spectra with larger number of counts were fit using both the $C$ and $\chi^{2}$ statistics.

We fit the {\sl Chandra} spectra for each epoch with a model that consists of a power law with neutral intrinsic absorption at the redshift of the source. 
Galactic column densities in the directions of RXJ1131, QJ0158, and SDSS1004 were fixed to N$_{H}$ = 3.60 $\times$ 10$^{20}$ cm$^{-2}$, 
N$_{H}$ = 1.88 $\times$ 10$^{20}$ cm$^{-2}$, and  N$_{H}$ = 1.13 $\times$ 10$^{20}$ cm$^{-2}$, respectively (Dickey \& Lockman 1990). 
Measurements of the differential X-ray absorption between images in lensed quasars SBS~0909+523, FBQS~0951+2635, and B~1152+199 by Dai \& Kochanek (2009) have been used to successfully constrain the dust-to-gas ratio of the lens galaxies and we plan to present the application of this method to our lens sample in a future paper.  We next added one or two Gaussian emission lines to the model and tested for the significance of the added line/s. 
The significance of the emission lines was determined by varying the energy and width of the iron line and its flux 
to calculate the $\chi$$^{2}$ of the fit as a function of the  Fe line energy and iron line flux. 
In several cases two emission lines are detected in a single spectrum.
We record all cases where one or more emission lines are detected above the 90\%, and 99\% confidence levels.
Note that all these lines are detected in single epochs and not from stacked observations.

The confidence levels between the iron line flux and energy were created using the \verb+steppar+ command in  \verb+XSPEC+.
As pointed out by Protassov et al. (2002), this approach may not apply for models near a boundary such as in cases where the line flux normalization is constrained to have only positive values. To account for this limitation we allowed the line flux normalization to obtain both positive and negative values. Protassov et al. (2002) proposed a more robust approach of estimating the significance of the shifted irons line based on Monte-Carlo simulations to determine the distribution of the $F$-statistic between different models.  We followed this approach and constructed the simulated probability density distribution of the $F$-statistic between spectral fits of models that included a simple absorbed power-law (null model), and one that included one or two Gaussian emission lines (alternative model). Specifically, for each observed spectrum we simulated 1000 data sets using the XSPEC \verb+fakeit+ command. We fit the null and alternative models to the 1000 simulated data sets and computed the $F$-statistic for each fit. Finally form the Monte-Carlo simulations we computed the probability of obtaining an $F$ value larger than the one obtained from the fits of the null and alternative models to the observed spectrum. In tables 5, 6, and 7 we provide the $F$-statistic  between the null and alternative models and the probability of exceeding this value as determined from the Monte-Carlo simulations.

In Figure 4 we show a typical example of the Monte Carlo simulated distribution of the $F$-statistic between fits of the null and alternative models to the observed spectrum of image C of RXJ1131 obtained in November  28, 2009 (obsid~=~11540). We find that in this spectrum the probability of obtaining an $F$ value larger than 2.69 is P = 0.018.
In all cases listed in Tables 5, 6, and 7 we  confirm the significance of the shifted lines and in all cases the significance inferred from the Monte-Carlo analysis is similar or larger than the lower limits provided by the 90\% and 99\% $\chi^{2}$ confidence contours.

In Figures 5, 6, and 7 we show typical examples of Fe~K${\alpha}$ lines detected in the spectra of individual images and epochs from RXJ1131, QJ0158, and SDSS1004, respectively. We also show the respective $\chi^{2}$ contours of the detected lines. 
Tables 5, 6, and 7 provide the line and continuum properties for all these detections.
For RX J1131 we have 78 line detections out of the 152 spectra (38 epochs $\times$ 4 images) at $>$~90\% confidence, of which 21 lines are detected at $>$~99\% confidence. 
For the 6  {\sl Chandra} observations of QJ0158 with exposure times of $\sim$19~ks, we detect 10 iron lines in 12 spectra (6 epochs $\times$ 2 images) at $>$~90\% confidence, of which 3 iron lines are detected at $>$~99\% confidence. 
For the 10 {\sl Chandra} observations of SDSS1004, %with exposure times of $\sim$ 25~ks 
we detect the iron line in 8 out of the 40 spectra (10 epochs $\times$ 4 images) at $>$~90\% confidence, of which 6 iron lines are detected at $>$~99\% confidence.
For several closely separated observations we detect energy shifts of the Fe~K${\alpha}$ line in consecutive and closely separated epochs most likely produced by the same caustic crossing.  For example, Figure 6 shows a likely caustic crossing in QJ0158 during December 2013.% and  7 show a likely  caustic crossing in RXJ1131. 

%\subsection{\sl Continuum and Line Variability of RXJ1131 During Microlensing Events}

In Figure 8 we show the generalized Doppler shift parameter $g$ of the Fe line as a function of the equivalent width (EW) of the iron line in RXJ1131.  We find a correlation between $g$ and the EW with a Kendall's rank correlation coefficient of $\tau$ = 0.28 significant at $>$~99.9\% confidence. One possible explanation of this correlation is that blueshifted line emission is Doppler boosted, resulting in the observed equivalent widths of the blueshifted lines being larger than the redshifted lines.

In Figure 9 we show the flux of the Fe line as a function of the flux of the continuum, where 
the line flux is the normalization of the Gaussian line component of the best-fit model, and 
the continuum flux is the normalization of the power-law component of the best-fit model calculated at 1~keV.
We find a strong correlation between the line and continuum fluxes, with a Kendall's rank correlation coefficient of $\tau$ = 0.5 that is  significant at $>$~99.9\% confidence. 
The relative flux of the X-ray continuum to the Fe~K$\alpha$ line flux depends on a variety of accretion disk parameters and 
geometries including the emissivity profile of the disk, the distance of the caustic from the black hole, the geometry of the corona, the geometry of the disk emission, the caustic crossing angle and the inclination angle. Variations of the X-ray continuum and the Fe~K$\alpha$ line flux during caustic crossing have been simulated in Popovic et al. (2006) for a variety of accretion disk parameters and geometries. These simulations indicate that once the magnification caustic has passed over the black hole, the microlensing magnifications of the line and continuum regions are similar and both the line and continuum decay in a similar manner with distance from the black hole. 
%The similar magnifications of the line and continuum emission that is expected once the caustic has crossed the center of the black hole 
This may explain part of the observed correlation between these quantities.

\section{The $g$- and ${\Delta}E$-DISTRIBUTIONS}

We have recently developed a new technique based on microlensing that provides a robust constraint on the disk inclination angle and on the location of the ISCO, which in turn may provide an estimate of the spin of the black hole. Our technique is very simple. The stars near each lensed image produce magnification patterns with a characteristic Einstein radius of

\begin{equation}
{R_{\rm E} =  D_{\rm OS} \left[  {{4G \langle M \rangle }\over{c^{2}}}  { {D_{\rm LS}}\over{D_{\rm OL}D_{\rm OS}   }}  \right ]^{1/2}}
\end{equation}

\noindent
where $\langle M \rangle$ is the mean mass of the lensing stars, the $D_{\rm ij}$ are the angular diameter distances, and the subscripts L, S, and O refer to the lens, source, and observer, respectively.

These patterns contain caustic curves on which the magnification diverges. As the observer, lens and source move, the quasar experiences a time varying magnification whose amplitude is determined by the size of the source, with larger sources showing lower amplitudes because they more heavily smooth the magnification patterns (e.g., Wyithe et al. 2000, 2002, Kochanek 2004). If the X-ray emission is dominated by the inner edge of the disk, then the characteristic source size is $r_{\rm s}$ $\sim$ 10$r_{\rm g}$ $\sim$ 10$^{15}$($M_{\rm BH}$/10$^{9}$M$_{\odot}$)~cm where $r_{\rm g}$ = G$M_{\rm BH}$/c$^{2}$. The effective source velocity across the pattern for our three quasars ranges between $v_{\rm e}$ $\sim$ 600 -- 785 km/s (see Mosquera \& Kochanek 2011) leading to two characteristic time scales for variability: the Einstein crossing time $R_{E}$/$v_{e}$, typically several years,  and the source crossing time $r_{\rm s}$/$v_{e}$, typically a few months (see Table 1). 
%Our strategy has been very successful, as we have detected X-ray microlensing variability in all 8 lenses we have studied (Morgan et al. 2008, 2012; Dai et al. 2010; Chen et al. 2011, 2012; Mosquera et al. 2013; Blackburne et al. 2014, 2015; MacLeod et al. 2015; Chartas et al. 2009, 2012, 2016).

As a caustic crosses the accretion disk it differentially magnifies the Fe~K$\alpha$ line emission to produce changes in the line profile. We will observe these as shifts in the line energy which we use to calculate the distribution of the fractional energy shifts $g$=$E_{\rm obs}$/$E_{\rm rest}$ (the ``$g$-distribution''). We first present an analytic estimate of the energy shift of the iron line caused by microlensing and compare these analytic estimates with the observed energy shifts. Numerical simulations of microlensing events are presented  later on in the section.
The observed energy, $E_{\rm obs}$, of a photon emitted near the event horizon of a supermassive black hole will be shifted with respect to the emitted rest-frame energy, $E_{\rm emit}$, due to general relativistic and Doppler effects. The ratio between the observed energy and the emitted rest-frame energy is often referred to as the generalized Doppler shift and defined as 

\begin{equation}
g=\frac { { E }_{ obs } }{ { E }_{ emit } } =\delta \sqrt { \frac { \Sigma \Delta  }{\rm A  }  } 
\end{equation}
\noindent
where the Doppler factor $\delta$ is
%and the quantities A,  $\Sigma$, $\Delta$  are given by
\begin{equation}
%\delta =\sqrt { \frac { 1-{ v }_{ \phi  }^{ 2 } }{ 1-{ v }_{ \phi  }\cos { { \theta  }_{ c } }  }  } 
%\delta =  \frac{ \sqrt { 1-{ v }_{ \phi  }^{ 2 } }{ 1-{ v }_{ \phi  }\cos { { \theta  }_{ c } }  }  }
\delta =\frac { \sqrt { 1-{ v }_{ \phi  }^{ 2 } }  }{ 1-{ v }_{ \phi  }\cos { { \theta  }_{ c } }  } 
\end{equation}

\noindent
$A={ \left( { r }^{ 2 }+{ a }^{ 2 } \right)  }^{ 2 }-{ a }^{ 2 }\Delta \sin^{ 2 }\theta$, $\Sigma ={ r }^{ 2 }{ { { +a }^{ 2 }\sin^{ 2 }\theta  } }$, 
and $\Delta ={ r }^{ 2 }{ { -2{ r }_{ g }r+a }^{ 2 } }$, where $\theta_{ c }$ is the angle between the direction of the orbital velocity $v_{\phi}$ of the emitting plasma and our line of sight neglecting the general relativistic effect of the bending of the photon trajectories and relativistic aberration (see appendix for details). In \S 4 we compare our analytic estimate of the generalized Doppler factor $g$ with the observed limits of the $g$-distributions of RXJ1131.

%\section{ NUMERICAL SIMULATIONS OF MICROLENSING EVENTS}

We next use numerical simulations to evaluate if the microlensing of the X-ray emission from a mass accreting supermassive black hole can indeed produce energy spectra similar to the observed ones. The simulations assume that RXJ1131 accretes through a standard geometrically thin,  optically thick accretion disk 
(Shakura \& Sunyaev 1973; Novikov \& Thorne 1973) described by the analytical general relativistic equations of Page \& Thorne (1974).
This assumption seems to be well justified for two reasons. Sluse et al. (2012) estimate that the black hole of RXJ1131 has a mass $M_{\rm BH}$ between 8~$\times~10^7~M_{\odot}$ and $2~\times 10^8~M_{\odot}$ and a bolometric luminosity of $L_{\rm Bol}$ $\approx$ 10$^{45}$ erg~s$^{-1}$. The inferred ratio of $L_{\rm Bol}$/$L_{\rm Edd}$ is therefore expected to range between 0.01 $-$ 0.42.  It is therefore likely that RXJ1131 is accreting in the regime  in which accretion is believed to be dominated by a geometrically thin, optically thick accretion disk (see e.g. the discussion in McKinney et al. 2014).
General relativistic (radiation) magnetohydrodynamical simulations indicate that the analytical equations describe accretion disks reasonably accurately (Noble et al. 2011; Kulkarni et al. 2011; Penna et al. 2012; Sadowski 2016). However, optical/UV observations of microlensing events in quasars indicate that accretion disks are larger than predicted by thin disk theory (Morgan et al. 2010), and similar discrepancies are found using measurements of continuum lags in the nearby Seyferts NGC2617 (Shappee et al. 2014) and NGC5548 (Edelson et al. 2015; Fausnaugh et al. 2016).

We use a general relativistic ray tracing code (Krawczynski 2012; Hoormann et al. 2016; Beheshtipour et al. 2016) to track photons of initially unspecified energy from a lamppost corona (Matt et al. 1991) to the observer, accounting for the possibility that the photons impinge on the accretion disk and either reflect or prompt the emission of  Fe-K$\alpha$ photons. 
The compactness of the corona has been observationally and independently confirmed via microlensing (e.g., Pooley et al. 2007; Chartas et al. 2009, 2016; Dai et al. 2010; Morgan et al. 2008; Mosquera et al. 2013; Blackburne et al. 2014, 2015; MacLeod et al. 2015) and reverberation studies (Fabian et al. 2009; De Marco et al. 2011; Kara et al. 2013, 2016; Cackett et al. 2014; Emmanoulopoulos et al. 2014; Uttley et al. 2014). The reverberation studies in many cases constrain the distance between the corona and central black hole to lie in the range of $3-10r_{g}$.

In our simulations the corona is located above the black hole at a radial Boyer Lindquist  (BL) coordinate $r=5\,r_{\rm g}$ slightly offset from the polar axis ($\theta=10^{\circ}$), and emits isotropically in its rest frame. The 0-component of the wavevector $k^{\mu}$ of the photon packet is proportional to the energies
of the photons, and we assume that the corona emits a power-law SED with a photon number power law of $dN/dE\propto E^{-\Gamma}$  with $\Gamma=1.75$.
After transforming the photon packet's wavevector into the global BL coordinates, the code integrates the geodesic equation until the photon packet either 
comes too close to the black hole horizon (when we assume it will enter the black hole), impinges on the accretion disk,  or arrives at a fiducial stationary observer at $r_{\rm obs}=10,000 \,r_{\rm g}$.

We assume that the accretion disk extends from the innermost stable circular orbit (ISCO) to 100~$r_{\rm g}$. 
We use simple prescriptions for treating the absorption, reflection and reprocessing in the disk's photosphere. If a photon packet hits the disk it is absorbed with probability $p_{\rm abs}$, 
it scatters with probability $p_{\rm es}=(1-p_{\rm abs}) R/(1+R)$, or it prompts the emission of a monoenergetic Fe-K$\alpha$ photon with probability $p_{\rm Fe-K\alpha}=(1-p_{\rm abs})/(1+R)$ 
, so that $p_{\rm es}/p_{\rm Fe-K\alpha}=R$.  Although we use $p_{\rm abs}=0.9$ and $R=1$, the results do not depend strongly on the specific choice (see Ross \& Fabian 2005; Garc{\'{\i}}a et al. 2013, for more detailed treatments). The Fe-K$\alpha$ photons are always emitted with an energy of 6.4~keV in the rest frame of the accretion disk with a statistical weight that depends on the net redshift or blueshift $g$ incurred between the emission of the photon in the corona and its absorption in the accretion disk. 
More specifically, the weight is given by $w=g^{\Gamma-1}$ with
\begin{equation}
g=\frac{u_{\hat{\mu}}k^{\hat{\mu}}}{u_{\check{\mu}}k^{\check{\mu}}}.
\end{equation}
Here $u^{\check{\mu}}$ and $u^{\hat{\mu}}$ denote the four velocities of the corona and the accretion disk plasmas, respectively, and $k^{\check{\mu}}$ and $k^{\hat{\mu}}$ denote the photon packet wave vectors in the rest frames of the corona plasma and accretion disk plasma (before the absorption of the photon packet), respectively. The scattering model is based on the classical treatment by Chandrasekhar (1960) of scattering by an indefinitely deep electron atmosphere. 
Tracking photon packets forward in time allows us to model multiple interactions of the photon packets with the accretion disk before or after a photon packet prompts the emission of a Fe-K$\alpha$ photon packet (see also Schnittman \& Krolik 2009, 2010, and references therein). Our treatment neglects the possibility that Fe-K$\alpha$ photons return to the accretion disk and may prompt 
the emission of another Fe-K$\alpha$ photon, but this is a very small correction. When a photon packet reaches the observer, its wavevector is transformed from the global BL frame into the
reference frame of a coordinate stationary observer.
Although we simulated 10 million photon packets for a range of black hole spins ($a=0, 0.1, 0.2, ..., 0.9, 0.95, 0.98, 0.998$)
we will show here only exemplary results for $a=0.3$. The results for other black hole spins will be presented in a forthcoming paper (Krawczynski et al. 2016).

In this paper we use the general parameterization of the microlensing magnification $\mu$ close to caustic folds above the magnification $\mu_0$ outside of the caustic (see e.g. Schneider et al. 1992; Chen et al. 2013):

\begin{equation}
\mu\,=\,1+\frac{K}{\mu_0 \sqrt{y_{\perp}}}H(y_{\perp})
\label{h:equA}
\end{equation}
with $y_{\perp}$ giving the position of the origin of the emission in the source plane along a coordinate axis 
perpendicular to the fold, $K$ is the caustic amplification factor,
and $H$ is the Heaviside function (i.e. $H(y_{\perp})=0$ for $y_{\perp}<0$ and $H(y_{\perp})=1$  for $y_{\perp}\ge 0$).
For microlensing by a random field of stars $K/\mu_0$ is given by (Witt et al. 1993; Chartas et al. 2002)

\begin{equation}
K/\mu_0\,\approx\, \beta \sqrt{\zeta_{\rm E}}
\end{equation}
where $\beta$ is a constant of order unity and $\zeta_{\rm E}$ is the Einstein radius for stars of average 
mass $\langle M \rangle $. In this paper, we show results for $\beta$ = 0.5 and $\zeta_{\rm E}=1640\,r_{\rm g}$, corresponding to $\langle M \rangle=0.25 M_{\odot}$ 
and $M_{\rm BH}=10^8 M_{\odot}$.
%$\langle M \rangle $
%$<\!\!M\!\!>$
For each simulated black hole spin, we simulate caustic crossings at crossing angles $\theta_{c}$  (the angle between the 
normal of the caustic fold and the black hole spin axis) between 0 and $360^{\circ}$ in $20^{\circ}$ steps and for 
$-30$~$r_{\rm g}$ to +30~$r_{\rm g}$ offsets of the caustic from the center of the black hole.
An angle of $\theta_{c}=0$ corresponds to a caustic fold perpendicular to the black hole spin axis 
with the positive side of the caustic (the side with $\mu>1$) pointing in the $\theta=0$ direction.

The left panel of Figure 10 shows a 2-D map of the surface brightness of the Fe-K$\alpha$ line emission
in the source plane after convolving it with a caustic magnification ($a=0.3,\,i=82.5^{\circ},\,\theta_{c}=\pi/2$).
An actual observer would see a different surface brightness distribution, as Figure 10 accounts 
for the flux magnification but not for the image distortion caused by the gravitational lensing.  
The magnification pattern of Equation (\ref{h:equA}) can clearly be recognized by a sudden jump from $\mu=1$ to
$\mu\gg1$ followed by the gradual return of the magnification to $\mu\approx 1$.  
The right panel shows the resulting Fe-K$\alpha$ energy spectrum. Similar to the {\sl Chandra} energy spectra shown in Figure 5, the simulated energy spectrum exhibits two distinct peaks. Scrutinizing similar maps and energy spectra 
for different caustic crossing angles and offsets reveals that double peaks appear naturally (but not only) 
when the positive side of the caustic magnifies lower surface brightness emission which is not as 
strongly Doppler boosted as the emission from the portions of the accretion disk approaching the observer
with near-relativistic speed.
%with a speed close to the speed of light. 

Figure 11 shows a 2-D map of the $g$-factors of the 
Fe-K$\alpha$ photons for the same parameters as Figure 10. The image clearly shows the highest blueshifts 
from regions of the accretion disk moving towards us with near-relativistic speed.

%a speed close to the speed of light. 
Note that the  gravitational redshift wins over the Doppler blueshift very close to the event horizon.  
The minimum and maximum $g$-values are 0.42 and 1.4, respectively, and are in good agreement 
with the observed values. We have explored the dependence of the observed $g$-values on the coronal height. We find that for larger coronal heights the $g$-distribution becomes narrower, as the inner edge of the disk is not illuminated as much as it is for low coronal heights, and thus the extreme $g$-factor emission is less intense. We conclude that for larger coronal heights the observed $g_{max}$ value for RXJ1131 would imply even higher inclinations.

We analyzed the energy spectra for all caustic crossing angles and offsets with an algorithm identifying the most prominent peaks and fitting them with Gaussians. The algorithm first finds the highest peak in the photon number
energy spectrum ($dN/dE$) and then searches for an additional peak which rises more than $10$\% 
above the valley between the two peaks. 
Figure 12 shows the distribution of all the peak energies  found in this way (singles and doubles).
%which closely resembles the observed distribution of Figures 7 and 8. 
Figure 13 presents all simulated and observed double peak energies.
Although a statistical analysis of the results is outside the scope of this paper, we will see in \S 4 that the simulated results
do resemble the observed ones at least on a qualitative level.

\section{RESULTS}

%Results from g distribution method

We detect shifted and broadened  Fe~K$\alpha$ lines in almost every image of RXJ1131 that has an exposure time of $>$20~ks as shown in Figure 5 and Table 5. %(Chartas et al. 2012, 2016). 
Relativistically broadened Fe~K$\alpha$ lines detected in the spectra of unlensed AGN are produced from emission originating from the entire inner accretion disk. In contrast, the microlensed Fe~K$\alpha$ lines in the spectra of lensed AGN are produced from a relatively smaller region on the disk that is magnified as a microlensing caustic crosses the disk. We therefore expect microlensed Fe~K$\alpha$ lines to be in general narrower and with larger equivalent widths that those detected in unlensed AGN. The observed equivalent widths of Fe~K$\alpha$ lines in RXJ1131 are significantly affected by microlensing of both the direct continuum emission of the corona and the reflected line and continuum from the disk. The relative microlensing magnification of the direct and reflected emission were simulated in Popovic et al. (2006) for a variety of accretion disk parameters and geometries. Chen et al. (2012) compare the properties of a sample of lensed quasars with non-lensed ones and find that the equivalent widths of the Fe lines detected in lensed quasars are systematically higher than those found in non-lensed ones, and this can be explained as the result of microlensing of both the continuum and line emission.

The presence of these energy shifts is evidence that most of the shifted iron line emission detected in RXJ1131 does not originate from reflection from a torus or other distant material but from material near the event horizon of the black hole. The evolution of the energy and shape of the Fe~K$\alpha$ line during a caustic crossing depends on the ISCO, spin, inclination angle of the disk and caustic angle. The extreme shifts are produced when the microlensing caustic is near the ISCO of the black hole. Measurements of the distribution of the fractional energy shifts $g$~=~$E_{\rm obs}$/$E_{\rm rest}$ of the Fe~K$\alpha$ line due to microlensing therefore provide a powerful limit on the $g$-distribution that can be used to  estimate the ISCO, spin and inclination angle of the disk.\\

In Figure 14 we show the $g$-distribution (multiplied by the rest-frame energy of the Fe~K$\alpha$ line, $E_{\rm rest}$  = 6.4~keV) of the Fe~K$\alpha$ lines with $>$ 90\% and $>$99\% detections  in RXJ1131 for all images and epochs. One important feature of this iron line energy-shift distribution is the significant limits of the distribution at rest-frame energies of $E_{\rm min}$~=~3.78$_{-0.10}^{+0.16}$~keV and $E_{\rm max}$~=~8.24$^{-0.48}_{+0.15}$ keV, where the the error bars for both $E_{\rm min}$ and $E_{\rm max}$ are at the 90\% confidence level, for iron lines detected at $>$ 90\% confidence. These limits represent the most extremely redshifted and blueshifted Fe~K$\alpha$  lines. If we interpret the largest energy-shifts as being due to X-ray emission originating close to the ISCO, we obtain upper limits on the size of the ISCO and inclination angle of RXJ1131. In Figure 15  we show the $g$-distribution for the individual images of RXJ1131. The apparent differences of the distributions between images can be a result of several factors including the differences in the frequency of microlensing along different lines of sight, the differences in the S/N of the spectra (D being the faintest image) and differences in caustic crossing angles.  The observed $g$-distributions of Figures 14 and 15 
closely resemble the simulated distribution of all peaks shown in Figure 12.

The energy of photons emitted from a ring at the ISCO are bounded by maximum and minimum values,  $g_{\rm max}$ and $g_{\rm min}$,  of the generalized Doppler shift. The maximum blueshift of the Fe line places a constraint on the inclination angle.  Specifically, for RXJ1131 the measured generalized Doppler factor $g_{\rm max}$ = 1.29 $\pm$ 0.04 (90\% confidence) constrains the inclination angle to be $\simgt$ 64$^{\circ}$, for any values of the spin parameter and caustic crossing angle, as shown in Figure 16, assuming that the thin accretion disc extends all the way down to the ISCO. 
If we also require that the measured $g_{\rm min}$ and $g_{\rm max}$ shifts are produced by photons emitted from the same radius, then the  inclination angle is required to be $i$~$\simgt$~76$^{\circ}$ for photons emitted from a radius of $r$ $\sim$ 8.5 $r_{g}$ for any values of the spin parameter and caustic crossing angle, as also shown in Figure 16.  We conclude that the current observed values of $g_{\rm min}$ and $g_{\rm max}$ place an upper limit on the ISCO radius of $r_{\rm ISCO}$~$\simlt$~8.5$r_{\rm g}$ and a disk inclination angle of $i$ $\simgt$  76$^{\circ}$. 

The $g_{\rm min}$ value provides an estimate of the radius of the emission region closest to the center of the black hole. If one assumes that the innermost emission region is near the ISCO radius one can obtain a constraint on the spin of the black hole based on the relation between ISCO and spin. However, recent 3D MHD simulations of thin accretion disks indicate slight bleeding of the iron line emission to the region inside of the ISCO (e.g., Reynolds \& Begelman 1997). Reynolds \& Fabian (2008) have attempted to estimate the systematic error on inferred black hole spin using the relativistic iron line method to infer the true spin of a black hole. They find that the systematic errors can be significant for slowly spinning black holes but become appreciably smaller as one considers more rapidly rotating black holes. In a future paper we plan to perform an analysis of the systematic errors on inferred black hole spin derived from the $g$-distribution method and from the distribution of energy separations of double peaked lines caused by the possible bleeding of the iron line emission to the region inside of the ISCO.

We also detect energy shifts in the Fe line in QJ0158 and SDSS1004, but they have not been monitored as frequently as RXJ1131. In Figure 17 we show the $g$-distributions of the Fe~K$\alpha$ lines with $>$~90\% confidence detections in QJ0158 and SDSS1004  for all images and epochs.  These $g$-distributions of QJ0158 and SDSS1004 are too sparsely populated to provide statistically significant constraints on the ISCO radii with the available data
but demonstrate that these lenses also have microlensed Fe emission.

In several observations we detect two shifted Fe lines, as shown in Figures 6 and 7. Specifically, the number of detected double lines in the spectra of images A, B, C, and D of RXJ1131 are 9, 5, 3, 0, respectively. All the cases are detected at the $>$ 90\% confidence level. In the following, we call energy spectra with one peak ``singles'' and energy spectra with two peaks ``doubles''.
We searched for possible correlations between the energies of the doubles. A moderately significant correlation, with a Kendall's rank correlation coefficient of $\tau$ = 0.6, and a significance of $>$~98\% confidence,  is detected between the observed $E_{\rm min}$ and $E_{\rm max}$ values in image A.  The observed energies of the double lines detected in the spectra of image A and the spectra of all images are shown in Figures 19 and 20, respectively. 
We performed a straight-line least-squares fit to the data with the FITEXY routine (Press et al. 1992), which takes into account errors in both coordinates.
Based on the magnification maps of \rxj\ (see Figures 4-6 of Dai et al. 2010) we expect a limited range of caustic crossing angles over $\sim$ 10 years. 
Any change in the caustic crossing angle over time will contribute to the scatter in the observed $E_{\rm min}$ and $E_{\rm max}$ values of doubles.
The scatter in the  $E_{\rm min}$ and $E_{\rm max}$ values detected in all images is even larger, as we would expect from the differences in the caustic structures and their directions of motion in different images (Kendall's $\tau$ = 0.4, significant at $>$~97\% confidence).

In Figure 20 we show the ${\Delta}E$$-$distribution of the rest-frame energy separations for the Fe line pairs detected in all images of RXJ1131 at the $>$ 90\% confidence level, where ${\Delta}E$ = $E_{\rm max} - E_{\rm min}$ and  $E_{\rm max}$, $E_{\rm min}$ are the rest-frame energies of the
shifted lines. As we discussed in \S3, our numerical simulations show that the peak energy of the ${\Delta}E$$-$distribution depends strongly on the spin parameter.
Specifically,  our simulations indicate that the ${\Delta}E$-distributions of \rxj\ for input spin parameters of $a$ = 0 and $a$ =  0.98  peak at 1.8 $\pm$ 0.2~keV and  3.2 $\pm$ 0.2~keV, respectively.
We find that the maximum of the observed  ${\Delta}E$-distribution of \rxj\ peaks at $\sim$ 3.5 $\pm$ 0.2~keV, implying a spin parameter of $a$~$\simgt$~0.8.  
Additional observations of RXJ1131 will provide more representative and complete $g$ and  ${\Delta}E$-distributions and place tighter constraints on the disk inclination angle,
the ISCO radius and the spin.

\section{DISCUSSION AND CONCLUSIONS}

Our systematic spectral analysis of all the available {\sl Chandra} observations of  lensed quasars RXJ1131, QJ0158, and SDSS1004 has revealed the presence of a significant fraction of lines blueshifted and redshifted with respect to the energy of the expected Fe~K$\alpha$ fluorescent line. We interpret these energy shifts as being the result of ongoing microlensing in all the images. 
This seems logical given the prior detections of microlensing of the optical/UV (e.g., Blackburne et al. 2006; Morgan 2008; Fohlmeister et al. 2007; Fian et al. 2016, Motta et al. 2012) and X-ray continuum (e.g., Chartas et al. 2009, 2012, 2016; Dai et al. 2010; Chen et al. 2012) in all three sources.
We consider several alternative scenarios and examine whether they can explain the observed shifted iron lines.

(a) {\sl Non-microlensed emission from hot-spots and patches from an inhomogeneous disk.}\\
Redshifted Fe emission lines have been reported in observations of a few bright Seyfert galaxies (i.e., Iwasawa et al. 2004; Turner et al. 2004, 2006; Miller et al. 2006; Tombesi et al. 2007).
Correlated modulation of redshifted Fe line emission and the continuum were reported in NGC 3783 (Tombesi et al. 2007).
Specifically, the spectrum of NGC 3783 shows, in addition to a core Fe~K$\alpha$ line at 6.4~keV, a weaker redshifted wing and redshifted Fe emission line component.
The redshifted line and wing appear to show an intensity modulation on a 27~ks timescale similar to that of the 0.3-10~keV continuum.
Tombesi et al. (2007) argue that the lack of Fe line energy modulation disfavors the orbiting flare/spot interpretation for NGC 3783.
We note that the relative intensity of the core Fe line to the redshifted Fe line component in NGC 3783 is about a factor of 9.
The core Fe~K$\alpha$ line has an equivalent width of about 120~eV and the redshifted line of $\sim$13 eV.
Redshifted lines are also reported to be present in the spectrum of Mrk 766 (Turner et al. 2004; 2006), where 
a weak component of the Fe line with an equivalent width of 15$^{+6}_{-5}$~eV  shows a periodic variation of photon energy.
The proposed scenario by Turner et al. (2006) is that the energy variation is caused by a hot spot on the disk within $\sim$ 100~$r_{\rm g}$ orbiting with a period of about $\sim$~165~ks.
The average spectrum of Mrk~766 shows a broad iron line centre near 6.7~keV with an equivalent width of about 90~eV (possibly from reflection off an ionized disk)
and a narrower component at 6.4keV (possibly from reflection from distant material).

The shifted Fe lines detected in our gravitationally lensed quasar sample have very different properties from those reported in NGC 3783 and Mrk 766 while our our microlensing interpretation is consistent with the observed properties of the energy shifted lines.
Specifically,  the equivalent widths of the energy shifted lines in the lensed quasar sample range from EW = 500~eV to 3,000~eV compared to the EW=13~eV and EW=15~eV in the shifted lines detected in NGC 3783 and Mrk 766, respectively.  Non-microlensed emission from hot-spots and patches from an inhomogeneous disk lie below the detection threshold of the individual spectra of the lensed quasars.
The non-microlensed emission from the Fe~K$\alpha$ line at 6.4~keV is not detected in individual spectra of \rxj, but is detected in the stacked spectra of \rxj\ with 
E = 6.36$^{+0.07}_{-0.08}$~keV and EW = 154$^{+70}_{-80}$~eV (Chartas et al. 2012, 2016).  In the cases of Seyfert galaxies NGC~3783 and Mrk~766  the redshifted weak Fe lines are always accompanied by a
significantly stronger core component near 6.4~keV whereas this is not the case for the lensed quasars.
Another point supporting the microlensing interpretation is the detection of a significant number of double lines in \rxj, as predicted in our numerical simulations of magnification caustics crossing an accretion disk. 
The intensities of these doubles are not consistent with non-microlensed Fe emission from hot-spots and patches from an inhomogeneous disk.

(b)  {\sl Possible intrinsic absorption mimicking two apparent lines in \rxj.} \\
Our analysis of the 4$\times$38 spectra of \rxj\ does not find any significant intrinsic absorption of the continuum spectra (Chartas et al 2012)
and there are no absorption lines detected in the spectra that contain doubles.

(c) {\sl Possible ionization of accretion disk.} \\ 
In Figure 9 of Chartas et al. (2012)  we showed a stacked spectrum of image C of RXJ1131 covering a period of about seven years filtering for epochs where microlensing was not significant in this image.  A significant Fe~K$\alpha$ line is detected in the stacked spectrum of image C at an energy of 6.36$^{+0.07}_{-0.08}$~keV with an equivalent width of 154$^{+70}_{-80}$~eV. This detected energy of the Fe~K$\alpha$  line in image C is consistent with the presence of a  non-ionized disk in RXJ1131. The expected energy of the iron line for an ionized disk would be 6.67keV (He-like Fe) or 6.97keV (H-like iron). 
We conclude that the variability in the flux and energy of the iron line detected in the {\sl Chandra} observations of RXJ1131, QJ0158, and SDSS1004  is due to ongoing microlensing of all the images.

Large magnification events are typically inferred from the departure of the time-delay corrected flux-ratios of images from a constant and images that show significant uncorrelated variability are significantly affected by microlensing. We note, however, that a relatively large number of the shifted Fe lines were found in images that do not show significant variability of their flux ratios or significant uncorrelated variability. This implies that the line profiles found using stacked spectra taken over multiple epochs, even when excluding spectra from images that show variability of their flux ratios, are also distorted by microlensing and that applying the relativistic Fe~K$\alpha$ method as used to analyze local Seyferts will not lead to reliable results.  Any apparent broadening of the iron line in a stacked spectrum of a lensed quasar is a combination of both microlensing and the relativistic blurring seen in unlensed Seyferts.
The average stacked spectrum of a microlensed  quasar differs significantly, especially near the Fe K$\alpha$ line, from that of an unlensed one.
The main reason for this difference is that the reflection and direct(coronal) components of a stacked spectrum of a microlensed quasar are not a simple uniform magnification of the reflection and direct components of a unlensed quasars.
Caustic magnification patterns (e.g. see Figures $12-16$ of Kochanek  2004)  move along a certain direction with respect to the source (that differs between images) and the caustic magnification is not uniform.  Stacked spectra of lensed quasars will thus not result in uniformly magnified reflection and direct spectral components  but will produce spectra that have been selectively magnified by caustics moving along a limited range of caustic crossing angles. The caustic magnification factor K will also vary between caustic crossings making the stacked spectrum deviate even more than a uniformly magnified quasar spectrum.

% \section{CONCLUSIONS}
We summarize our main conclusions from the X-ray observations of RXJ1131, QJ0158, and SDSS1004 as follows:

(1) Redshifted and blueshifted Fe lines with rest-frame EWs ranging between 500~eV and 3,000~eV are detected in the individual epoch spectra of lensed quasars RXJ1131, QJ0158, and SDSS1004.
We interpret these energy shifts as the result of microlensing of Fe line emission within $\sim$ 20 $r_{\rm g}$ of the black hole.

(2)  The $g$-distribution of the observed energy shifts in RXJ1131 is compared to analytic and numerical models and both models provide similar constraints on accretion disk parameter. Specifically, 
the maximum value of $g_{\rm max}$= 1.29$^{+0.04}_{-0.04}$ constrains the inclination angle to be $i \simgt 76^{\circ}$ and the minimum value $g_{\rm min}=0.59$  constrains  $r_{\rm ISCO} \simlt 8.5 r_{g}$. 
One of the strengths of the $g$-distribution method is that the energies of the shifted Fe lines are more robustly detected than the extreme weak red wings of the relativistic Fe~K$\alpha$ line 
that are found mostly in nearby Seyfert galaxies. For example, the energies of the shifted lines are not sensitive to the modeling of the underlying continuum, while, the shape of the relativistic Fe~K$\alpha$ red wing is very sensitive to the continuum model. The $g$-distribution method can be applied to infer the inclination angles, ISCO radii and spins of distant quasars where the relativistic Fe~K$\alpha$ red wing is typically to faint to constrain these parameters in distant quasars. One of the weaknesses of the $g$-distribution method is that there are additional model parameters related to describing the gravitational lens.

(3) Several spectra show two shifted Fe lines, which we refer to as doubles. The peak energies of the doubles are moderately correlated. 
Our numerical simulations reproduce the double lines during caustic crossings and we find that the distribution of the
separations of the peak energies is strongly dependent on the spin parameter.
The maximum of the observed  ${\Delta}E$-distribution of \rxj\ peaks at $\sim$ 3.5 $\pm$ 0.2~keV suggesting a high spin parameter.
Although a statistical analysis of the results still needs to be performed, inspection of the mean energy, peak energy, and shape of the distribution indicates a rather high value of the spin $a \simgt 0.8$.
The available ${\Delta}E$-distribution of \rxj\  is sparsely populated and additional observations are required to better constrain the peak energy of this distribution and infer the spin parameter more accurately.

(4) We find several correlations in the microlensed spectra of RXJ1131 with results summarized in Table 8.
Specifically, we find a  correlation between the rest-frame equivalent width of the iron lines and the generalized Doppler shift parameter $g$ of the iron line ($\tau$~=~0.28, $P > 99.9\%$ ).  The flux of the shifted Fe line is found to be correlated with the flux of the continuum for Fe K$\alpha$ lines detected at $>$ 90\% confidence in all images of RXJ1131 ($\tau$~=~0.5, $P > 99.9\%$ ). The energies of the doubles in image A are also found to be correlated ($\tau$~=~0.6, $P > 98\%$ ).

(5) Our numerical simulations of microlensing caustic crossings reproduce the observed distribution of energy of single and double shifts in microlensed spectra, and predict the correlations observed in the spectra.

Scheduled future monitoring observations with {\sl Chandra} of the lensed quasars RXJ1131,  QJ0158, SDSS1004, and Q 2237+0305 with sufficiently long exposure times to improve the significance of the detections will provide more representative and complete distributions of the generalized Doppler shift $g$ values of the Fe~K$\alpha$ line in these objects. The $g$-distribution method and modeling of doubles is expected to provide robust constraints on inclination angle, the ISCO radii and spins of the black holes of these distant lensed quasars. 
With the Large Synoptic Survey Telescope coming online in the near future we expect $\sim$ 4,000 new lensed systems to be discovered, opening up the possibility of measuring black hole and accretion disk parameters over a wide range of redhsifts and quasar Eddington ratios  $L_{Bol}/L_{\rm Edd}$.

\acknowledgments
We acknowledge financial support from NASA via the Smithsonian Institution grants SAO GO4$-$15112X, GO3$-$14110A/B/C, GO2-13132C, GO1-12139C, and GO0-11121C. CWM gratefully acknowledges support from NSF Award AST-1211146. CSK is supported by NSF grant AST-1515876.

\clearpage
\appendix

\section{Calculating the Generalized Doppler Shifts $g$} %\label{App:AppendixA}

The generalized Doppler shifts $g$ were calculated for the case of a spinning black hole (Kerr 1963) using the formalism described 
in Bardeen et al. (1972), Muller \& Camenzind (2004), and Karas \& Sochora (2010). The azimuthal velocity is

\begin{equation}
{ v }_{ \phi  }=\tilde { \Omega  } \frac { \left( \Omega -{ \Omega  }_{ 0 } \right)  }{ \sqrt { \frac { \Delta \Sigma  }{ A  }  }  } 
\end{equation}

\noindent
where $\tilde { \Omega  }$ and ${ \Omega  }_{ 0 }$ are given by

\begin{equation}
\tilde { \Omega  } =\sqrt { \frac { A  }{ \Sigma  }  } \sin { \theta  }  \hspace{0.3cm}{\rm  and}
\end{equation}

\begin{equation}
{ \Omega  }_{ 0 }=\frac { 2a }{ A } { r }_{ g }r
\end{equation}

\noindent
For $r$ $>$ $r_{\rm ms}$ a Keplerian velocity profile is assumed and 

\begin{equation}
{ \Omega  }=\frac { { { r }_{ g }^{ 1/2 } } }{ { r }^{ 3/2 }+a{ r }_{ g }^{ 1/2 } } 
\end{equation}

\noindent
For $r$ $\leq$ $r_{\rm ms}$ constant specific angular momentum is assumed and the angular velocity $\Omega$ is given by

\begin{equation}
{ \Omega  }={ \Omega  }_{ 0 }+\frac { \Delta \Sigma  }{ A  } { \tilde { \Omega  }  }^{ -2 }\frac { { \lambda  }_{ ms } }{ 1-{ \Omega  }_{ 0 }{ \lambda  }_{ ms } } 
\end{equation}

\noindent
where 

\begin{equation}
{ \lambda  }_{ ms }=\frac { { \widetilde { \Omega  }  }_{ ms }^{ 2 }\left( { \Omega  }_{ ms }-{ \Omega  }_{ 0,ms } \right)  }{ \left[ \left( \frac { \Delta \Sigma  }{ A  }  \right) +{ \Omega  }_{ 0,ms }{ \widetilde { \Omega  }  }_{ ms }^{ 2 }\left( { \Omega  }_{ ms }-{ \Omega  }_{ 0,ms } \right)  \right]  } 
\end{equation}

\noindent
All quantities with the subscript ms are calculated at the marginal stable orbit.

\noindent
For these calculations, we assumed the radial and toroidal velocity components  $v_{r}$ and v$_{\theta}$  of the radiating plasma of the accretion disk in the  Zero Angular Momentum Observer frame to be relatively small.
For a given inclination angle $i$, the angle between the direction of the orbital velocity $v_{\phi}$ of the plasma and our line of sight 
is assumed to range between a minimum and maximum value that will depend on the inclination angle $i$, and the 
angle between the caustic direction of motion and the projection of our line-of-sight onto the disk plane.

\clearpage
\begin{table}
\caption{Log of Observations of Quasar \rxj\ }
\scriptsize
\begin{center}
\begin{tabular}{clccrlrrrr}
 & & && & & && & \\ \hline\hline
         &        & & {\it Chandra}                                      & Exposure & & & &      \\
Epoch & Observation & JD${}^{a}$  &  Observation  &   Time   & $t_{\rm f}$${}^{b}$& $N_{\rm A}$$\tablenotemark{c}$ & $N_{\rm B}$$\tablenotemark{c}$ & $N_{\rm C}$$\tablenotemark{c}$ & $N_{\rm D}$ $\tablenotemark{c}$   \\
& Date           &     (days)        &          ID                &   (ks)       &(s)& counts   & counts & counts & counts   \\
&   &    &    &   &  &  & & &\\
\hline
1&2004 April 12             &3108           &  4814     & 10.0 &3.14&       425$_{-22}^{+22}$ & 2950$_{-54}^{+54}$ &  839$_{-29}^{+29}$ & 211$_{-15}^{+15}$  \\
2&2006 March 10          & 3805        &  6913    & 4.9    &0.741&     393$_{-20}^{+20}$  & 624$_{-25}^{+25}$  & 204$_{-14}^{+14}$ &  103$_{-10}^{+10}$    \\
3&2006 March 15           & 3810        &  6912    &  4.4   &0.741&       381$_{-20}^{+20}$  & 616$_{-25}^{+25}$ & 233$_{-15}^{+15}$  & 93$_{-10}^{+10}$  \\
4&2006 April 12               & 3838        &  6914    &  4.9   &0.741&     413$_{-20}^{+20}$   & 507$_{-23}^{+23}$  & 146$_{-12}^{+12}$   & 131$_{-12}^{+12}$  \\
5&2006 November 10    & 4050       &  6915     &  4.8  &0.741&     3708$_{-61}^{+61}$   & 1411$_{-38}^{+38}$ & 367$_{-19}^{+19}$   & 155$_{-13}^{+13}$  \\
6&2006 November 13    &4053        &  6916    &  4.8   &0.741&     3833$_{-62}^{+62}$  & 1618$_{-40}^{+40}$  & 415$_{-20}^{+20}$   & 115$_{-11}^{+11}$  \\
7&2006 December 17     & 4087&7786       & 4.88          &0.841& 3541$_{-99}^{+102}$   & 1443$_{-58}^{+59}$     &417$_{-28}^{+29}$      & 117$_{-13}^{+15}$\\
8&2007 January 01          &4102 &7785       & 4.70              &0.441& 2305$_{-74}^{+75}$    & 1082$_{-49}^{+51}$     & 312$_{-25}^{+26}$     & 108$_{-13}^{+15}$\\
9&2007 February 13        & 4145 &7787       & 4.71             &0.441& 2451$_{-78}^{+79}$    & 1116$_{-49}^{+51}$     & 301$_{-24}^{+25}$     & 169$_{-17}^{+18}$\\
10&2007 February 18      & 4150&7788       & 4.43           &0.441& 2232$_{-73}^{+75}$    &  950$_{-45}^{+47}$     & 252$_{-22}^{+23}$     & 115$_{-14}^{+15}$\\
11&2007 April 16             &4207  &7789       & 4.71                &0.441& 2328$_{-74}^{+74}$    & 1202$_{-53}^{+54}$     & 344$_{-26}^{+27}$     &  99$_{-12}^{+14}$\\
12&2007 April 25              &4216 &7790       & 4.70                &0.441& 2043$_{-70}^{+67}$    & 1063$_{-48}^{+50}$     & 363$_{-26}^{+28}$     & 142$_{-16}^{+17}$\\
13&2007 June 04             &4256 &7791       & 4.66              &0.441& 2079$_{-69}^{+70}$    & 1480$_{-59}^{+60}$     & 373$_{-27}^{+28}$     & 108$_{-13}^{+15}$\\
14&2007 June 11              &4263 &7792       & 4.68              &0.441& 2254$_{-72}^{+73}$    & 1466$_{-56}^{+58}$     & 337$_{-25}^{+26}$     & 129$_{-15}^{+16}$\\
15&2007 July 24               &4306 &7793       & 4.67                &0.441& 1958$_{-68}^{+69}$    & 1324$_{-55}^{+57}$     & 353$_{-25}^{+27}$     &  81$_{-11}^{+12}$\\
16&2007 July 30               & 4312&7794       & 4.67                &0.441& 2725$_{-79}^{+80}$    & 1844$_{-67}^{+69}$     & 496$_{-31}^{+32}$     &  100$_{-12}^{+13}$\\
17&2008 March 16           &4542 &9180       &14.32            &0.741& 5557$_{-117}^{+118}$  & 4347$_{-104}^{+105}$   &1337$_{-51}^{+53}$     & 351$_{-24}^{+25}$\\
18&2008 April 13             &4570 &9181       &14.35               &0.741& 8199$_{-147}^{+147}$  & 5654$_{-117}^{+118}$   &1453$_{-52}^{+52}$     & 377$_{-24}^{+25}$\\
19&2008 April 23             & 4580&9237       &14.31               &0.741& 6786$_{-130}^{+130}$  & 4927$_{-111}^{+112}$   &1279$_{-51}^{+51}$     & 232$_{-20}^{+21}$\\
20&2008 June 01            &4619 &9238       &14.24              &0.741& 4647$_{-106}^{+108}$  & 3252$_{-88}^{+90}$     & 878$_{-41}^{+43}$     & 463$_{-27}^{+28}$\\
21&2008 July 05              &4653    &9239       &14.28               &0.741& 5587$_{-118}^{+119}$  & 3584$_{-94}^{+95}$     &1001$_{-44}^{+44}$     & 635$_{-32}^{+33}$\\
22&2008 November 11  &4782&9240       &14.30       &0.741& 5135$_{-113}^{+115}$  & 3085$_{-85}^{+87}$     & 885$_{-42}^{+44}$     & 488$_{-29}^{+30}$\\
23&2009 November 28  &5164&11540       &27.52      &0.741&36024$_{-340}^{+342}$  & 7357$_{-128}^{+126}$   &2420$_{-68}^{+68}$     &3827$_{-81}^{+82}$\\
24&2010 February 09      &5237&11541       &25.62        &0.741& 26850$_{-290}^{+281}$  & 5814$_{-117}^{+117}$   &2059$_{-81}^{+86}$     &2437$_{-88}^{+79}$\\
25&2010 April 17              &5304&11542       &25.67             &0.741& 20935$_{-245}^{+246}$  & 6124$_{-119}^{+120}$   &1962$_{-62}^{+62}$     &2813$_{-69}^{+70}$\\
26&2010 June 25             &5373&11543       &24.62           &0.741&18521$_{-228}^{+230}$  & 5445$_{-111}^{+111}$   &1522$_{-54}^{+54}$     &1487$_{-51}^{+51}$\\
27&2010 November 11   &5512&11544       &25.56     &0.741& 27077$_{-467}^{+298}$  & 6316$_{-158}^{+120}$   &1821$_{-64}^{+57}$     &1730$_{-59}^{+56}$\\
28&2011 January 21        &5583&11545       &24.62        &0.741& 5689$_{-164}^{+153}$  & 3175$_{-123}^{+111}$   &1008$_{-53}^{+47}$     & 982$_{-46}^{+47}$\\
29&2011 February 25        &5618& 12833   &  13.61            &0.441& $5412_{-165}^{+159}$ &   $4375_{-128}^{+131}$ &   1161$_{-52}^{+52}$ &     $821_{-41}^{+44}$ \\
30&2011 November 9         &5875       &  12834  &   13.61           & 0.441         &    2206$_{-77}^{+77}$                  &   2929$_{-104}^{+104}$          &  633$_{-37}^{+37}$     & 210$_{-18}^{+18}$ \\
31&2012 April 11                 &6229       &  13962  &   13.67           &  0.441        &       2557$_{-99}^{+99}$              &     3660$_{-107}^{+107}$        &   730$_{-39}^{+39}$    &  480$_{-27}^{+27}$\\
32&2012 November 7         &6239       &   13963 &    14.57                 & 0.441          &    1521$_{-73}^{+73}$           &     2174$_{-105}^{+105}$       &  504$_{-33}^{+33}$     &  270$_{-22}^{+22}$\\
33&2012 November 18       &6250       &   14507 &    9.13                 &0.441    &     738$_{-59}^{+59}$              &    1068$_{-70}^{+70}$        &   262$_{-27}^{+27}$     & 169$_{-17}^{+17}$ \\
34&2012 December 12       &6274       &   14508 &    9.13                 & 0.441     &     1163$_{-73}^{+203}$            &    1518$_{-158}^{+98}$        &  399$_{-47}^{+34}$      & 187$_{-19}^{+49}$ \\
35&2013 November 30       &6627       &   14509  &   9.13                  &0.441   &    1420$_{-80}^{+100}$            &   1649$_{-111}^{+89}$         & 649$_{-44}^{+41}$       & 347$_{-29}^{+36}$ \\
36&2014 January 3             &6661       &   14510  &   8.75                  &0.441    &      1374$_{-76}^{+69}$          &   1344$_{-89}^{+88}$        & 441$_{-30}^{+32}$        & 322$_{-30}^{+34}$ \\
37&2014 June 13                &6822       &  14511    &   9.13                  &0.441     &      1352$_{-70}^{+75}$         &   1795$_{-122}^{+95}$         &  525$_{-36}^{+36}$      & 296$_{-26}^{+31}$ \\
38&2014 July 7                   & 6851      &  14512    &   10.04                   & 0.441      &    806$_{-68}^{+76}$          &  1132$_{-92}^{+88}$           &  392$_{-42}^{+42}$      & 317$_{-29}^{+31}$  \\

\hline \hline
\end{tabular}
\end{center}
${}^{a}$ Julian Date $-$ 2450000.
${}^{b}$ {ACIS frame-time}. 
${}^{c}${Background-subtracted source counts for events with energies in the 0.2--10~keV band.
The counts for images A and B  of \rxj\ are corrected for pile-up. Images C and D are not affected by pile-up.}\\
\end{table}

\clearpage
\begin{table}
\caption{Log of Observations of Quasar Q~J0158$-$4325} 
\scriptsize
\begin{center}
\begin{tabular}{clccrlrrrr}
 & & && & & && & \\ \hline\hline
         &        & & {\it Chandra}                                      & Exposure & & & &      \\
Epoch & Observation & JD${}^{a}$  &  Observation  &   Time   & $t_{\rm f}$${}^{b}$& N$_{\rm soft,A}$$\tablenotemark{c}$ & N$_{\rm soft, B}$$\tablenotemark{c}$ & N$_{\rm hard,A}$$\tablenotemark{c}$ & N$_{\rm hard,B}$ $\tablenotemark{c}$   \\
& Date           &     (days)        &          ID                &   (ks)       &(s)& counts   & counts  & counts  & counts \\
&   &    &    &   &  &  & & &\\
\hline

1&2009 November 06             &5142           &  11556     & 5.03 &1.74&       103$\pm 10$ & 43$\pm 7$ &  35$\pm 6$ & 11$\pm 3$  \\
2&2010 January 12          & 5209        &  11557    & 5.02    &1.74&     125$\pm 11$  & 61$\pm 8$  & 35$\pm 6$ &  19$\pm 4$    \\
3&2010 March 10           & 5266        &  11558    &  5.04   &1.74&       146$\pm 12$  & 48$\pm 7$ & 38$\pm 6$  & 8$\pm 3$  \\
4&2010 May 23               & 5340        &  11559    &  4.94   &1.74&     131$\pm 11$   & 38$\pm 6$  & 38$\pm 6$   & 10$\pm 3$  \\
5&2010 July 28              & 5406       &  11560     &  4.95  &1.74&     144$\pm 12$   & 34$\pm 6$ & 35$\pm 6$   & 7$\pm 3$  \\
6&2010 October 06    &5476        &  11561    &  4.94   &1.74&     122$\pm 11$  & 39$\pm 6$  & 28$\pm 5$   & 14$\pm 4$  \\
7&2013 March 26     & 6378&14483       & 18.62          &1.74& 375$\pm 19$   & 100$\pm 10$     &110$\pm 10$      & 24$\pm 5$\\
8&2013 April 24          &6407 &14484       & 18.62              &1.74& 314$\pm 18$    & 93$\pm 10$     & 90$\pm 10$     & 34$\pm 6$\\
9&2013 December 5        & 6632 &14485       & 18.62             &1.74& 458$\pm 21$    & 99$\pm 10$     & 119$\pm 11$     & 31$\pm 6$\\
10&2013 December 28      & 6655&14486       & 18.61           &1.74& 339$\pm 18$    &  134$\pm 12$     & 138$\pm 12$     & 45$\pm 7$\\
11&2014 May 29             &6807  &14487       & 18.62                &1.74& 557$\pm 24$    & 137$\pm 13$     & 166$\pm 13$     &  59$\pm 8$\\
12&2014 June 10              &6819 &14488       & 18.62                &1.74& 520$\pm 23$    & 163$\pm 12$     & 150$\pm 12$     & 55$\pm 7$\\

\hline \hline
\end{tabular}
\end{center}
${}^{a}$ Julian Date $-$ 2450000.
${}^{b}$ {ACIS frame-time}. 
${}^{c}${Background-subtracted source counts for events with energies in the 0.2--10~keV band.}\\
\end{table}

\clearpage
\begin{table}
\caption{Log of Observations of Quasar \sdss\ } 
\scriptsize
\begin{center}
\begin{tabular}{clccrrrrr}
 & & && & & & & \\ \hline\hline
Epoch    & Observation  & & {\it Chandra}                                      &   & & &      \\
              & Date              & JD${}^{a}$  &  ObsID{}$^{b}$  &   Time{}$^{c}$   &  $N_{\rm A}$$\tablenotemark{d}$ & $N_{\rm B}$$\tablenotemark{d}$ & $N_{\rm C}$$\tablenotemark{d}$ & $N_{\rm D}$$\tablenotemark{d}$   \\
              &                       &     (days)        &                         &   (ks)       & counts   & counts & counts & counts   \\
&   &    &    &    &  & & &\\
\hline
1&2010 March 8             &5264           &  11546     & 5.96 &              53$\pm 7$(16$\pm 4$) & 82$\pm 9$(19$\pm4$)           &  66$\pm 8$(20$\pm4$) & 90$\pm 9$(20$\pm4$)  \\
2&2010 June 19          & 5367        &  11547    & 5.96    &                 44$\pm 7$(16$\pm 4$)  & 44$\pm 7$(15$\pm 4$)         & 97$\pm 10$(29$\pm 5$) &  84$\pm 9$(14$\pm 4$)    \\
3&2010 September 23           & 5463        &  11548    &  5.96   &      51$\pm 7$(13$\pm 4$)  & 65$\pm 8$(17$\pm 4$)         & 66$\pm 8$(22$\pm 5$)  & 58$\pm 8$(20$\pm 4$)  \\
4&2011 January 30       & 5592        &  11549    &  5.96   &               29$\pm 5$(7$\pm 3$)   & 36$\pm 6$(14$\pm 4$)       & 115$\pm 11$(27$\pm 5$)   & 85$\pm 9$(28$\pm 5$)  \\
5&2013 January 27    & 6320       &  14495     &  24.74  &               192$\pm 14$(60$\pm 8$) & 425$\pm 21$(118$\pm 11$)& 360$\pm 19$(94$\pm 10$)   & 223$\pm 15$(74$\pm 9$)  \\
6&2013 March 1    &6353        &  14496    &  24.74   &                   164$\pm 13$(88$\pm 9$)  & 414$\pm 20$(139$\pm 12$)& 338$\pm 18$(124$\pm 11$)   & 184$\pm 14$(83$\pm 9$)  \\
7&2013 October 5     & 6571 & 14497       & 24.13          &             179$\pm 13$(87$\pm 9$)   & 355$\pm 19$(97$\pm 10$)   &356$\pm 19$(114$\pm 11$)      & 182$\pm 14$(48$\pm 7$)\\
8&2013 November 16   &6613  &14498       & 23.75              &      171$\pm 13$(53$\pm 7$)    & 358$\pm 19$(98$\pm 10$) & 406$\pm 20$(92$\pm 10$)     & 250$\pm 16$(83$\pm 9$)\\
9&2014 April 29        & 6777  &14499       & 23.32             &          139$\pm 12$(65$\pm 8$)    & 284$\pm 17$(94$\pm 10$)  & 245$\pm 16$(82$\pm 9$)     & 151$\pm 12$(59$\pm 8$)\\
10&2014 June 2      & 6811 &14500       & 24.74                         & 132$\pm 11$(56$\pm 8$)    &  422$\pm 21$(132$\pm 11$)& 261$\pm 16$(85$\pm 9$)     & 138$\pm 12$(45$\pm 7$)\\
\hline \hline
\end{tabular}
\end{center}
\tablecomments{The ACIS-S frame-time for all observations is 3.1 s.
${}^{a}$ Julian Date $-$ 2450000.
${}^{b}$ Observation ID.
${}^{c}$ Effective exposure time after applying filters.
${}^{d}$ Soft Band: 0.2--2 keV(hard band: 2--10 keV counts).}
\end{table}

\clearpage
\begin{table}
\caption{Properties of Sample } 
\scriptsize
\begin{center}
\begin{tabular}{lcccccccccc}
\hline
\multicolumn{1}{c} {Object} &
\multicolumn{1}{c} {$z_{s}$}  &
\multicolumn{1}{c} {$z_{l}$}  &
\multicolumn{1}{c} {$L_{\rm Bol}/L_{\rm Edd}$} &
\multicolumn{1}{c} {$\log(M_{\rm BH})$} &
\multicolumn{1}{c} {$\log(R_{\rm E})$} &
\multicolumn{1}{c} {$\log(r_{\rm g})$}  &
\multicolumn{1}{c} {$R_{E}/v_{e}$ }  &
\multicolumn{1}{c} {10$r_{g}/v_{e}$}  &
\multicolumn{1}{c} {$v_{e}$} &
\multicolumn{1}{c} {$\mu$}  \\
\multicolumn{1}{c} {} &
\multicolumn{1}{c} {} &
\multicolumn{1}{c} {} &
\multicolumn{1}{c} {} &
\multicolumn{1}{c} {$M_\odot$} &
\multicolumn{1}{c} {cm} &
\multicolumn{1}{c} {cm} &
\multicolumn{1}{c} {years} &
\multicolumn{1}{c} {months}  &
\multicolumn{1}{c} {km/s}  &
\multicolumn{1}{c} {}  \\ \hline
RXJ1131     &  0.658 & 0.295 & 0.01-0.42   &  7.9-8.3        & 16.4 &13.1-13.5   & 11.1  &  0.64-1.6  & 720 & 57    \\
QJ0158       & 1.29   & 0.317  & 0.4   & 8.2                          & 16.5  &13.4          & 18.0 &  1.5   & 600 & 5       \\
SDSS1004  & 1.73   &0.68     & 0.05 & 8.6                          & 16.4  & 13.8         & 9.4   &  2.9 & 785 & 70      \\
\hline \hline
\end{tabular}
\end{center}
\tablecomments{$M_{\rm BH}$ and $R_{\rm E}$ are the black hole mass and Einstein radius ($\langle$M$\rangle$ = 0.3$M_{\odot}$) with $r_{\rm g}$ = G$M_{\rm BH}$/$c^{2}$.
$R_{E}/v_{e}$  and $r_{g}/v_{e}$  are the crossing times given the effective velocity  $v_{e}$ (see Mosquera \& Kochanek 2011). The ACIS count-rates apply for of the combined images  and represent average values based on available observations.  $\mu$ is the total flux magnification of the background quasar. For the estimate of the Eddington ratios we assumed a 2--10~keV bolometric correction factor of $\kappa_{\rm 2-10keV}$~$\sim$~30.}
\end{table}

\clearpage
\begin{table}
\caption{Properties of Shifted Fe K$\alpha$ Line and Continuum in \rxj} 
\scriptsize
\begin{center}
\begin{tabular}{lccccrrrrrcrr}
\hline
\multicolumn{1}{c} {ID$^{a}$} &
\multicolumn{1}{c} {Im$^{b}$}  &
\multicolumn{1}{c} {$E_{\rm Fe}$$^{c}$}  &
\multicolumn{1}{c} {$\sigma_{\rm Fe}$$^{d}$} &
\multicolumn{1}{c} {$EW_{\rm Fe}$$^{e}$} &
\multicolumn{1}{c} {$N_{\rm Fe}$$^{f}$} &
\multicolumn{1}{c} {$N_{\rm cont}$$^{g}$}  &
\multicolumn{1}{c} {$\Gamma$}  &
\multicolumn{1}{c} {$cstat$$^{h}$}  &
\multicolumn{1}{c} {$dof$$^{h}$} &
\multicolumn{1}{c} {P${}^{j}$}  &
\multicolumn{1}{c} {F${}^{k}$} &
\multicolumn{1}{c} {P$_{F}{}^{l}$} \\
\multicolumn{1}{c} {} &
\multicolumn{1}{c} {} &
\multicolumn{1}{c} {keV} &
\multicolumn{1}{c} {keV} &
\multicolumn{1}{c} {keV} &
\multicolumn{1}{c} {${}^{c}$} &
\multicolumn{1}{c} {${}^{d}$} &
\multicolumn{1}{c} {} &
\multicolumn{1}{c} {}  &
\multicolumn{1}{c} {}  &
\multicolumn{1}{c} {}  &
\multicolumn{1}{c} {}  &
\multicolumn{1}{c} {}  \\ \hline
4814 & 1 & 3.96$_{-0.14}^{+0.13}$     &  0.15$_{-0.15}^{+0.13}$        &    1.66$_{-0.97}^{+0.46}$    &   4.5$_{-2.6}^{+3.3}$      & 7.3$_{-1.3}^{+1.5}$      &  1.48$_{-0.19}^{+0.19}$  & 118.44  &163  & 1            &6.41 & $<$ 0.001 \\
4814 &   2 &   2.60$_{-0.05}^{+0.05}$ &     $<$     0.12                        &    0.26$_{-0.09}^{+0.06}$    &  9.6$_{-4.8}^{+4.8}$       &   46.7$_{-3.1}^{+3.3}$  &  1.35$_{-0.06}^{+0.06}$       &   491.61  & 491 &  1    & 3.94 & 0.013 \\
6912 &   2 &   3.46$_{-0.25}^{+0.09}$ &     $<$     0.25                        &    0.35$_{-0.19}^{+0.25}$    &   4.6$_{-3.6}^{+5.0}$      &   47.0$_{-5.2}^{+5.7}$  &  1.76$_{-0.12}^{+0.12}$      &  217.63  & 264  & 1      & 2.57 & 0.033 \\
6913 &   1 &   2.94$_{-0.27}^{+0.05}$ &     0.03$_{-0.03}^{+0.53}$     &     0.61$_{-0.36}^{+0.37}$    &   4.2$_{-3.2}^{+6.5}$     &  18.8$_{-3.4}^{+3.9}$   &  1.76$_{-0.20}^{+0.20}$           &   116.63  & 152  & 1 & 4.35 & 0.003\\
6913 &   1 &   3.81$_{-0.52}^{+0.20}$ &     $<$      0.12                       &    1.11$_{-0.66}^{+0.73}$     &   4.7$_{-4.0}^{+6.0}$     &  18.8$_{-3.4}^{+3.9}$   & 1.76$_{-0.20}^{+0.20}$             &   116.63  & 152  & 1 & 4.34 & 0.004\\
6913 &   4 &   2.42$_{-0.04}^{+0.04}$ &     $<$      0.15                       &    1.01$_ {-0.48}^{+0.53}$   &    3.5$_{-2.2}^{+3.3}$     &   6.7$_{-1.7}^{+2.1}$    &  1.76$_{-0.29}^{+0.29}$        &  52.03  &  77  &  0       & 7.81 & $<$ 0.001 \\
6914 &   1 &   4.59$_{-0.65}^{+0.44}$ &     0.45$_{-0.45}^{+1.13}$     &    2.52$_{-1.44}^{+1.62}$    &   12.4$_{-8.0}^{+9.3}$    &  23.6$_{-3.7}^{+4.2}$    &  1.66$_{-0.17}^{+0.17}$        &  146.05  & 191  & 1     & 2.99 & 0.03 \\
6914 &   2 &   2.77$_{-0.06}^{+0.07}$ &     0.01$_{-0.01}^{+0.17}$   &    0.73$_{-0.31}^{+0.20}$    &   9.4$_{-4.4}^{+10.2}$     &   33.7$_{-4.2}^{+4.7}$    & 1.81$_{-0.14}^{+0.14}$       &  175.40  & 234  & 0        & 11.96 & $<$ 0.001\\
6914 &   4 &   3.63$_{-0.67}^{+0.68}$ &     $<$      0.60                       &   1.00$_{-0.60}^{+1.00}$     &  2.1$_{-1.6 }^{+2.7}$       &   9.1$_{-2.1}^{+2.5}$    &  1.83$_{-0.26}^{+0.27}$       &   92.55   & 95  &  1         &2.24 & 0.1\\
6915 &   1 &   2.59$_{-0.15}^{+0.11}$ &     0.22$_{-0.16}^{+0.17}$     &    0.59$_{-0.17}^{+0.19}$     &  48.4$_{-18.9}^{+20.3}$  &  144.0$_{-9.2}^{+9.7}$ &  1.62$_{-0.06}^{+0.06}$      &  445.59  & 476 &  0         & 7.43 & $<$ 0.001\\
6915 &   1 &   4.35$_{-0.10}^{+0.10}$ &     0.08$_{-0.08}^{+0.09}$     &    0.31$_{-0.16}^{+0.18}$     &  11.1$_{-7.9}^{+9.5}$      &  144.0$_{-9.2}^{+9.7}$ &  1.62$_{-0.06}^{+0.06}$     &  445.59  & 476  & 1          & 7.42 & $<$ 0.001\\
6915 &  2 &  3.81$_{-0.03}^{+0.05}$ &    $<$ 0.07                            &      0.49$_{-0.38}^{+0.15}$   & 8.1$_{-4.3}^{+5.5}$       & 91.6$_{-7.3}^{+7.7}$        &  1.91$_{-0.09}^{+0.09}$      &  333.73  & 343 &  0          & 6.65 & 0.001\\
6916 &  1 &  2.28$_{-0.06}^{+0.10}$ &    $<$ 0.30                            &      0.16$_{-0.05}^{+0.07}$   &15.8$_{-10.5}^{+10.8}$  & 147.0$_{-9.3}^{+9.8}$      &1.58$_{-0.07}^{+0.06}$        & 440.07  & 479 &  1           &  4.30 & 0.001\\
6916 &  1 &  2.95$_{-0.07}^{+0.09}$ &    $<$ 0.50                              &      0.24$_{-0.07}^{+0.09}$   &16.2$_{-10.5}^{+10.5}$   & 147.0$_{-9.3}^{+9.8}$   & 1.58$_{-0.07}^{+0.06}$     &440.07  & 479  & 1              & 4.36 & 0.002\\
7785 &  2 &  2.69$_{-0.03}^{+0.03}$ &    $<$ 0.10                              &      0.48$_{-0.11}^{+0.14}$   &12.8$_{-5.8}^{+7.0}$       & 72.0$_{-6.3}^{+6.7}$     & 1.86$_{-0.10}^{+0.10}$         & 295.14  & 312  & 0          & 9.40 & $<$0.001\\
7785 &  2 &  3.18$_{-0.28}^{+0.05}$ &    $<$ 0.10                              &      0.35$_{-0.21}^{+0.15}$   &7.0$_{-4.4}^{+5.6}$         & 72.0$_{-6.3}^{+6.7}$     &1.86$_{-0.10}^{+0.10}$          &  295.14  & 312  & 1        & 9.40 & $<$0.001\\
7786 &  2 &  3.42$_{-0.09}^{+0.15}$ &    $<$ 0.20                              &      0.35$_{-0.19}^{+0.15}$   &9.7$_{-6.6}^{+9.7}$         &89.0$_{-6.7}^{+7.1}$      &1.70$_{-0.08}^{+0.08}$     &  331.99  & 384 &  1              & 4.64 & 0.003\\ 
7787 &  1 &  3.07$_{-0.23}^{+0.28}$ &    0.15$_{-0.15}^{+0.25}$   &      0.24$_{-0.16}^{+0.15}$    &12.0$_{-11.4}^{+13.4}$    & 142.0$_{-9.0}^{+9.5}$     & 1.74$_{-0.07}^{+0.07}$     &   421.39  & 439  & 1             & 4.10 & 0.023\\
7787 &  1 &  4.20$_{-0.05}^{+0.04}$ &    $<$ 0.08                            &      0.32$_{-0.14}^{+0.20}$   &9.4$_{-5.5}^{+6.8}$          & 142.0$_{-9.0}^{+9.5}$    & 1.74$_{-0.07}^{+0.07}$    &   421.39  & 439 &  0             & 4.10 & 0.012\\
7788 &  1 &  3.92$_{-0.23}^{+0.30}$ &    $<$ 1.10                              &      0.21$_{-0.11}^{+0.10}$   &13.6$_{-5.4}^{+6.9}$        &136.0$_{-9.0}^{+9.4}$   & 1.72$_{-0.07}^{+0.07}$     &   362.37  & 423 &  1             &2.46  & 0.033\\
7789 &  1 &  2.49$_{-0.07}^{+0.08}$ &    $<$ 0.40                              &      0.17$_{-0.09}^{+0.07}$   &11.4$_{-7.4}^{+8.6}$        &146.0$_{-9.3}^{+9.7}$   & 1.82$_{-0.07}^{+0.07}$     &   330.18  & 429  & 1             & 2.50 & 0.097\\
7789 &  2 &  2.87$_{-0.04}^{+0.05}$ &    $<$ 0.11                            &      0.34$_{-0.15}^{+0.12}$   & 9.6$_{-5.3}^{+6.6}$         & 69.8$_{-6.1}^{+6.5}$      & 1.73$_{-0.09}^{+0.09}$     & 300.51  & 339  & 0               & 4.97   & 0.006  \\
7790 &  3 &  2.42$_{-0.20}^{+0.20}$ &   0.24$_{-0.24}^{+0.21}$    &      1.04$_{-0.44}^{+0.36}$   &13.0$_{-8.1}^{+9.6}$        & 21.7$_{-3.3}^{+3.7}$        & 1.69$_{-0.16}^{+0.16}$     &  25.96   & 29 &   1                &4.46 & 0.007\\
7791 &  2 &  3.21$_{-0.05}^{+0.06}$ &    $<$  0.10                             &      0.23$_{-0.13}^{+0.14}$   &5.82$_{-4.42}^{+5.65}$      &  99.6$_{-7.5}^{+8.0}$ & 1.89$_{-0.08}^{+0.08}$   & 294.67  & 340 &  1                 & 2.08 & 0.04\\
7792 &  1 &  2.55$_{-0.06}^{+0.07}$ &    $<$  0.50                             &      0.20$_{-0.09}^{+0.10}$   & 11.3$_{-7.6}^{+9.1}$       &145.0$_{-9.9}^{+10.0}$ & 1.92$_{-0.08}^{+0.08}$      &   342.32  & 407 &  1            & 2.46 & 0.073\\
7792 &  2 &  3.25$_{-0.12}^{+0.13}$ &    0.13$_{-0.13}^{+0.18}$   &      0.47$_{-0.21}^{+0.27}$   & 11.6$_{-7.6}^{+9.4}$       & 101.9$_{-7.7}^{+ 8.1}$     & 1.91$_{-0.08}^{+0.09}$     &  333.74  & 363 &  1               & 2.81 & 0.1\\
7793 &  1 &  3.75$_{-0.06}^{+0.05}$ &    $<$ 0.16                            &      0.22$_{-0.15}^{+0.12}$   &6.7$_{-5.0}^{+6.3}$         & 122.9$_{-8.4}^{+8.9}$    &1.75$_{-0.07}^{+0.07}$     &  343.24 & 400 &  1                  & 3.49 & 0.013\\
7793 &  2 &  2.59$_{-0.07}^{+0.09}$ &    0.04$_{-0.04}^{+0.11}$   &      0.38$_{-0.22}^{+0.16}$   & 8.2$_{-6.0}^{+6.5}$          & 84.4$_{-6.9}^{+7.3}$       & 1.83$_{-0.09}^{+0.09}$     &  312.71  & 346 &  1                & 2.44 & 0.023\\
7793 &  2 &  4.97$_{-0.60}^{+0.09}$ &    $<$ 0.55                            &      0.46$_{-0.28}^{+0.31}$   & 4.9$_{-3.8}^{+13.4}$       &84.4$_{-6.9}^{+7.3}$       &1.83$_{-0.09}^{+0.09}$     &   312.71  & 346 &  1                & 3.47 & 0.0033\\
7793 &  3 &  2.63$_{-0.05}^{+0.05}$ &    $<$ 0.12                            &      0.91$_{-0.54}^{+0.48}$   &3.1$_{-2.2}^{+3.4}$          &7.1$_{-1.8}^{+2.2}$         & 1.73$_{-0.29}^{+0.29}$     &   85.62  &  82  &  1                 & 4.45  & 0.014\\
7794 &  2 &  4.37$_{-0.07}^{+0.07}$ &    $<$ 0.75                            &      0.41$_{-0.17}^{+0.23}$   &8.0$_{-5.7}^{+17.9}$        & 114.1$_{-8.0}^{+8.4}$     & 1.79$_{-0.07}^{+0.08}$    &  375.19  & 401 &  1                 & 4.10 & 0.0033\\
7794 &  3 &  4.79$_{-0.39}^{+0.40}$ &    $<$ 0.14                            &      0.65$_{-0.40}^{+0.51}$   &3.0$_{-2.5}^{+3.9}$          & 38.0$_{-4.6}^{+5.0}$       &1.89$_{-0.13}^{+0.14}$     &   166.94  & 221  & 1                & 3.76 & 0.01\\
9180 &  1 &  2.81$_{-0.10}^{+0.07}$ &    0.05$_{-0.05}^{+0.43}$   &      0.11$_{-0.06}^{+0.04}$   &5.8$_{-4.2}^{+7.5}$          & 101.6$_{-4.4}^{+4.5}$       & 1.62$_{-0.04}^{+0.04}$   &  558.90  & 596  & 1                   & 2.17 & 0.1\\
9180 &  3 &  3.53$_{-0.07}^{+0.06}$ &    $<$  0.17                          &       0.23$_{-0.14}^{+0.12}$   & 2.0$_{-1.5}^{+1.9}$         &33.7$_{-2.5}^{+2.6}$       & 1.77$_{-0.08}^{+0.08}$    &   368.30  & 385 &  1                  & 2.60  & 0.02\\
9181 &  1 &  2.35$_{-0.10}^{+0.10}$ &     0.15$_{- 0.10}^{+0.11}$    &      0.20$_{-0.07}^{+0.06}$   &18.2$_{-9.5}^{+10.7}$       & 142.1$_{-5.2}^{+5.4}$   &1.65$_{-0.04}^{+0.04}$   &  632.52  & 638  & 0                    & 6.16 & 0.001\\
9181 &  1 &  2.80$_{-0.05}^{+0.05}$ &     $<$ 0.12                          &      0.11$_{-0.06}^{+0.04}$    &7.2$_{-4.3}^{+4.3}$          & 142.1$_{-5.2}^{+5.4}$     & 1.65$_{-0.04}^{+0.04}$      &632.52  & 638 &  1                  & 6.14 & 0.001\\
9181 &  4 &  3.70$_{-0.30}^{+0.05}$ &     $<$  0.12                         &      0.59$_{-0.36}^{+0.36}$    &1.5$_{-1.0}^{+1.3}$          &12.0$_{-1.4}^{+1.6}$        &1.84$_{-0.13}^{+0.13}$          &   192.77  & 227 &  1            & 2.2 & 0.02\\
\hline \hline
\end{tabular}
\end{center}
\tablecomments{{}$^{a}${\it Chandra} Observation ID,  
{}$^{b}$Image numbers 1, 2, 3, and 4 correspond to images A, B, C, and D of \rxj,  
{}$^{c}$Observed-frame energy of the shifted Fe line, 
{}$^{d}$ Observed-frame energy width of the shifted Fe line, 
{}$^{e}$Rest-frame equivalent width of the shifted Fe line, 
{}$^{f}$Flux of the shifted Fe line in units of  $\times$ 10$^{-6}$~photons~cm$^{-2}$~s$^{-1}$, 
{}$^{g}$Flux density of the continuum in units of $\times$ 10$^{-5}$~photons~keV$^{-1}$~cm$^{-2}$~s$^{-1}$ at 1~keV, 
{}$^{h}$ Cash statistic, 
{}$^{i}$degrees of freedom, 
{}$^{j}$P values of 0 and 1 correspond to confidence detection levels of the shifted Fe line of  $>$~99\% and  $>$~90\% respectively.
{}$^{k}$$F$-statistic between the null and alternative model.
{}$^{l}$The probability of exceeding this $F$-value as determined from the Monte-Carlo simulations.}
\end{table}

%10$^{-6}$ $\times$~photons~cm$^{-2}$~s$^{-1}$
% 10$^{-5}$ $\times$~photons~keV$^{-1}$~cm$^{-2}$~s$^{-1}$

\clearpage
\setcounter{table}{4}
\begin{table}
\caption{(Continued)} 
\scriptsize
\begin{center}
\begin{tabular}{lccccrrrrrcrr}
\hline
\multicolumn{1}{c} {ID$^{a}$} &
\multicolumn{1}{c} {Im}  &
\multicolumn{1}{c} {$E_{\rm Fe}$}  &
\multicolumn{1}{c} {$\sigma_{\rm Fe}$} &
\multicolumn{1}{c} {$EW_{\rm Fe}$} &
\multicolumn{1}{c} {$N_{\rm Fe}$} &
\multicolumn{1}{c} {$N_{\rm cont}$}  &
\multicolumn{1}{c} {$\Gamma$}  &
\multicolumn{1}{c} {cstat}  &
\multicolumn{1}{c} {$dof$} &
\multicolumn{1}{c} {P} &
\multicolumn{1}{c} {F${}^{k}$} &
\multicolumn{1}{c} {P$_{F}{}^{l}$} \\
\multicolumn{1}{c} {} &
\multicolumn{1}{c} {} &
\multicolumn{1}{c} {keV} &
\multicolumn{1}{c} {keV} &
\multicolumn{1}{c} {keV} &
\multicolumn{1}{c} {${}^{c}$} &
\multicolumn{1}{c} {${}^{d}$} &
\multicolumn{1}{c} {} &
\multicolumn{1}{c} {}  &
\multicolumn{1}{c} {}  &
\multicolumn{1}{c} {}  &
\multicolumn{1}{c} {}  &
\multicolumn{1}{c} {}  \\ \hline
9237 &  2 &  3.16$_{-0.05}^{+0.06}$ &     $<$  0.13                        &      0.12$_{-0.06}^{+0.07}$    & 4.0$_{-2.8}^{+3.5}$        &90.2$_{-4.0}^{+4.2}$& 1.65$_{-0.04}^{+0.04}$      &  560.80  & 584 &  1       & 2.07 & 0.037\\
9237 &  3 &  3.84$_{-0.05}^{+0.06}$ &    0.05$_{-0.05}^{+0.08}$   &      0.59$_{-0.22}^{+0.19}$    & 3.8$_{-1.9}^{+2.4}$        &30.7$_{-2.4}^{+2.5}$& 1.81$_{-0.08}^{+0.08}$     &  322.8   & 353  & 0         & 7.88 & $<$ 0.001\\
9237 &  4 &  3.46$_{-0.28}^{+0.15}$ &    0.19$_{-0.08}^{+0.32}$   &      1.61$_{-0.60}^{+0.59}$    &3.3$_{-1.7}^{+2.1}$         &6.8$_{-1.1}^{+1.2}$&1.7$_{-0.17}^{+0.18}$    &  183.75  & 195 &  0              & 5.56 &$<$ 0.001 \\
9238 &  3 &  2.72$_{-0.10}^{+0.10}$ &    0.13$_{-0.10}^{+0.13}$     &     0.72$_{-0.22}^{+0.27}$    &5.8$_{-3.0}^{+3.7}$          & 21.4$_{-2.0}^{+2.2}$& 1.85$_{-0.10}^{+0.10}$     & 253.35  & 316  & 0      & 6.93 & 0.056\\
9238 &  3 &  3.80$_{-0.09}^{+0.10}$ &    0.10$_{- 0.10}^{+ 0.11}$       &      0.65$_{-0.29}^{+0.26}$    &2.8$_{-1.8}^{+2.2}$           & 21.4$_{-2.0}^{+2.2}$&1.85$_{-0.10}^{+0.10}$    &  253.35  & 316  & 0 & 2.88 & 0.033\\
9239 &  2 &  2.41$_{-0.29}^{+0.07}$ &    0.05$_{-0.05}^{+0.28}$   &      0.13$_{-0.05}^{+0.07}$    &4.9$_{-3.6}^{+8.0}$           & 63.5$_{-3.4}^{+3.5}$ & 1.65$_{-0.05}^{+0.05}$   &  503.17  & 514  & 1      & 3.25 & 0.053\\
9240 &  2 &  3.06$_{-0.29}^{+0.05}$ &    $<$ 0.10                              &      0.22$_{-0.07}^{+0.08}$    & 5.0$_{-2.9}^{+2.9}$          & 62.4$_{-3.6}^{+3.6}$& 1.72$_{-0.06}^{+0.06}$   &  498.60  & 495 & 1    & 2.97 & 0.046\\
9240 &  2 &  3.96$_{-0.14}^{+0.17}$ &    0.20$_{-0.11}^{+0.24}$     &      0.55$_{-0.22}^{+0.24}$    & 8.1$_{-4.5}^{+7.5}$           & 62.4$_{-3.6}^{+3.6}$&  1.72$_{-0.06}^{+0.06}$ &  498.60  & 495  & 1      & 4.18 & 0.026\\
9240 &  4 &  3.34$_{-0.04}^{+0.04}$ &    $<$  0.06                           &      0.46$_{-0.18}^{+0.19}$    &1.8$_{-1.1}^{+1.5}$           & 12.4$_{-1.5}^{+1.6}$& 1.73$_{-0.12}^{+0.12}$    &  228.56  & 260  & 1     & 3.75& 0.063\\
11540 &  2 &   3.27$_{-0.05}^{+0.04}$ &     $<$ 0.10                            &         0.09$_{-0.03}^{+0.04}$ &  3.2$_{-2.0}^{+2.2}$    &  91.4$_{-3.0}^{+3.1}$     & 1.63$_{-0.03}^{+0.03}$   &  678.26 & 681&  1    & 3.52 & 0.03\\
11540 &  2 &   4.28$_{-0.05}^{+0.05}$ &     $<$ 0.12                          &         0.16$_{-0.12}^{+0.23}$ &   3.4$_{-2.0}^{+2.0}$   &   91.4$_{-3.0}^{+3.1}$    & 1.63$_{-0.03}^{+0.03}$  &  678.26 & 681&  1       & 3.01 & 0.057 \\
11540 &  3 &   3.93$_{-0.16}^{+0.08}$ &     0.05$_{-0.05}^{+0.14}$   &         0.21$_{-0.15}^{+0.10}$ &   2.2$_{-1.6}^{+2.1}$   &   35.7$_{-2.1}^{+2.3}$        &  1.61$_{-0.06}^{+0.06}$     & 411.79&  496&  1 & 2.69 & 0.02\\
11542 &  1 &   2.36$_{-0.05}^{+0.05}$&      0.13$_{-0.05}^{+0.06}$    &        0.22$_{-0.04}^{+0.04}$ &   3.0$_{-9.0}^{+9.9}$    &   149.7$_{-4.1}^{+4.1}$      & 1.39$_{-0.02}^{+0.02}$   &   963.88 & 822&  0 & 8.30 & $<$ 0.001 \\
11542 &  1 &   2.94$_{-0.09}^{+0.09}$ &     0.11$_{-0.07}^{+0.20}$      &         0.22$_{-0.04}^{+0.05}$ &   1.2$_{-6.3}^{+9.7}$   &  149.7$_{-4.1}^{+4.1}$       & 1.39$_{-0.02}^{+0.02}$    &  963.88 & 822&  0 & 2.63 & 0.01\\
11542 &  2 &   3.18$_{-0.05}^{+0.03}$ &     $<$ 0.10                          &         0.13$_{-0.05}^{+0.06}$ &   3.6$_{-2.1}^{+2.1}$    &  62.1$_{-2.6}^{+2.6}$    & 1.58$_{-0.04}^{+0.04}$   &  633.02&  641&  1        & 2.82 & 0.02\\
11543 &  1 &   2.31$_{-0.04}^{+0.04}$ &     $<$ 0.13                           &         0.06$_{-0.03}^{+0.02}$  &  8.0$_{-4.6}^{+4.6}$   &   151.2$_{-4.4}^{+4.4}$      & 1.44$_{-0.03}^{+0.03}$   & 233.51 & 184&  1     &2.56  & 0.01\\
11543 &  1 &   2.62$_{-0.04}^{+0.04}$ &     $<$ 0.07                           &         0.08$_{-0.02}^{+0.02}$  & 8.7$_{-4.1}^{+4.2}$    &  151.2$_{-4.4}^{+4.4}$       & 1.44$_{-0.03}^{+0.03}$    &  233.51&  184 & 0   & 4.79 & 0.01\\
11543 &  3 &   2.62$_{-0.06}^{+0.05}$ &    $<$ 0.10                             &         0.17$_{-0.08}^{+0.11}$  &  2.4$_{-1.7}^{+2.0}$   &   30.6$_{-2.2}^{+2.3}$    & 1.75$_{-0.07}^{+0.07}$    &  354.98 & 408 & 1      & 2.66 & 0.06\\
11543 &  3 &   4.30$_{-0.05}^{+0.04}$ &     $<$ 0.10                            &         0.34$_{-0.13}^{+0.14}$ &   2.0$_{-1.2}^{+1.6}$   &   30.6$_{-2.2}^{+2.3}$    &  1.75$_{-0.07}^{+0.07}$   &  354.98 & 408 & 1      & 1.91 & 0.01\\
11544 &  3 &   3.83$_{-0.05}^{+0.05}$ &     $<$ 0.28                           &         0.25$_{-0.12}^{+0.13}$  &  2.3$_{-1.4}^{+1.7}$    &   35.6$_{-2.4}^{+2.5}$     &  1.71$_{-0.07}^{+0.07}$    &  104.89 & 130 & 1    & 3.72 & 0.003\\
11545 &  3 &   2.96$_{-0.06}^{+0.06}$ &     $<$ 0.10                             &         0.28$_{-0.13}^{+0.12}$ &   2.1$_{-1.3}^{+1.7}$    &  20.2$_{-1.8}^{+2.0}$   &  1.75$_{-0.09}^{+0.09}$  &   319.40 & 348&  0      & 3.51& 0.016\\
11545 &  3 &   4.44$_{-0.14}^{+0.13}$ &     0.20$_{-0.14}^{+0.13}$      &         1.30$_{-0.46}^{+0.52}$  &   4.8$_{-2.3}^{+2.8}$    &   20.2$_{-1.8}^{+2.0}$  &  1.75$_{-0.09}^{+0.09}$  &    319.40 & 348 & 0     &6.09 & 0.004\\
12833 &  1 &   3.97$_{-0.05}^{+0.04}$ &     $<$ 0.2                            &          0.21$_{-0.06}^{+0.07}$  &  6.4$_{-3.2}^{+3.6}$   &  121.5$_{-5.3}^{+5.5}$      & 1.69$_{-0.04}^{+0.04}$     &  492.66 & 577&  0    & 4.80 & 0.005\\
12833 &  2 &   4.22$_{-0.33}^{+0.20}$ &     0.13$_{-0.13}^{+0.23}$   &          0.25$_{-0.13}^{+0.16}$  &  4.9$_{-4.0}^{+5.0}$    &   108.0$_{-5.0}^{+5.2}$     & 1.82$_{-0.04}^{+0.04}$   &  497.50 & 542 & 1      &2.93  & 0.037\\
12833 &  3 &   4.84$_{-0.16}^{+0.12}$ &     0.1$_{-0.10}^{+0.24}$       &          0.74$_{-0.34}^{+0.33}$  &  4.0$_{-2.7}^{+3.8}$    &   40.7$_{-3.6}^{+3.8}$     & 1.84$_{-0.09}^{+0.09}$   &  337.96 & 346&  1      & 2.98 & 0.036\\
12834 &  2 &   3.84$_{-0.08}^{+0.07}$ &     0.06$_{-0.06}^{+0.10}$     &          0.22$_{-0.11}^{+0.11}$  &  4.5$_{-3.0}^{+3.7}$    &   78.5$_{-4.3}^{+4.5}$     & 1.69$_{-0.05}^{+0.05}$   &  496.96 & 522 & 1       &3.58 & 0.023\\
12834 &  4 &   4.42$_{-0.06}^{+0.05}$ &     $<$ 0.15                         &          0.66$_{-0.41}^{+0.46}$  &  1.0$_{-0.8}^{+1.2}$    &  9.0$_{-1.4}^{+1.6}$      & 1.82$_{-0.16}^{+0.16}$   &   160.98 & 184 & 1          & 3.11 & 0.04\\
13962 &  1 &   2.62$_{-0.28}^{+0.21}$ &     $<$ 0.80                            &          0.13$_{-0.10}^{+0.10}$  &  3.6$_{-3.3}^{+4.0}$  &   71.5$_{-4.6}^{+4.9}$     & 1.81$_{-0.07}^{+0.07}$  &  423.80 & 454 & 1         & 5.10 & 0.001\\
13962 &  1 &   3.91$_{-0.27}^{+0.30}$ &     0.38$_{-0.21}^{+0.28}$   &          0.80$_{-0.31}^{+0.43}$  &  11.5$_{-6.8}^{+8.5}$   &   71.5$_{-4.6}^{+4.9}$     & 1.81$_{-0.07}^{+0.07}$  &   423.80&  454&  0        & 5.10 & 0.002\\
13962 &  3 &   3.83$_{-0.18}^{+0.23}$ &     0.25$_{-0.20}^{+0.30}$       &          1.09$_{-0.28}^{+0.53}$ &   5.2$_{-2.9}^{+3.8}$     &   29.1$_{-2.8}^{+3.1}$     & 1.95$_{-0.11}^{+0.11}$  &  267.15&  324&  0      & 3.97 & 0.003\\
13962 &  4 &   3.90$_{-0.07}^{+0.07}$ &     $<$ 0.12                         &          0.38$_{-0.22}^{+0.20}$ &  1.4$_{-1.1}^{+1.5}$     &  20.1$_{-2.2}^{+2.4}$    & 1.87$_{-0.11}^{+0.11}$   &    241.22&  278&  1           & 3.00 & 0.053\\
13963 &  1 &   2.39$_{-0.08}^{+0.07}$ &     0.03$_{-0.03}^{+0.08}$  &            0.21$_{-0.11}^{+0.11}$ &    3.8$_{-2.9}^{+3.2}$   &   33.2$_{-3.2}^{+3.3}$      & 1.73$_{-0.10}^{+0.10}$   &   401.43 & 403&  1        &  4.32& 0.007\\
13963 &  1 &   3.29$_{-0.06}^{+0.07}$ &     $<$ 0.30                            &            0.22$_{-0.10}^{+0.07}$  &  2.8$_{-1.9}^{+2.4}$     &  33.2$_{-3.2}^{+3.3}$      &  1.73$_{-0.10}^{+0.10}$  &    401.43 & 403&  1    &4.32   & 0.01\\
13963 &  2 &   2.30$_{-0.09}^{+0.09}$ &     0.08$_{-0.08}^{+0.10}$    &            0.19$_{-0.11}^{+0.10}$ &   7.3$_{-4.6}^{+5.8}$     &  57.8$_{-3.7}^{+3.9}$       &  1.65$_{-0.06}^{+0.06}$   &   463.75 & 506&  1     & 3.94 & 0.0033\\
13963 &  2 &   3.92$_{-0.06}^{+0.07}$ &     0.02$_{-0.02}^{+0.10}$    &            0.19$_{-0.01}^{+0.01}$  &   3.1$_{-2.2}^{+2.8}$    & 57.8$_{-3.7}^{+3.9}$        &  1.65$_{-0.06}^{+0.06}$    &   463.75 & 506&  1    & 3.95 & 0.001\\
13963 &  3 &   2.42$_{-0.30}^{+0.10}$ &      0.04$_{-0.04}^{+0.40}$    &           0.23$_{-0.17}^{+0.15}$ &   2.2$_{-1.8}^{+5.2}$     &  15.8$_{-1.9}^{+2.1}$       &  1.68$_{-0.12}^{+0.12}$   &   240.66 & 294&  1      & 3.01 & 0.013\\
14507 &  1 &   3.82$_{-0.27}^{+0.27}$ &     $<$ 0.14                         &            0.39$_{-0.19}^{+0.18}$  &  3.3$_{-2.2}^{+3.0}$     &  26.4$_{-3.2}^{+3.6}$         &  1.59$_{-0.12}^{+0.12}$    &  268.03 & 282&  1      &3.28 & 0.03\\
14507 &  2 &   3.60$_{-0.05}^{+0.05}$ &     0.05$_{-0.05}^{+0.08}$  &            0.54$_{-0.15}^{+0.20}$  &  6.7$_{-3.3}^{+4.2}$     & 42.9$_{-4.0}^{+4.3}$        &  1.70$_{-0.09}^{+0.09}$   & 309.99 & 361&  0           & 6.85& $<$ 0.001\\
14507 &  3 &   4.30$_{-0.29}^{+0.11}$ &     0.05$_{-0.05}^{+0.33}$  &            0.35$_{-0.03}^{+0.04}$  &  2.3$_{-1.7}^{+3.4}$     &  14.5$_{-2.3}^{+2.7}$       &   1.77$_{-0.16}^{+0.17}$    &   150.66&  198&  1        & 3.93& 0.002\\
14508 &  3 &   4.24$_{-0.22}^{+0.23}$ &     0.30$_{-0.30}^{+1.00 }$         &            1.59$_{-0.77}^{+0.80}$  &   6.2$_{-3.5}^{+4.1}$    &  19.4$_{-2.7}^{+3.0}$       &   1.74$_{-0.14}^{+0.14}$    &   195.38&  249&  1 & 6.20 & 0.01\\
14510 &  4 &   4.63$_{-0.21}^{+0.22}$ &     0.30$_{-0.10}^{+0.50}$         &           2.43$_{-1.11}^{+1.04}$  &   6.0$_{-3.2}^{+3.8}$    &  20.1$_{-2.9}^{+3.3}$         &   1.91$_{-0.15}^{+0.15}$     &  46.62  & 61 &  1 & 6.81 & $<$ 0.001\\

\hline \hline
\end{tabular}
\end{center}
%\tablecomments{}
\end{table}

\clearpage
\begin{table}
\caption{Properties of Shifted Fe K$\alpha$ Line and Continuum in \qj} 
\scriptsize
\begin{center}
\begin{tabular}{lccrrrrrrrrcrr}
\hline
\multicolumn{1}{c} {ID$^{a}$} &
\multicolumn{1}{c} {Im$^{b}$}  &
\multicolumn{1}{c} {$E_{\rm Fe}$$^{c}$}  &
\multicolumn{1}{c} {$\sigma_{\rm Fe}$$^{d}$} &
\multicolumn{1}{c} {$EW_{\rm Fe}$$^{e}$} &
\multicolumn{1}{c} {$N_{\rm Fe}$$^{f}$} &
\multicolumn{1}{c} {$N_{\rm cont}$$^{g}$}  &
\multicolumn{1}{c} {$\Gamma$}  &
\multicolumn{1}{c} {$cstat$$^{h}$}  &
\multicolumn{1}{c} {$dof$$^{h}$} &
\multicolumn{1}{c} {P${}^{j}$}  &
\multicolumn{1}{c} {F${}^{k}$} &
\multicolumn{1}{c} {P$_{F}{}^{l}$} \\
\multicolumn{1}{c} {} &
\multicolumn{1}{c} {} &
\multicolumn{1}{c} {keV} &
\multicolumn{1}{c} {keV} &
\multicolumn{1}{c} {keV} &
\multicolumn{1}{c} {${}^{c}$} &
\multicolumn{1}{c} {${}^{d}$} &
\multicolumn{1}{c} {} &
\multicolumn{1}{c} {}  &
\multicolumn{1}{c} {}  &
\multicolumn{1}{c} {}  &
\multicolumn{1}{c} {}  &
\multicolumn{1}{c} {}  \\ \hline

14487 &  2 &   3.63$_{-0.04}^{+0.04}$ &     0.01$_{-0.01}^{+0.20}$ &        1.14$_{-0.62}^{+0.70}$ &        2.1$_{-1.6}^{+2.4}$ &       2.2$_{-0.3}^{+0.4}$ &       1.94$_{-0.23}^{+0.23}$ &       101.83 &   111 &    1     & 3.56 & 0.040\\
14486 &  1 &   3.41$_{-0.15}^{+0.21}$ &     0.39$_{-0.39}^{+0.60}$ &        2.25$_{-0.73}^{+0.98}$ &        7.1$_{-3.4}^{+4.3}$   &     4.8$_{-0.5}^{+0.5}$ &       1.92$_{-0.15}^{+0.16}$  &      157.29 &   170 &    0     & 9.43& $<$ 0.001\\
14485 &  1 &   2.93$_{-0.06}^{+0.06}$ &     0.14$_{-0.14}^{+0.14}$ &        1.01$_{-0.40}^{+0.36}$  &        6.8$_{-3.7}^{+4.6}$ &      6.7$_{-0.6}^{+0.6}$ &       2.19$_{-0.16}^{+0.17}$ &       166.84 &   191 &    0     & 6.50&$<$ 0.001\\
14485 &  1 &   4.37$_{-0.35}^{+0.57}$ &     1.10$_{-0.60}^{+1.50}$ &        3.86$_{-1.51}^{+1.61}$ &        10.3$_{-5.8}^{+6.6}$  &      6.7$_{-0.6}^{+0.6}$  &      2.19$_{-0.16}^{+0.17}$ &       166.84 &   191 &    0   & 6.50 &$<$ 0.001\\
14484 &  1 &   2.86$_{-0.06}^{+0.06}$ &     $<$ 1.9                          &       0.63$_{-0.49}^{+0.61}$ &         3.0$_{-1.9}^{+2.6}$  &      3.6$_{-0.3}^{+0.4}$ &       1.97$_{-0.15}^{+0.15}$ &       152.69 &   168 &    1     & 3.92& 0.023\\
14484 &  2 &   2.83$_{-0.26}^{+0.13}$ &     0.3$_{-0.3}^{+1.1}$ &              2.21$_{-1.06}^{+1.24}$ &        4.2$_{-2.7}^{+3.6}$  &      1.3$_{-0.2}^{+0.3}$ &       1.92$_{-0.29}^{+0.30}$ &        68.33 &    80 &     1      &4.92 & 0.006\\
11561 &  1 &   2.27$_{-0.35}^{+0.39}$ &     0.4$_{-0.4}^{+0.8}$  &             2.19$_{-0.96}^{+1.12}$ &       14.8$_{-9.4}^{+11.9}$  &        4.2$_{-0.6}^{+0.7}$ &       2.26$_{-0.28}^{+0.30}$ &        55.67 &    81 &    1   & 6.63 & 0.003\\
11561 &  2 &   2.84$_{-0.31}^{+0.31}$ &     0.5$_{-0.4}^{+1.3}$  &             4.41$_{-2.36}^{+2.63}$ &       10.5$_{-8.2}^{+10.9}$   &       2.2$_{-0.5}^{+0.6}$ &       2.13$_{-0.50}^{+0.45}$ &        30.53 &    41  &    1  &2.91 & 0.037\\
11560 &  1 &   1.59$_{-0.06}^{+0.06}$ &     0.12$_{-0.12}^{+0.30}$ &        0.58$_{-0.24}^{+0.26}$ &       11.7$_{-7.0}^{+8.6}$   &       4.7$_{-0.7}^{+0.8}$ &       1.81$_{-0.23}^{+0.23}$ &       106.62 &   105  &   1   & 4.72 & 0.005\\
11557 &  1 &   1.73$_{-0.03}^{+0.04}$ &     0.01$_{-0.01}^{+0.13}$   &      0.55$_{-0.25}^{+0.28}$ &       10.4$_{-6.6}^{+8.7}$  &        5.5$_{-0.8}^{+0.9}$ &       1.94$_{-0.25}^{+0.26}$ &        76.04 &    88  &    1   & 5.07 & 0.008\\

\hline \hline
\end{tabular}
\end{center}
\tablecomments{{}$^{a}${\it Chandra} Observation ID,  
{}$^{b}$Image numbers 1 and 2  correspond to images A and B of \qj,  
{}$^{c}$Observed-frame energy of the shifted Fe line, 
{}$^{d}$ Observed-frame energy width of the shifted Fe line, 
{}$^{e}$Rest-frame equivalent width of the shifted Fe line, 
{}$^{f}$Flux of the shifted Fe line in units of  $\times$ 10$^{-6}$~photons~cm$^{-2}$~s$^{-1}$, 
{}$^{g}$Flux density of the continuum in units of $\times$ 10$^{-5}$~photons~keV$^{-1}$~cm$^{-2}$~s$^{-1}$ at 1~keV, 
{}$^{h}$ Cash statistic, 
{}$^{i}$degrees of freedom, 
{}$^{j}$P values of 0 and 1 correspond to confidence detection levels of the shifted Fe line of  $>$~99\% and  $>$~90\% respectively.
{}$^{k}$$F$-statistic between the null and alternative model.
{}$^{l}$The probability of exceeding this $F$-value as determined from the Monte-Carlo simulations.}
\end{table}

\clearpage
\begin{table}
\caption{Properties of Shifted Fe K$\alpha$ Line and Continuum in \sdss} 
\scriptsize
\begin{center}
\begin{tabular}{lccrrrrrrrrcrr}
\hline
\multicolumn{1}{c} {ID$^{a}$} &
\multicolumn{1}{c} {Im$^{b}$}  &
\multicolumn{1}{c} {$E_{\rm Fe}$$^{c}$}  &
\multicolumn{1}{c} {$\sigma_{\rm Fe}$$^{d}$} &
\multicolumn{1}{c} {$EW_{\rm Fe}$$^{e}$} &
\multicolumn{1}{c} {$N_{\rm Fe}$$^{f}$} &
\multicolumn{1}{c} {$N_{\rm cont}$$^{g}$}  &
\multicolumn{1}{c} {$\Gamma$}  &
\multicolumn{1}{c} {$cstat$$^{h}$}  &
\multicolumn{1}{c} {$dof$$^{h}$} &
\multicolumn{1}{c} {P${}^{j}$}  &
\multicolumn{1}{c} {F${}^{k}$} &
\multicolumn{1}{c} {P$_{F}{}^{l}$} \\
\multicolumn{1}{c} {} &
\multicolumn{1}{c} {} &
\multicolumn{1}{c} {keV} &
\multicolumn{1}{c} {keV} &
\multicolumn{1}{c} {keV} &
\multicolumn{1}{c} {${}^{c}$} &
\multicolumn{1}{c} {${}^{d}$} &
\multicolumn{1}{c} {} &
\multicolumn{1}{c} {}  &
\multicolumn{1}{c} {}  &
\multicolumn{1}{c} {}  &
\multicolumn{1}{c} {}  &
\multicolumn{1}{c} {}  \\ \hline

11549  & 4 &   1.18$_{-0.03}^{+0.04}$ &      $<$0.1                               &      0.60$_{-0.25}^{+0.27}$    &       3.6$_{-2.2}^{+2.8}$  &      10.8$_{-3.9}^{+5.6}$   & 1.59$_{-0.29}^{+0.28}$       &         57.68  &    80   &    0         & 6.76  & 0.004\\
14495 &  2 &   2.12$_{-0.21}^{+0.05}$ &       $<$0.16                            &      0.61$_{-0.31}^{+0.23}$    &       1.8$_{-1.1}^{+1.4}$  &      21.0$_{-3.5}^{+2.9}$    &  1.89$_{-0.13}^{+0.13}$     &        152.36  &   185   &      0     &6.03  & 0.003\\
14495 &  4 &   2.33$_{-0.13}^{+0.14}$ &       0.30$_{-0.10}^{+0.13}$    &       4.36$_{-1.08}^{+1.30}$   &        4.4$_{-1.7}^{+1.8}$  &      15.3$_{-3.6}^{+4.7}$   &  2.17$_{-0.21}^{+0.23}$       &      124.95  &   150  &     0       & 14.3  & $<$ 0.001\\
14496 &  2 &   3.10$_{-0.07}^{+0.06}$ &       $<$0.12                            &       0.74$_{-0.34}^{+0.34}$   &       1.2$_{-0.8}^{+1.0}$  &      19.4$_{-3.3}^{+3.9}$     &  1.79$_{-0.13}^{+0.13}$      &     185.30  &   200   &    1        &4.12  & 0.02\\
14496 &  3 &   2.36$_{-0.07}^{+0.07}$ &       $<$0.20                            &        1.71$_{-0.56}^{+0.60}$  &        2.8$_{-1.5}^{+1.5}$  &     16.5$_{-4.0}^{+5.1}$     &    1.93$_{-0.19}^{+0.21}$    &     159.62  &   196  &     0       & 9.07 &$<$ 0.001 \\
14498 &  1 &   1.80$_{-0.36}^{+0.05}$ &      $<$0.37                             &         0.58$_{-0.24}^{+0.26}$ &        1.0$_{-0.6}^{+0.7}$  &       8.3$_{-2.2}^{+3.0}$   &  1.81$_{-0.21}^{+0.21}$      &        94.73  &   124   &    0         & 7.59  &$<$ 0.001 \\
14498 &  1 &   2.23$_{-0.08}^{+0.05}$ &       $<$0.11                            &         1.02$_{-0.50}^{+0.51}$  &       1.2$_{-0.8}^{+1.0}$   &      8.3$_{-2.2}^{+3.0}$   & 1.81$_{-0.21}^{+0.21}$       &        94.73  &   124   &    0         & 7.54 & $<$ 0.001\\
14500 &  2 &   1.81$_{-0.04}^{+0.05}$ &      $<$0.10                             &          0.30$_{-0.12}^{+0.12}$ &       1.3$_{-0.8}^{+1.0}$  &      23.2$_{-4.0}^{+4.8}$   &  1.84$_{-0.13 }^{+0.13}$     &      178.21  &   204  &     1        & 4.59  & 0.008\\

\hline \hline
\end{tabular}
\end{center}
\tablecomments{{}$^{a}${\it Chandra} Observation ID,  
{}$^{b}$Image numbers 1, 2, 3, and 4 correspond to images A, B, C, and D of \sdss,  
{}$^{c}$Observed-frame energy of the shifted Fe line, 
{}$^{d}$ Observed-frame energy width of the shifted Fe line, 
{}$^{e}$Rest-frame equivalent width of the shifted Fe line, 
{}$^{f}$Flux of the shifted Fe line in units of  $\times$ 10$^{-6}$~photons~cm$^{-2}$~s$^{-1}$, 
{}$^{g}$Flux density of the continuum in units of $\times$ 10$^{-5}$~photons~keV$^{-1}$~cm$^{-2}$~s$^{-1}$ at 1~keV, 
{}$^{h}$ Cash statistic, 
{}$^{i}$degrees of freedom, 
{}$^{j}$P values of 0 and 1 correspond to confidence detection levels of the shifted Fe line of  $>$~99\% and  $>$~90\% respectively.
{}$^{k}$$F$-statistic between the null and alternative model.
{}$^{l}$The probability of exceeding this $F$-value as determined from the Monte-Carlo simulations.}
\end{table}

\clearpage
\begin{table}
\caption{Correlation Results of \rxj.} 
\scriptsize
\begin{center}
\begin{tabular}{lccccc}
\hline
\multicolumn{1}{c} {TEST} &
\multicolumn{1}{c} {QUANTITY I}  &
\multicolumn{1}{c} {QUANTITY II}  &
\multicolumn{1}{c} {$N_{\rm sample}$${}^{a}$} &
\multicolumn{1}{c} {STATISTIC} &
\multicolumn{1}{c} {CHANCE PROBABILITY} \\
\multicolumn{1}{c} {} &
\multicolumn{1}{c} {} &
\multicolumn{1}{c} {} &
\multicolumn{1}{c} {} &
\multicolumn{1}{c} {} &
\multicolumn{1}{c} {}  \\ \hline
Spearman's ($\rho$)       &  $E_{\rm min}$${}^{b}$   & $E_{\rm max}$${}^{b}$   & 9   &   0.78        &  0.015    \\
Kendalls's  ($\tau$)         &  $E_{\rm min}$${}^{b}$   & $E_{\rm max}$${}^{b}$   & 9   &   0.61         &  0.022     \\

Spearman's ($\rho$)       &  $E_{\rm min}$${}^{c}$   & $E_{\rm max}$${}^{c}$    & 17   &  0.52        & 0.032      \\
Kendalls's  ($\tau$)         &  $E_{\rm min}$${}^{c}$    & $E_{\rm max}$${}^{c}$    & 17   &   0.39        &  0.031     \\

Spearman's ($\rho$)       &  $EW_{\rm Fe}$  & $g$                   & 78    & 0.42         &  1.6 $\times$ 10$^{-4}$    \\
Kendalls's  ($\tau$)         &  $EW_{\rm Fe}$  & $g$                   & 78    &  0.28         & 2.4 $\times$ 10$^{-4}$     \\

Spearman's ($\rho$)       &  $N_{\rm Fe}$  & $N_{\rm cont}$   &  78  &  0.69      &    4.4 $\times$ 10$^{-12}$      \\
Kendalls's  ($\tau$)         &  $N_{\rm Fe}$  & $N_{\rm cont}$   &   78 &  0.50              &     $<$ 1 $\times$ 10$^{-12}$  \\

\hline \hline
\end{tabular}
\end{center}
\tablecomments{ Spearman's ($\rho$) and Kendalls's ($\tau$) rank correlations of two sample populations labelled QUANTITY~I and QUANTITY~II.
 ${}^{a}$  The number of elements in the sample.
 ${}^{b}$ Rest-frame energies of the shifted double Fe lines in image A detected at the  $>$ 90\% confidence level.
${}^{c}$ Rest-frame energies of the shifted double Fe lines in all images detected at the  $>$~90\% confidence level.}
\end{table}

\clearpage
 \begin{figure}
   \includegraphics[width=15cm]{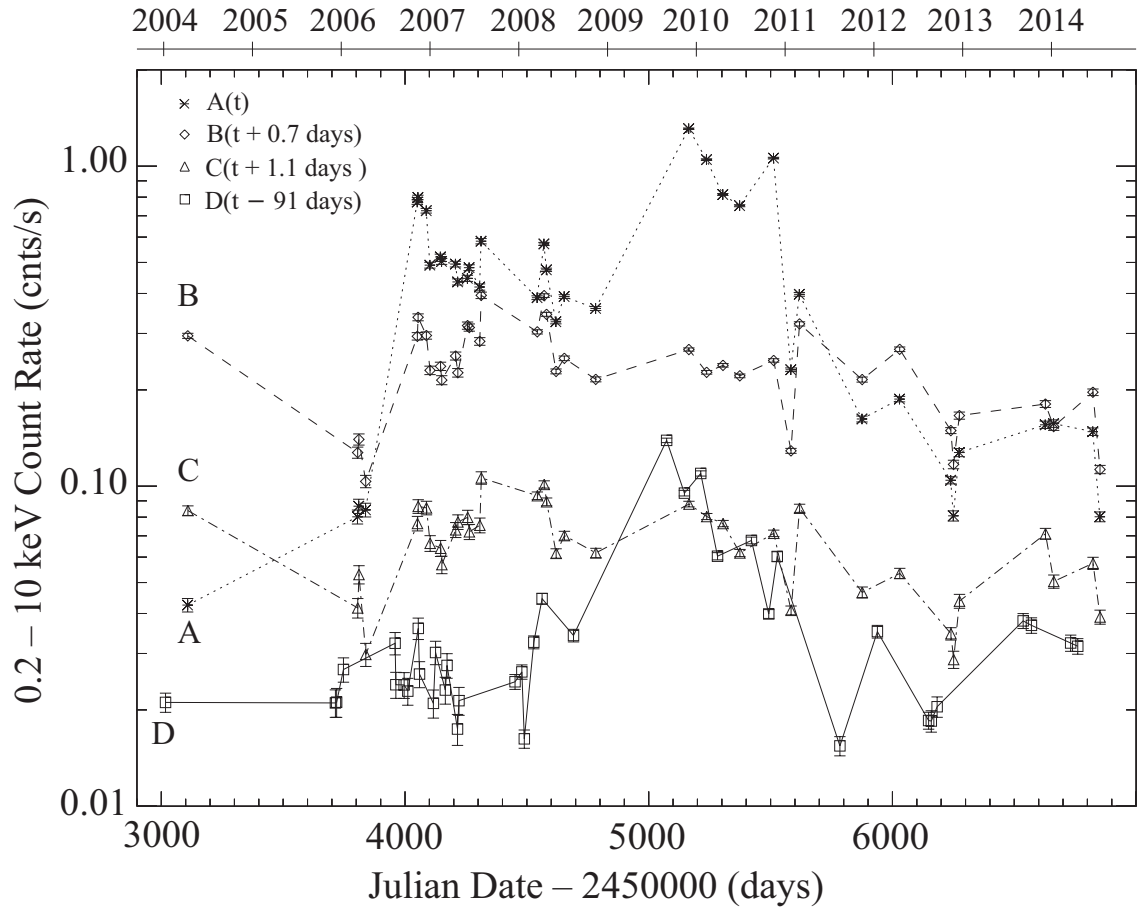}
        \centering
\caption[]{
The total (0.2--10~keV) light-curves of images A, B, C and D of \rxj\ shifted by the time delays estimated by 
Tewes et al. (2012). %Chantry et al. (2010).  
The total counts for images A and B have been corrected for pile-up effects while pile-up 
is unimportant for images C and D. The new X-ray data begin after epoch 29 (February 2011, $\hbox{JD}-2450000=5618$). 
}
\end{figure}

\clearpage
\begin{figure}
   \includegraphics[width=15cm]{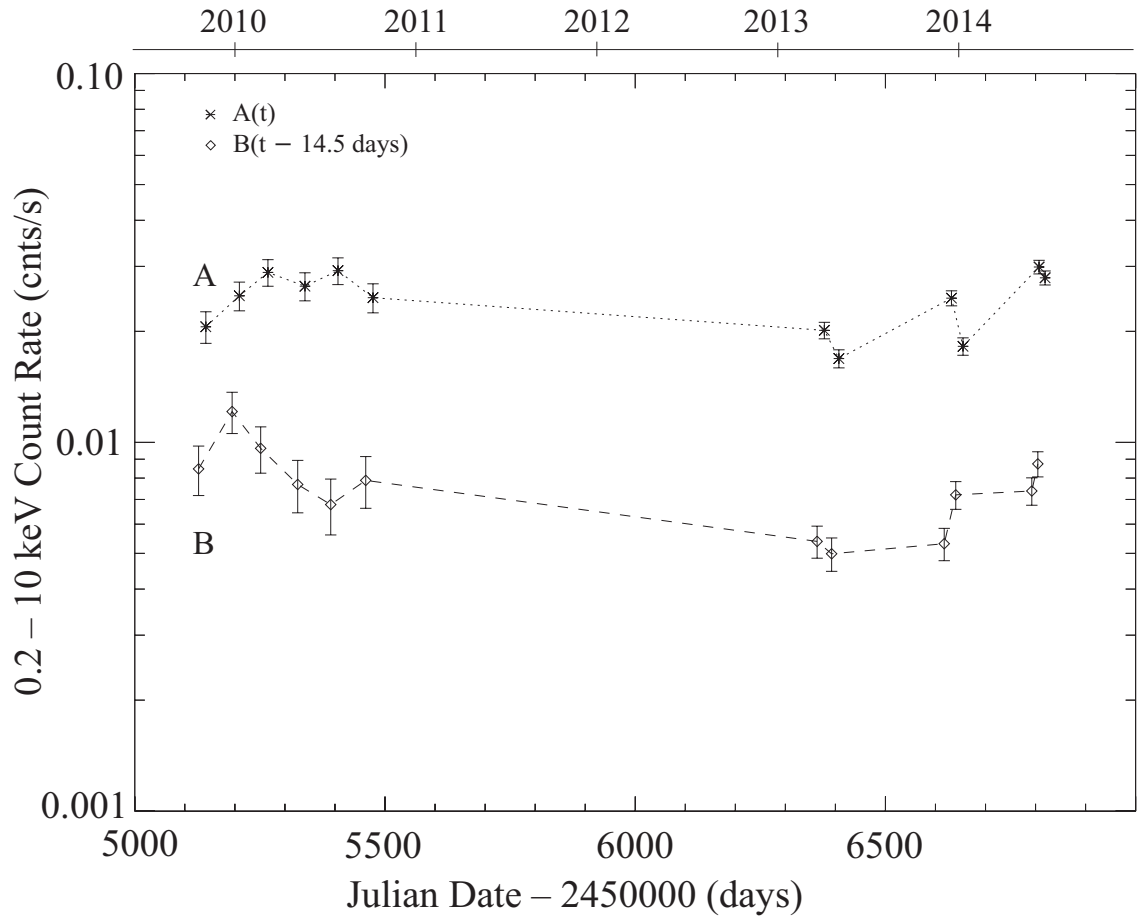}
        \centering
\caption[]{
The total (0.2--10~keV) light-curves of images A and B  of \qj\ shifted by the time delay of ${\Delta}$t$_{A-B}$ = $-$14.5 days estimated by 
Faure et al. (2009).  
The new X-ray data begin after epoch 6 (October 2010, $\hbox{JD}-2450000=5476$). 
%Data from epochs 1 -- 6 were presented in Chen et al. (2012). 
}
\end{figure}

\clearpage
\begin{figure}
   \includegraphics[width=15cm]{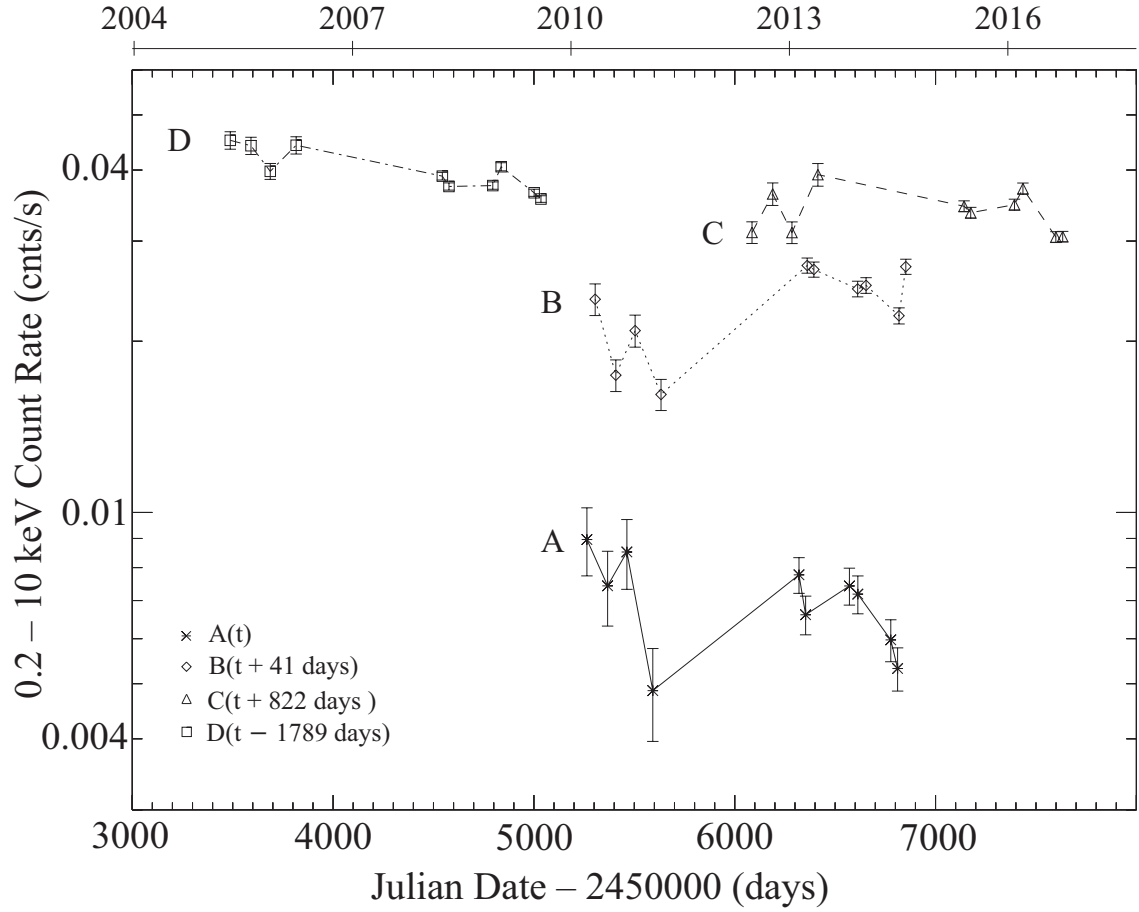}
        \centering
\caption[]{
The total (0.2--10~keV) light-curves of images A, B, C, and D  of \sdss\ shifted by the time delays estimated by 
Fohlmeister et al. (2008, 2016). We have added offsets of 0.01 cnts~s$^{-1}$, 0.02 cnts~s$^{-1}$, and 0.03 cnts~s$^{-1}$  to the light-curves of the images B, C, and D, respectively, for clarity. The new X-ray data begin after epoch 5 (March 2013, $\hbox{JD}-2450000=6353$). 
%We note that the published time delay for image D is only a limit.
% and the true delay may be actually larger \Delta_{AB} $\sim$
%Data from epochs 1 -- 5 were presented in Chen et al. (2012). 
}
\end{figure}

\clearpage
\begin{figure}
   \includegraphics[width=15cm]{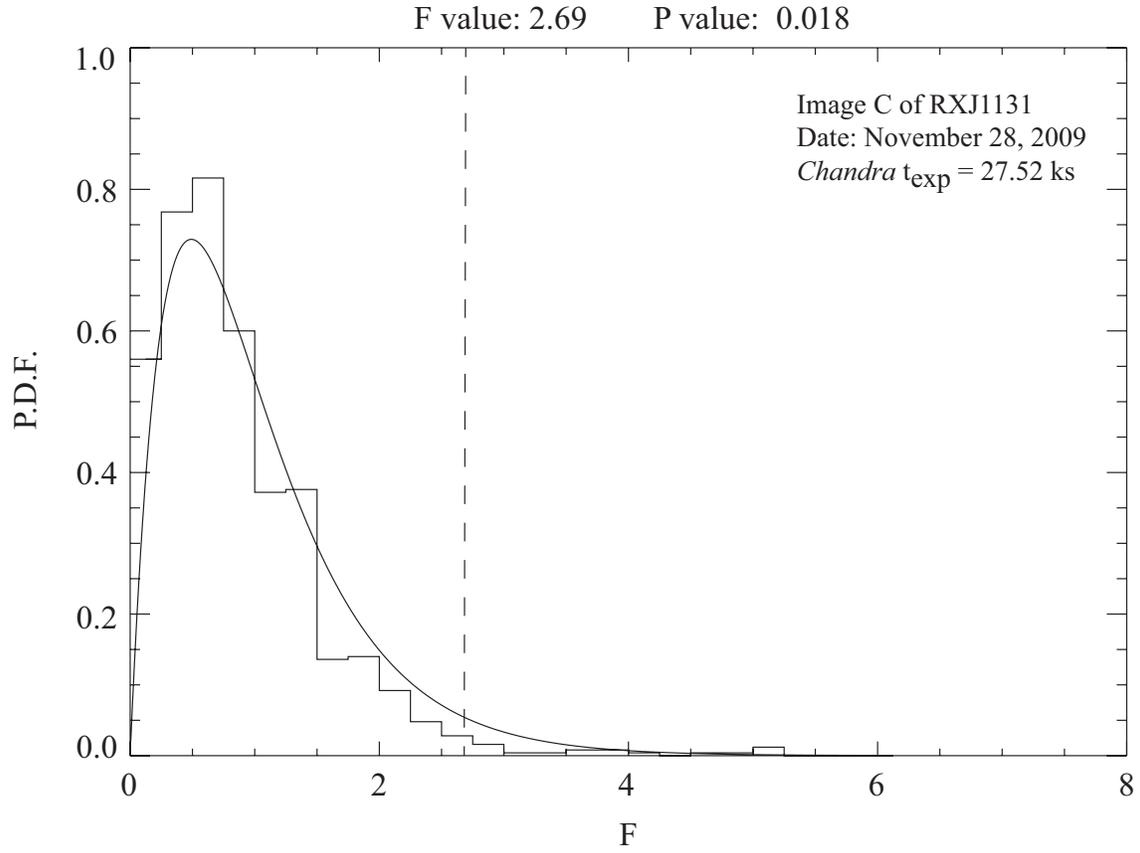}
        \centering
\caption[]{Monte-Carlo simulated (histogram) and theoretical (smooth curve) probability density distributions of the $F$-statistic between fits of
models that included a simple absorbed power-law (null model), and one that included one Gaussian emission line (alternative model)
 to the observed spectrum of image C of RXJ 1131 obtained in November 28, 2009 (obsid = 11540). We find that the probability of obtaining an $F$ value larger than 2.69 is P = 0.018.
 }
\end{figure}

\clearpage
\begin{figure}
   \includegraphics[width=15cm]{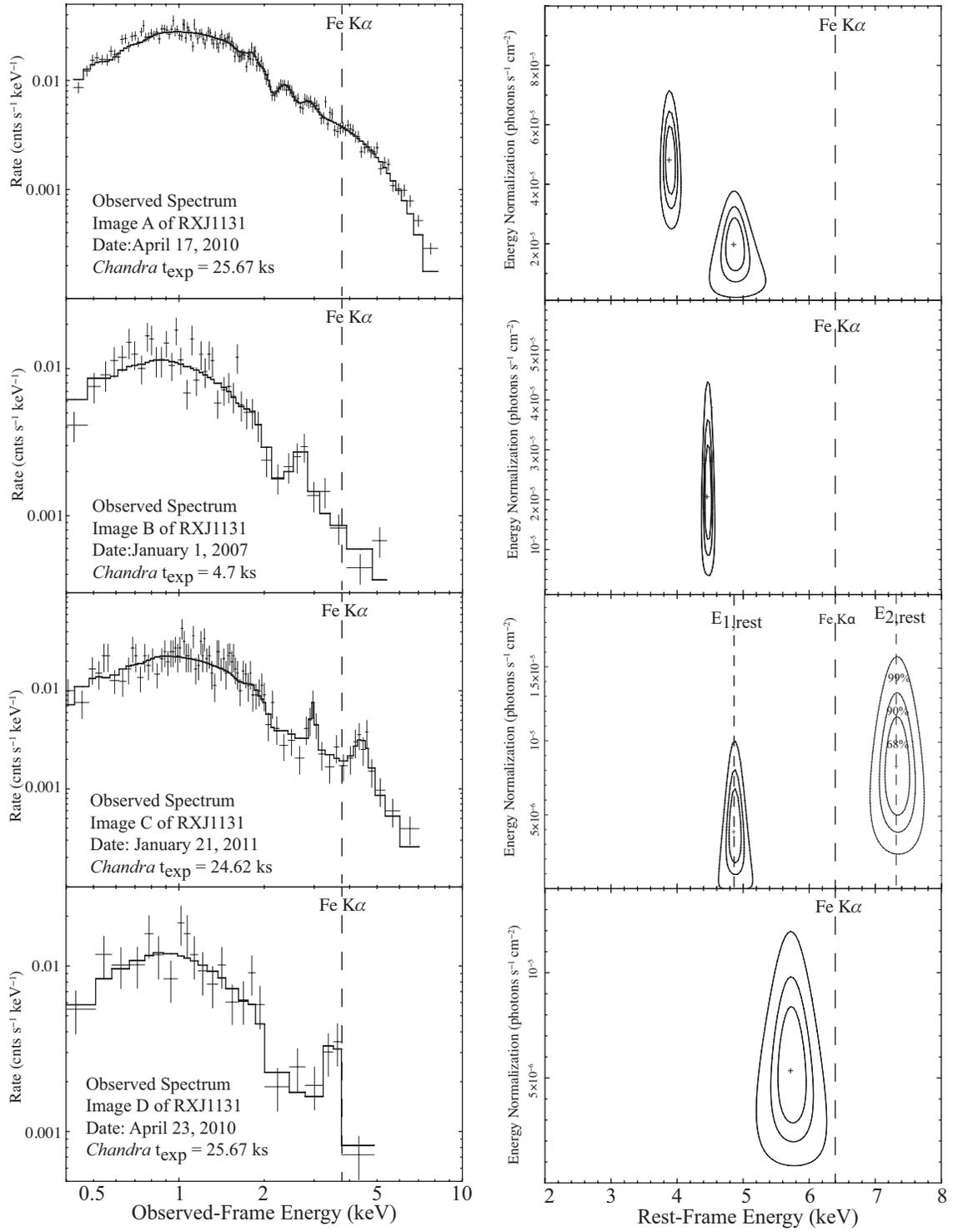}
        \centering
\caption[]{(left) Spectra of images A,B,C, and D of RXJ1131 at 4 different epochs showing the shifted iron lines. The best-fit model is comprised of a power law,  one or two Gaussian lines,  plus Galactic and intrinsic absorption. (right) The 68\%, 90\%, and 99\%  $\chi^{2}$ confidence contours for the line energies and flux normalizations.
 }
\end{figure}

\clearpage
\begin{figure}
   \includegraphics[width=15cm]{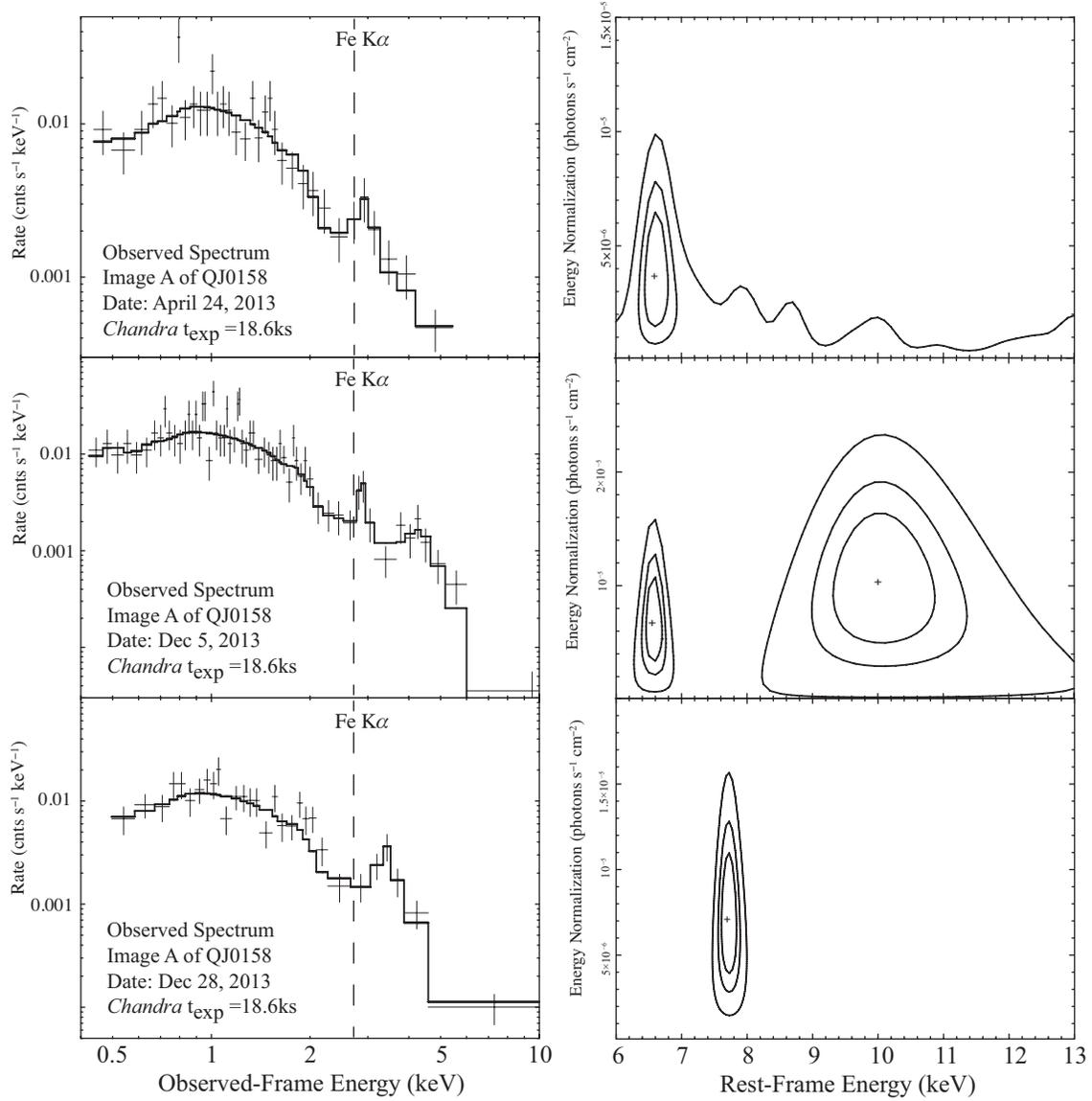}
        \centering
\caption[]{(left) Spectra of image A of QJ0158 at 3 different epochs showing the shifted iron lines. The best-fit models are comprised of a power law, and one or two Gaussian lines,  plus Galactic and intrinsic absorption. (right) The 68\%, 90\%, and 99\%  $\chi^{2}$ confidence contours for the line energies and flux normalizations.
 }
\end{figure}

\clearpage
\begin{figure}
   \includegraphics[width=15cm]{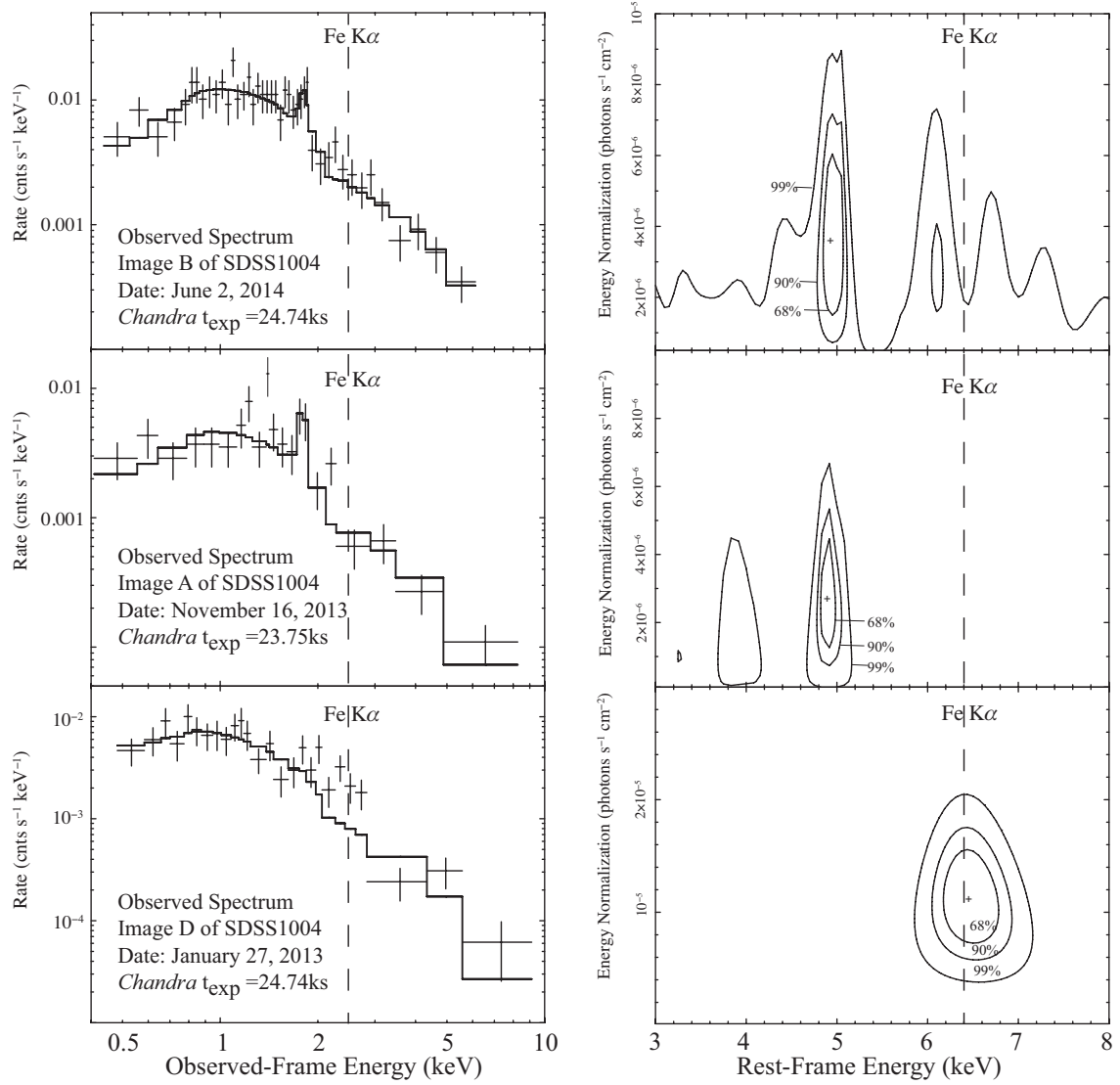}
        \centering
\caption[]{(left) Spectra of images B, A, and D of SDSS1004 at 3 different epochs showing shifted iron lines. The best-fit models are comprised of a power law, and a Gaussian line, plus Galactic and intrinsic absorption. The model for image D does not include the Gaussian line to better show the residual line emission near the instrumental edge at $\sim$ 2~keV. (right) The 68\%, 90\%, and 99\%  $\chi^{2}$ confidence contours for the line energies and flux normalizations.
 }
\end{figure}

\clearpage
\begin{figure}
   \includegraphics[width=15cm]{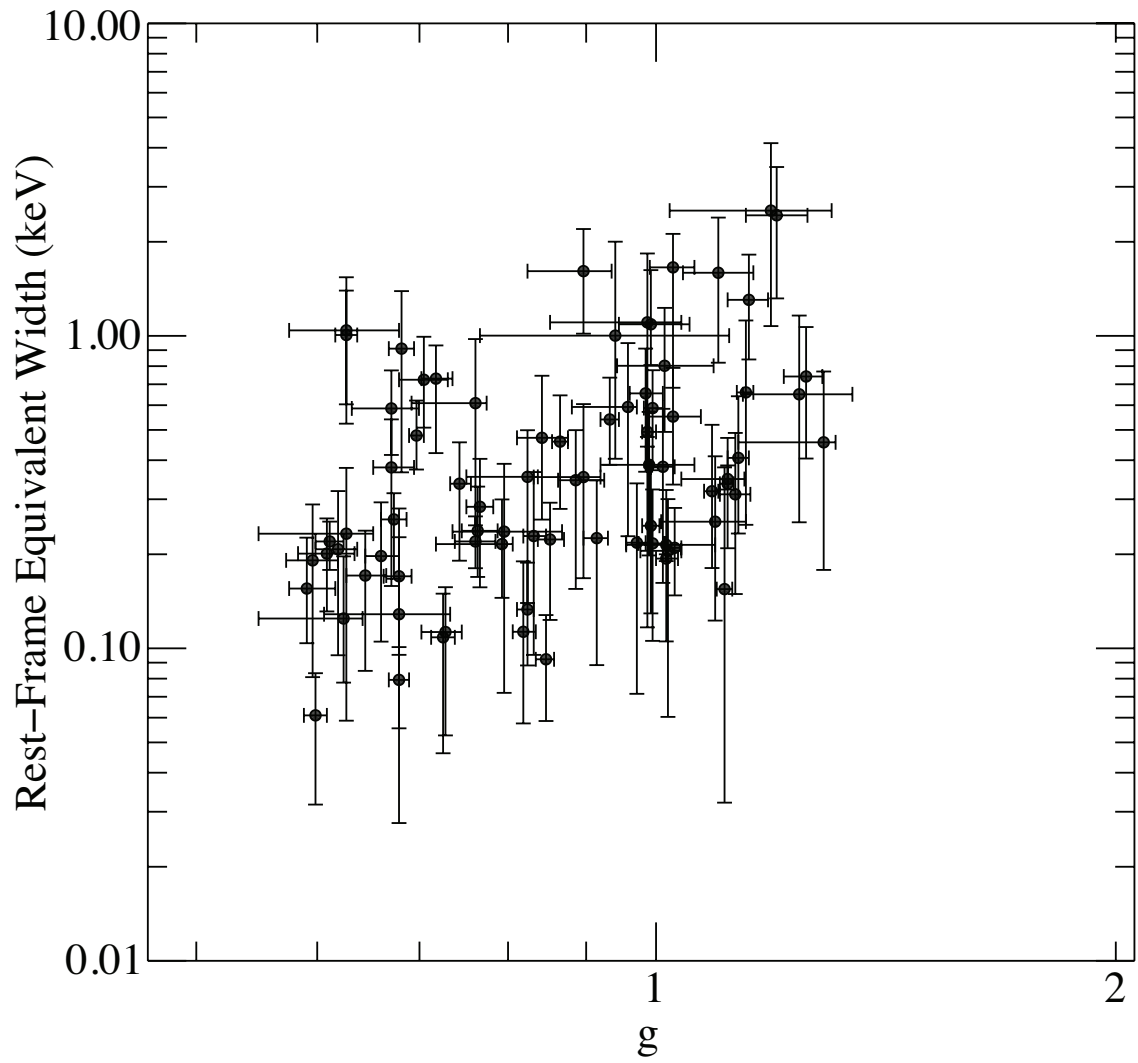}
        \centering
\caption[]{The Fe line equivalent width as a function of the generalized Doppler shift parameter $g$ for all images of RXJ1131. Only cases where the Fe line is detected at $>$ 90\% confidence are shown.
}
\end{figure}

\clearpage
\begin{figure}
   \includegraphics[width=15cm]{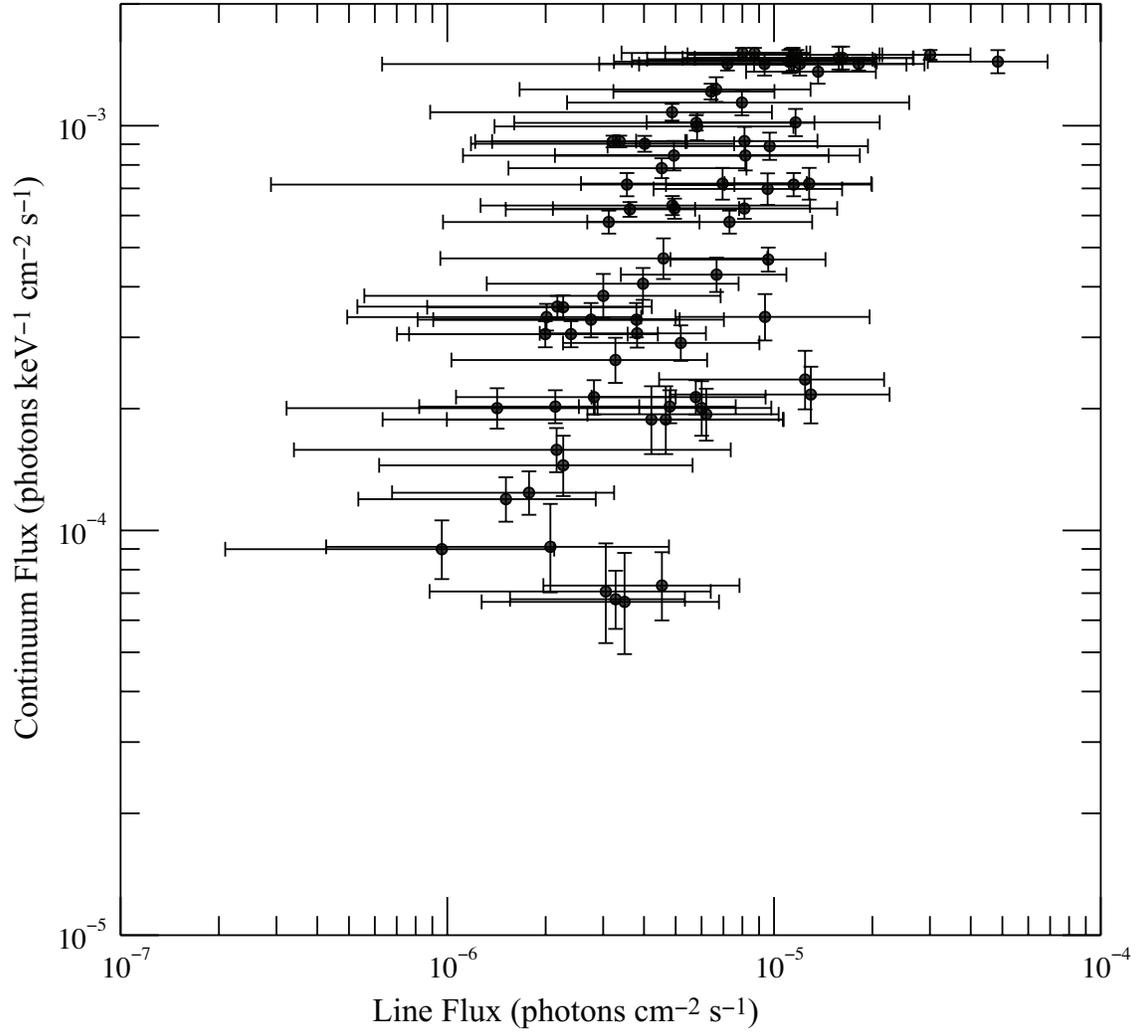}
        \centering
\caption[]{The flux of the Fe~K$\alpha$ line as a function of the continuum flux for Fe~K$\alpha$ lines detected at $>$ 90\% confidence in RXJ1131.
The line flux is the normalization of the Gaussian line component of the best-fit model.
The continuum flux is the normalization of the power-law component of the best-fit model calculated at 1~keV. }
\end{figure}

\clearpage
\begin{figure}
 \includegraphics[width=16cm]{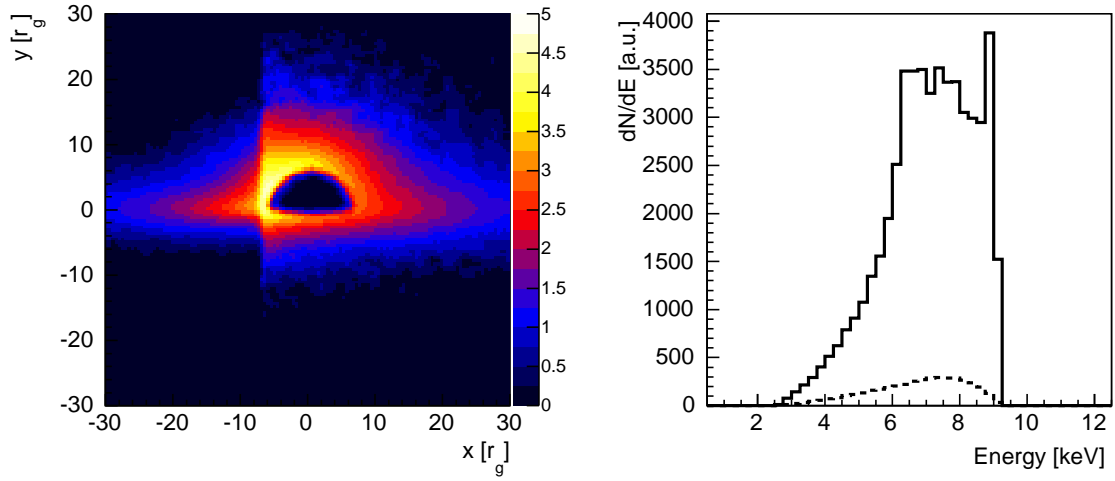}
        \centering
\caption{The left panel shows an image of the surface brightness of the Fe-K$\alpha$ line emission
as seen by an observer at 10$^{4}~r_{\rm g}$ and an inclination of $i=82.5^{\circ}$
from a black hole of spin $a$=0.3 for a caustic crossing angle of $\theta_{c}=\pi/2$.
The surface brightness scale is  logarithmic with an arbitrary absolute scale.
%The expressions $k^{\tilde{\mu}}$ refer to the components of the
%photon packet's wavevector in the reference frame of a coordinate stationary observer at a Boyer Lindquist 
%radial coordinate $r=r_{\rm r}=10^{4}\, r_{\rm g}$. 
The surface brightness exhibits a left-right asymmetry owing to the motion of the
accretion disk plasma towards to (left) or away from (right)  the observer. The right panel shows the resulting 
energy spectrum of the Fe-K$\alpha$ emission in the rest frame of the source.}
\end{figure}

\clearpage
\begin{figure}
 \includegraphics[width=16cm]{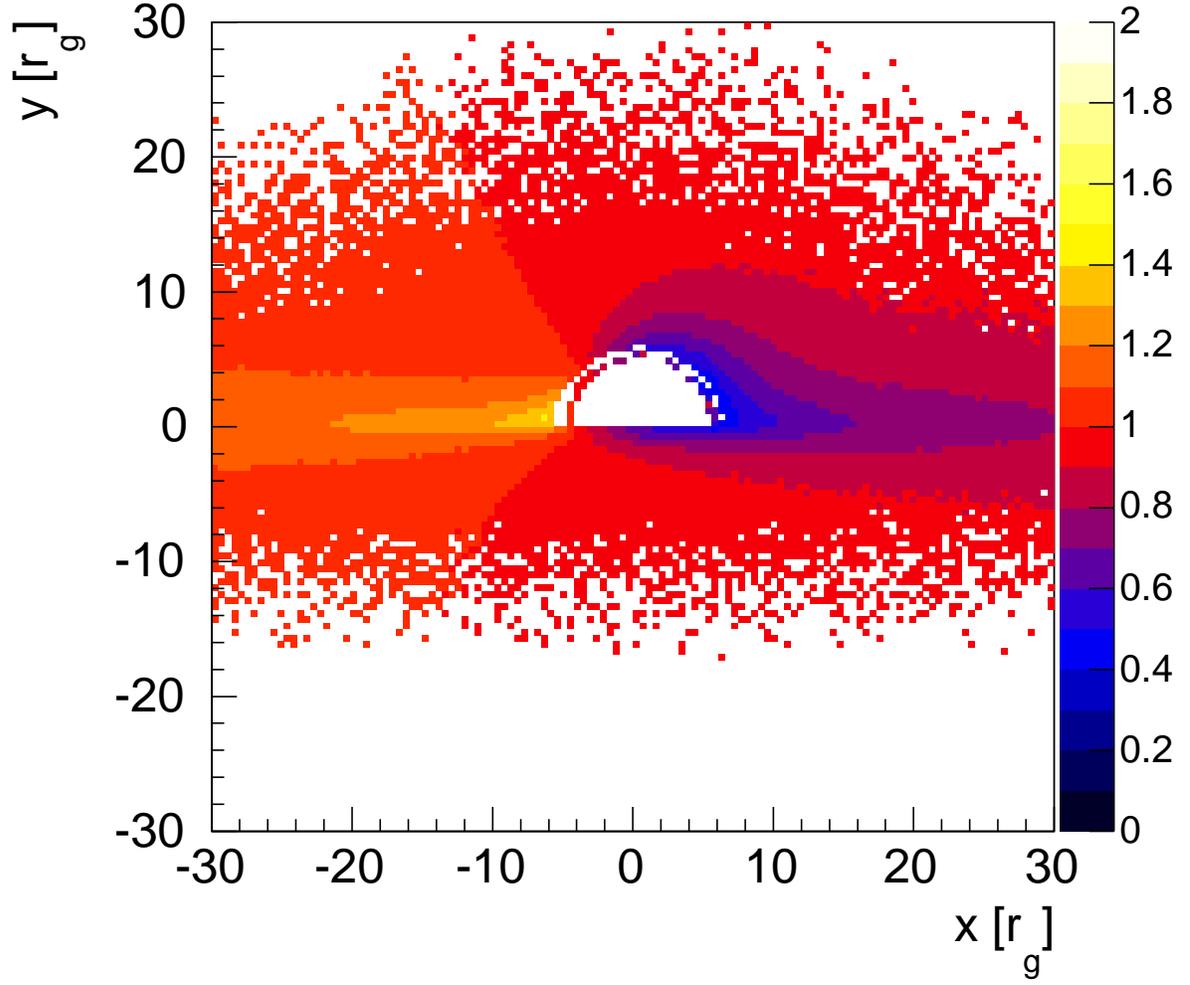}
        \centering
\caption{Map of the $g$-factors of the Fe-K$\alpha$ emission originating in the accretion disk.
The net $g$-factors result from the motion of the accretion disk plasma relative to the observer shown in Figure 9 and the propagation
of the photon through the curved spacetime for the same parameters as were used in Figure 9. }%Same parameters as in Figure 9.}
\end{figure}

\clearpage
\begin{figure}
 \includegraphics[width=16cm]{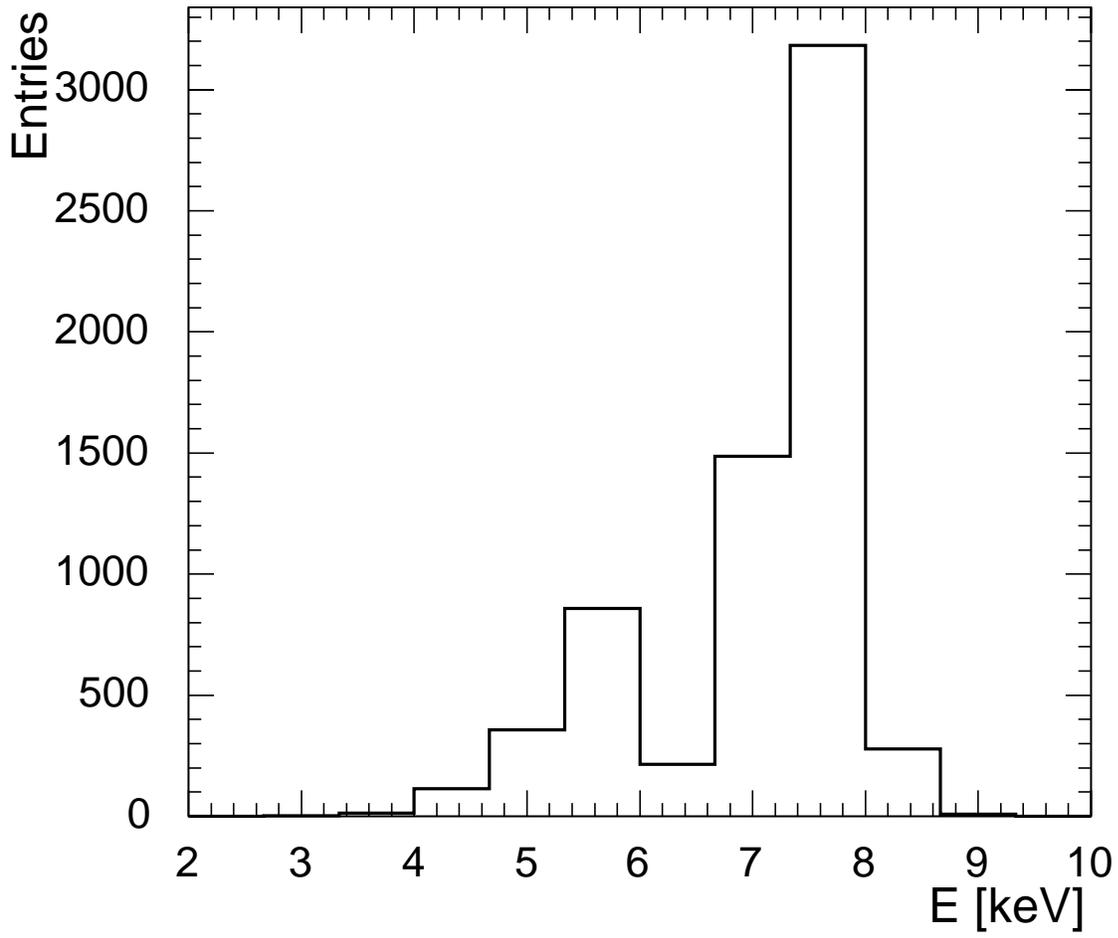}
        \centering
\caption{Simulated distribution of the {\it single and double peak} Fe line energies for a black hole 
with a spin of $a=0.3$ seen at an inclination of $i=82.5^{\circ}$.}
\end{figure}

\clearpage
\begin{figure}
 \includegraphics[width=16cm]{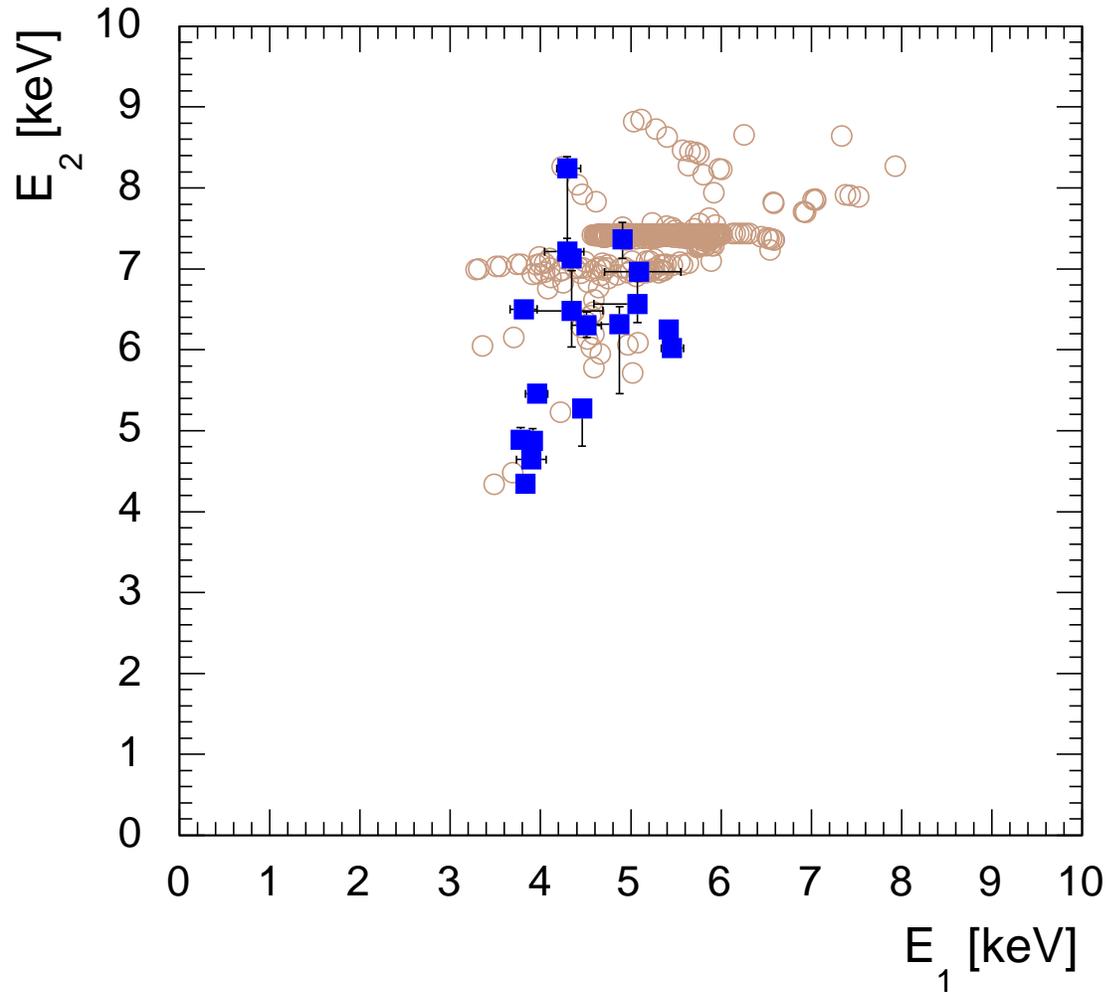}
        \centering
\caption{Scatter between the peak energies of simulated (open circles) and observed (solid squares) double-peaked line profiles. The simulations are for a  black hole with a spin of $a=0.3$ seen at an inclination of $i=82.5^{\circ}$.}
\end{figure}

\clearpage
\begin{figure}
   \includegraphics[width=15cm]{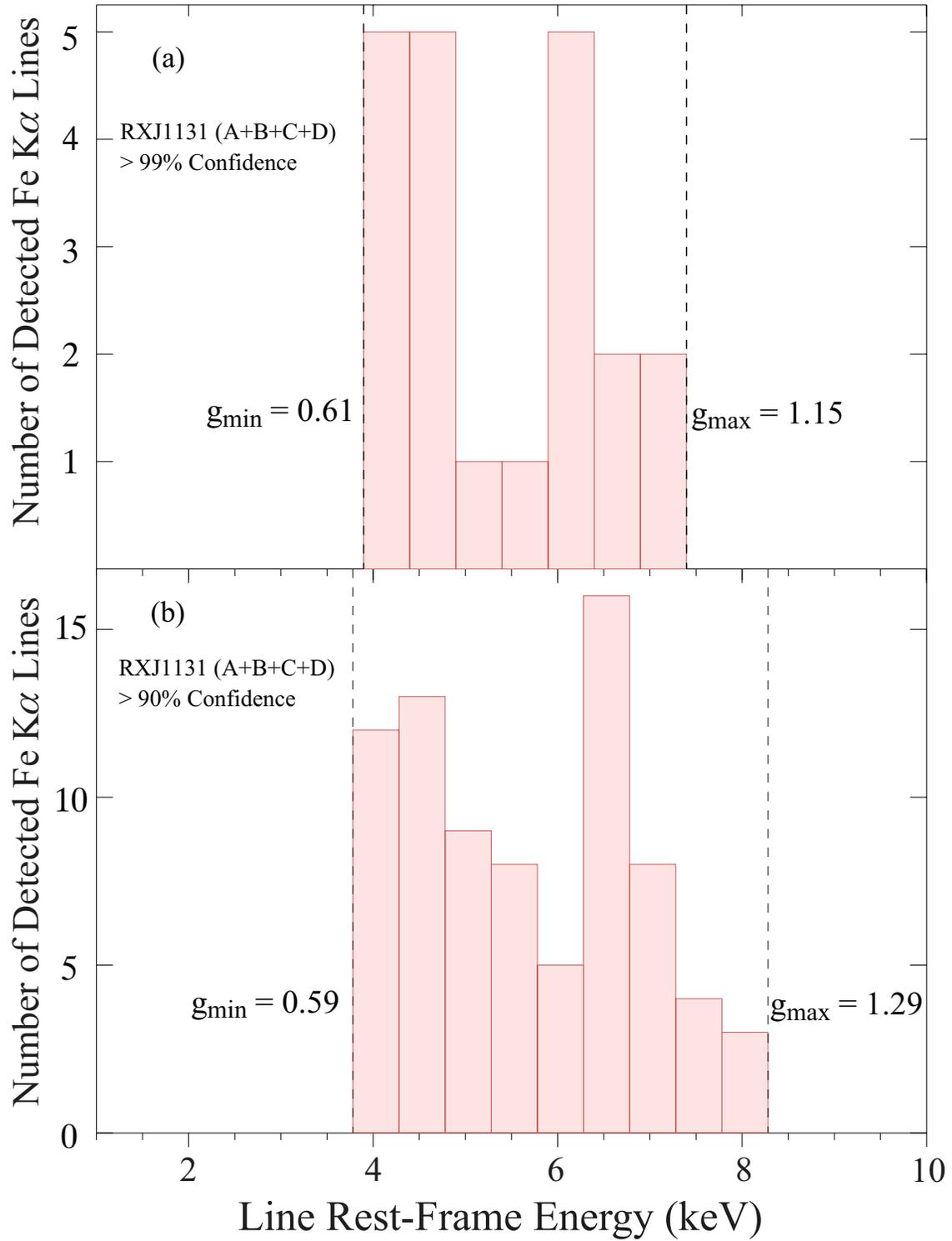}
        \centering
\caption[]{ Distribution of the Fe~K$\alpha$ line energies for all images and all 38 epochs of data for RXJ1131. Only cases where the iron line is detected at $>$~99\% (panel a) and at $>$~90\% confidence (panel~b) are shown. The vertical lines mark the extreme limits of the distribution used to determine upper limits on the ISCO and inclination angle.
 }
\end{figure}

\clearpage
\begin{figure}
   \includegraphics[width=15cm]{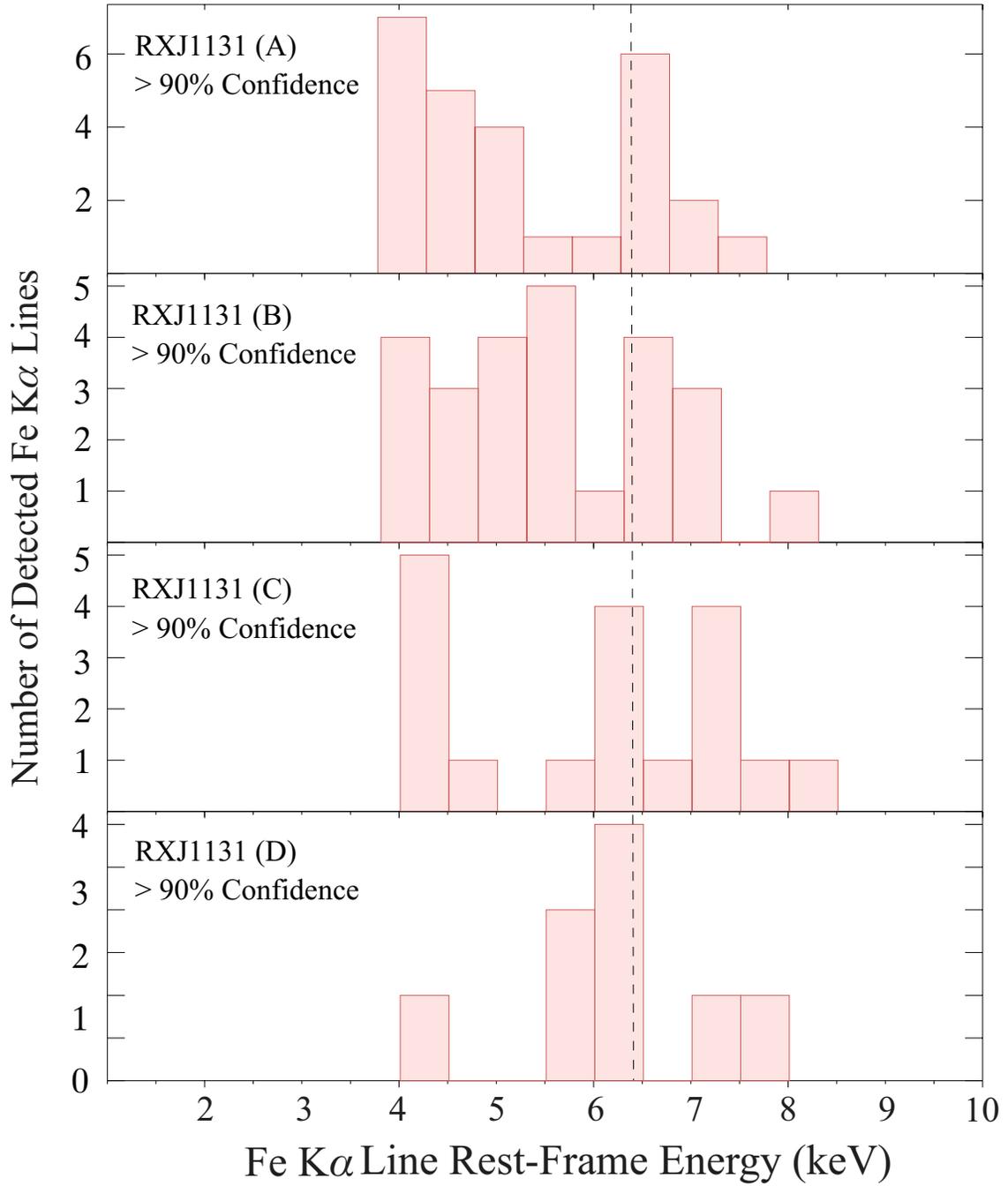}
        \centering
\caption[]{ Distribution of the Fe~K$\alpha$ line energies for the  individual images A, B, C, and D images and all 38 epochs of data for RXJ1131. Only cases where the iron line is detected  at $>$~90\% confidence are shown. 
 }
\end{figure}

\clearpage
\begin{figure}
   \includegraphics[width=14cm]{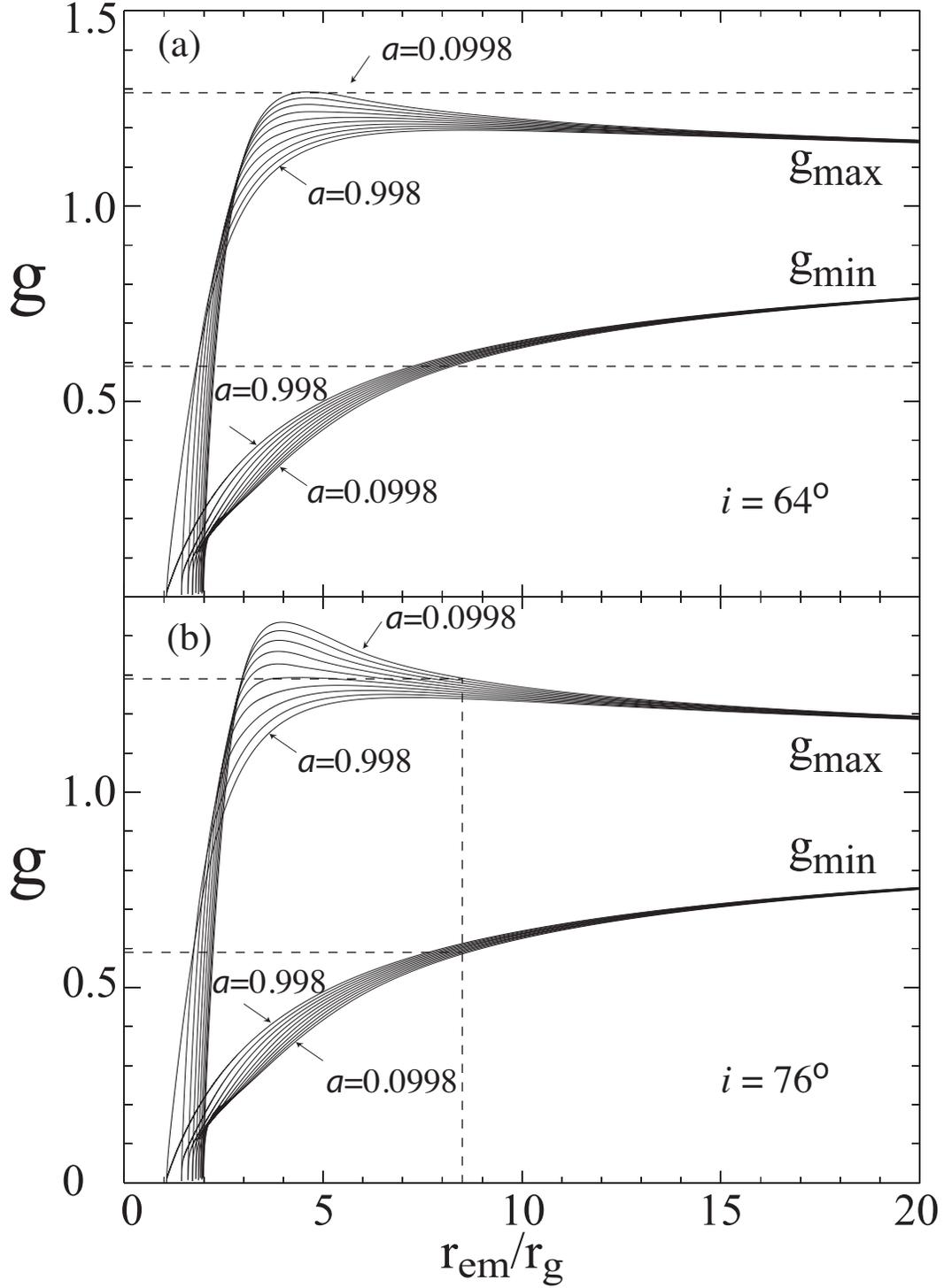}
        \centering
\caption[]{Extremal shifts of the Fe~K$\alpha$ line energy for spin values ranging between 0.098 and 0.998 in increments of 0.1. Horizontal  lines represent the observed values of  $g$~=~$E_{\rm obs}$/$E_{\rm rest}$ of the most redshifted and blueshifted Fe~K$\alpha$ lines from all 38 epochs and all images of RXJ1131. The extreme $g$ values are for Fe~K$\alpha$  lines detected at $>$ 90\% confidence.  
(a) The extremal shifts for an inclination angle of $i$ = 64$^{\circ}$.
(b) The extremal shifts for an inclination angle of $i$ = 76$^{\circ}$. The inner radius of the accretion disk is constrained to be $r_{\rm ISCO}$ $<$ 8.5$r_{\rm g}$
  }
\end{figure}

\clearpage
\begin{figure}
   \includegraphics[width=15cm]{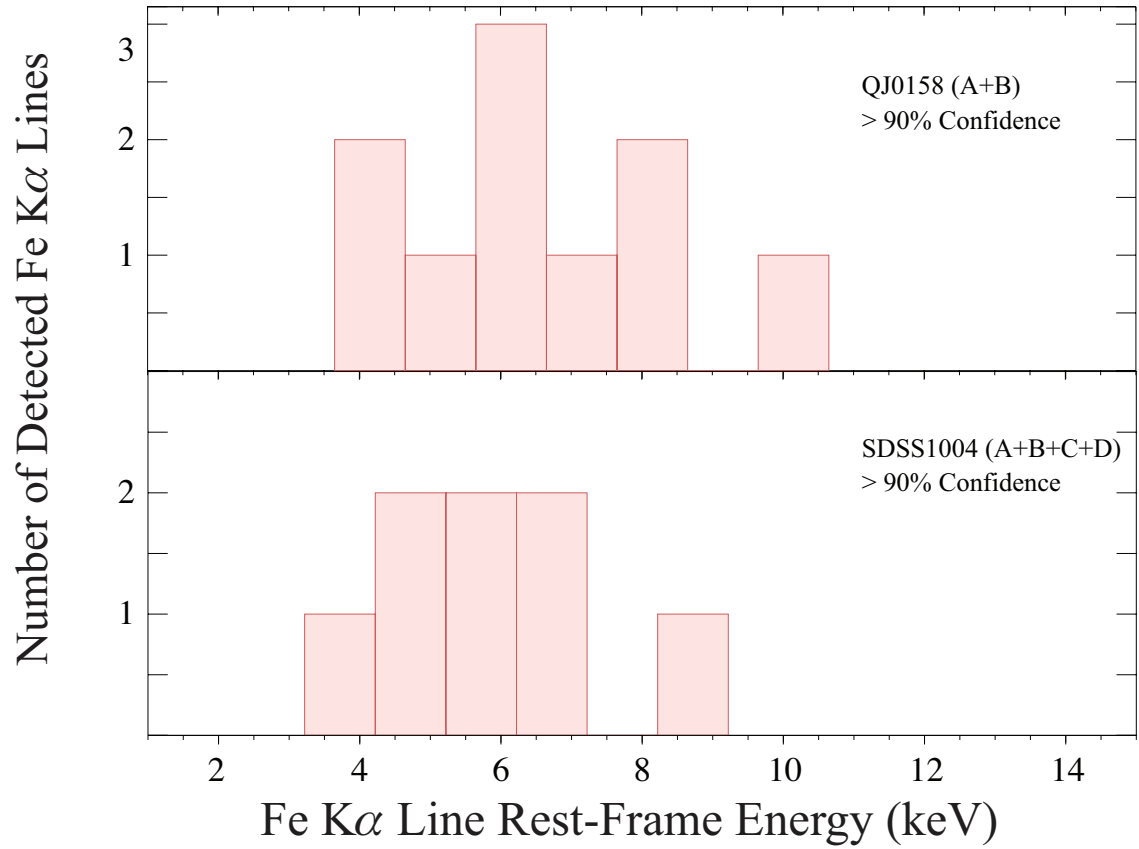}
        \centering
\caption[]{ Distribution of the Fe~K$\alpha$ line energies for all images of QJ0158 (top, 12 epochs) and SDSS1004 (bottom, 10 epochs).  Only cases where the iron line is detected at $>$~90\% confidence  are shown.
 }
\end{figure}

\clearpage
\begin{figure}
   \includegraphics[width=16cm]{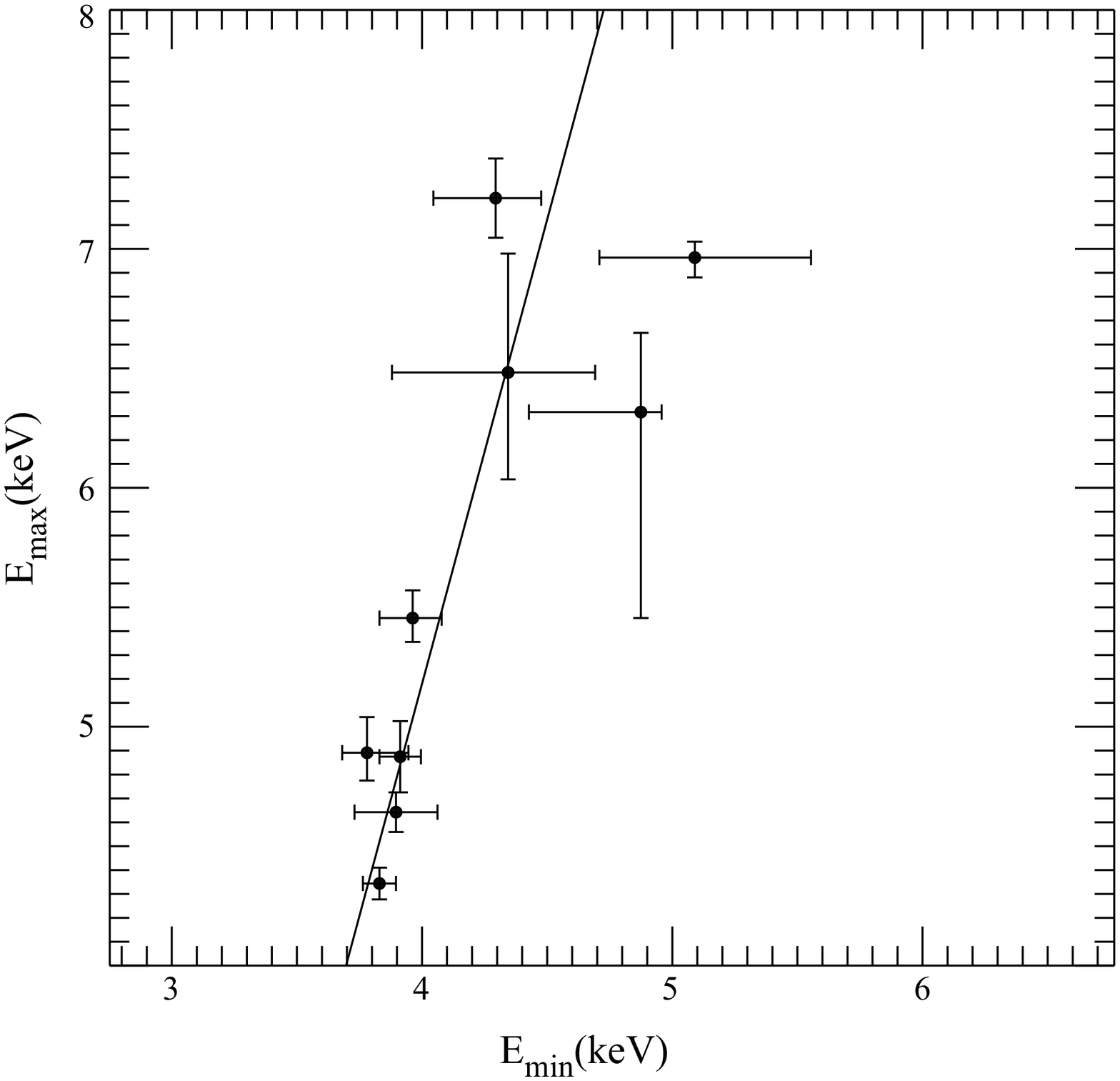}
        \centering
\caption[]{The rest-frame energies of the shifted double Fe lines detected in image A of RXJ1131 at the $>$ 90\% confidence level. 
We also show the straight-line least-squares fit to the data in the solid line.
 }
\end{figure}

\clearpage
\begin{figure}
   \includegraphics[width=16cm]{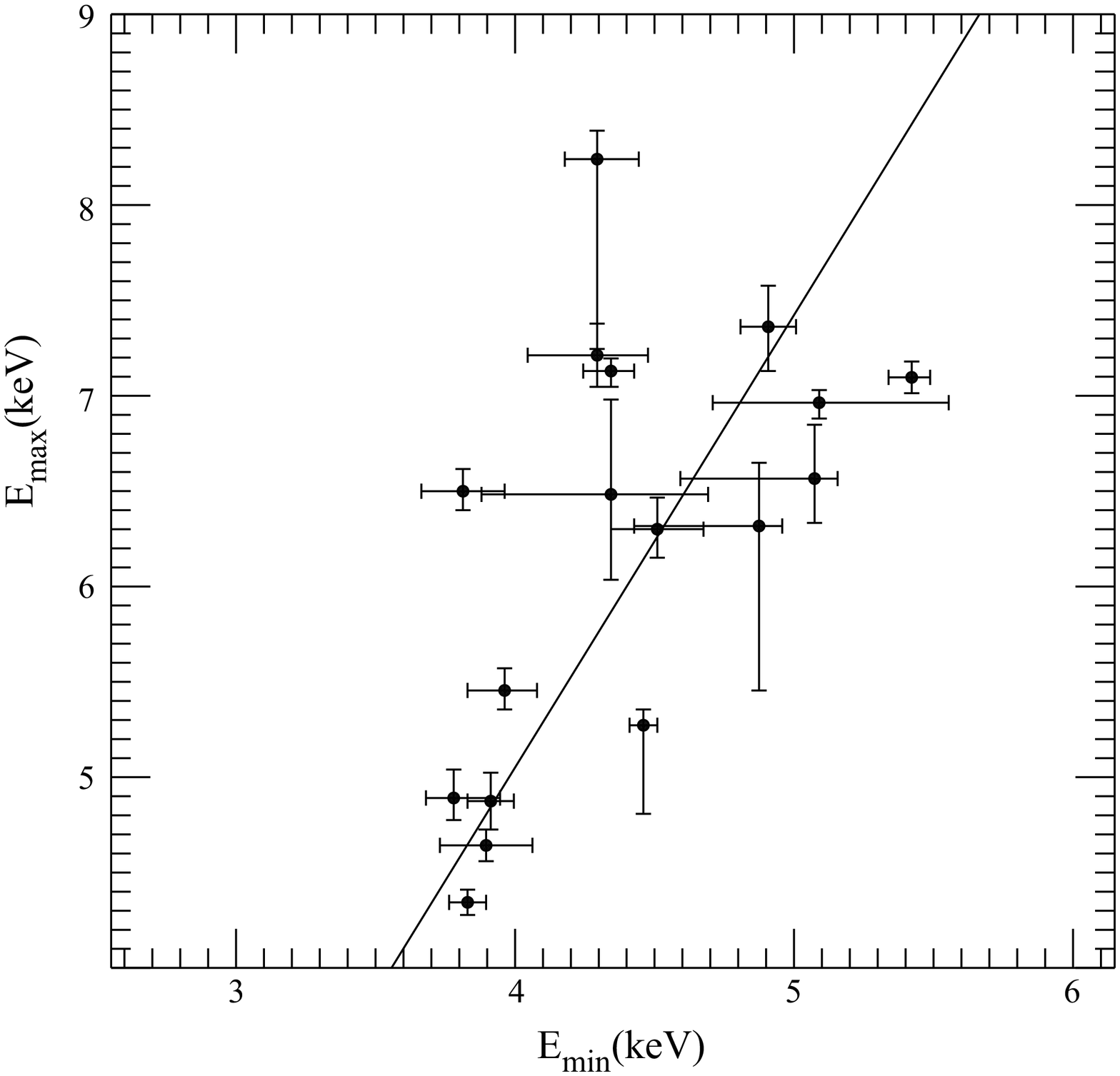}
        \centering
\caption[]{The rest-frame energies of the shifted double Fe lines detected in all images of RXJ1131 at the $>$ 90\% confidence level.  
We also show the straight-line least-squares fit to the data in the solid line.
}
\end{figure}

\clearpage
\begin{figure}
   \includegraphics[width=16cm]{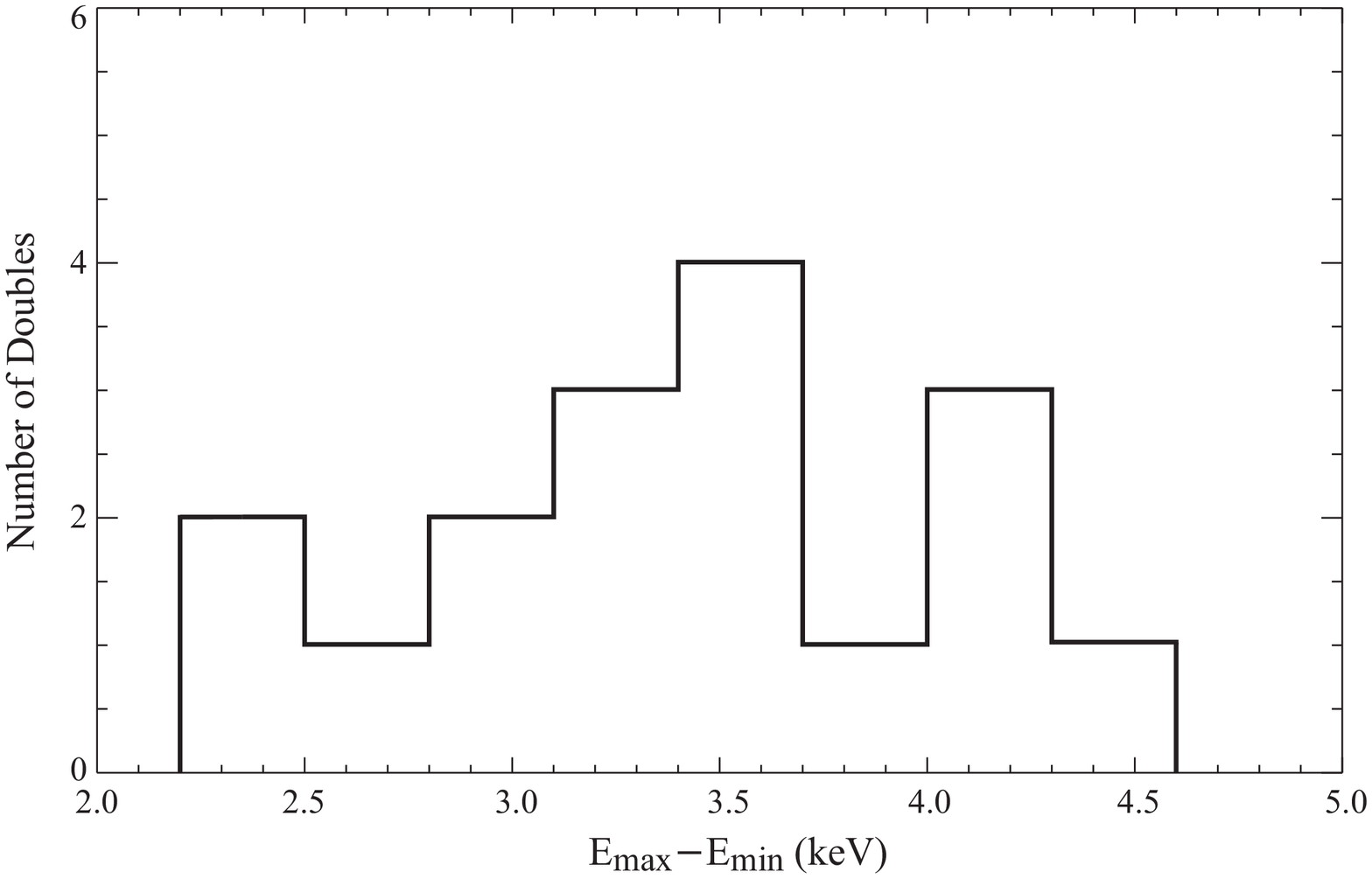}
        \centering
\caption[]{The observed distribution of the rest-frame energy separations for shifted double Fe lines  detected in all images of RXJ1131 at the $>$ 90\% confidence level.  
}
\end{figure}


\clearpage


\begin{references}

%\reference{} Assef, R.~J., Denney, K.~D., Kochanek, C.~S., et al.\ 2011, \apj, 742, 93 \\

%\reference{} Faure, C., \\Anguita, T., Eigenbrod, A., et al.\ 2009, \aap, 496, 361 

%\reference{}  Tewes, M., Courbin, F., Meylan, G., et al.\ 2013, \aap, 556, A22 \\

\reference{} Bardeen, J.~M., Press, W.~H., \& Teukolsky, S.~A.\ 1972, \apj, 178, 347 \\

\reference{}  Beheshtipour, B., Hoormann, J., Krawczynski, H.\ 2016, ApJ (in press). \\

\reference{} Blackburne, J.~A., Pooley, D., \& Rappaport, S.\ 2006, \apj, 640, 569 \\

\reference{}  Blackburne, J.~A., Kochanek, C.~S., Chen, B., Dai, X., \& Chartas, G.\ 2014, \apj, 789, 125 \\

\reference{} Blackburne, J.~A., Kochanek, C.~S., Chen, B., Dai, X., \& Chartas, G.\ 2015, \apj, 798, 95 \\

\reference{} Cackett, E.~M., Zoghbi, A., Reynolds, C., et al.\ 2014, \mnras, 438, 2980 \\

\reference{} Chandrasekhar, S.\ 1960, New York: Dover, 1960 \\

\reference{} Chartas, G., Falco, E., Forman, W., et al.\ 1995, \apj, 445, 140  \\

\reference{} Chartas, G., Agol, E., Eracleous, M., et al.\ 2002, \apj, 568, 509 \\

\reference{} Chartas, G., Kochanek, C.~S., Dai, X., Poindexter, S., \& Garmire, G.\ 2009, \apj, 693, 174 \\

\reference{} Chartas, G., Kochanek, C.~S., Dai, X., et al.\ 2012, \apj, 757, 137 \\

\reference{} Chartas, G., Rhea, C., Kochanek, C., et al.\ 2016, Astronomische Nachrichten, 337, 356 \\

\reference{} Chen, B., Dai, X., Kochanek, C.~S., et al.\ 2011, \apjl, 740, L34 \\

\reference{} Chen, B., Dai, X., Kochanek, C.~S., et al.\ 2012, \apj, 755, 24 \\

\reference{} Dai, X., Chartas, G., Agol, E., Bautz, M.~W., \& Garmire, G.~P.\ 2003, \apj, 589, 100 \\

\reference{} Dai, X., \& Kochanek, C.~S.\ 2009, \apj, 692, 677 \\

\reference{} Dai, X., Kochanek, C.~S., Chartas, G., et al.\ 2010, \apj, 709, 278 \\

\reference{}  de Marco, B., Ponti, G., Uttley, P., et al.\ 2011, \mnras, 417, L98 \\

\reference{} Emmanoulopoulos, D., Papadakis, I.~E., Dov{\v c}iak, M., \& McHardy, I.~M.\ 2014, \mnras, 439, 3931 \\

\reference{} Fabian, A.~C., Rees, M.~J., Stella, L., \& White, N.~E.\ 1989, \mnras, 238, 729 \\

\reference{} Fabian, A.~C., Zoghbi, A., Ross, R.~R., et al.\ 2009, \nat, 459, 540 \\

\reference{}  Fohlmeister, J., Kochanek, C.~S., Falco, E.~E., et al.\ 2007, \apj, 662, 62 \\

\reference{} Fausnaugh, M.~M., Denney, K.~D., Barth, A.~J., et al.\ 2016, \apj, 821, 56 \\

\reference{} Fian, C., Mediavilla, E., Hanslmeier, A., et al.\ 2016, arXiv:1608.03831\\

\reference{} Garc{\'{\i}}a, J., Dauser, T., Reynolds, C.~S., et al.\ 2013, \apj, 768, 146 \\

\reference{} Hoormann, J.~K., Beheshtipour, B., \& Krawczynski, H.\ 2016, \prd, 93, 044020 \\

\reference{}  Kara, E., Alston, W.~N., Fabian, A.~C., et al.\ 2016, \mnras, 462, 511 \\

\reference{} Kara, E., Cackett, E.~M., Fabian, A.~C., Reynolds, C., \& Uttley, P.\ 2014, \mnras, 439, L26 \\

\reference{} Karas, V., \& Sochora, V.\ 2010, \apj, 725, 1507 \\

\reference{} Kerr, R.~P.\ 1963, Physical Review Letters, 11, 237 \\

\reference{} Kochanek, C.~S.\ 2004, \apj, 605, 58  \\

\reference{} Kochanek, C.~S., Dai, X., Morgan, C., Morgan, N., \& Poindexter, S.~C., G.\ 2007, Statistical Challenges in Modern Astronomy IV, 371, 43 \\

%\reference{} Kochanek, C.~S.\ 2006, Saas-Fee Advanced Course 33: Gravitational Lensing: Strong, Weak and Micro,  \\

\reference{} Krawczynski, H.\ 2012, \apj, 754, 133 \\

\reference{} Krawczynski, H., Chartas, G., Kislat, F., Beheshtipour, B.\ 2016, (in preparation). \\

\reference{} Kulkarni, A.~K., Penna, R.~F., Shcherbakov, R.~V., et al.\ 2011, \mnras, 414, 1183 \\

\reference{} MacLeod, C.~L., Morgan, C.~W., Mosquera, A., et al.\ 2015, \apj, 806, 258 \\

\reference{} Matt, G., Perola, G.~C., \& Piro, L.\ 1991, \aap, 247, 25 \\

\reference{} McKinney, J.~C., Tchekhovskoy, A., Sadowski, A., \& Narayan, R.\ 2014, \mnras, 441, 3177 \\

\reference{} Mosquera, A.~M., \& Kochanek, C.~S.\ 2011, \apj, 738, 96 \\

\reference{} Mosquera, A.~M., Kochanek, C.~S., Chen, B., et al.\ 2013, \apj, 769, 53 \\

\reference{} Motta, V., Mediavilla, E., Falco, E., \& Mu{\~n}oz, J.~A.\ 2012, \apj, 755, 82 \\

\reference{} Miller, L., Turner, T.~J., \& Reeves, J.~N.\ 2009, \mnras, 399, L69 \\

\reference{} Morgan, C.~W., Kochanek, C.~S., Dai, X., Morgan, N.~D., \& Falco, E.~E.\ 2008, \apj, 689, 755-761 \\

\reference{} Morgan, C.~W., Kochanek, C.~S., Morgan, N.~D., \& Falco, E.~E.\ 2010, \apj, 712, 1129 \\

\reference{} Morgan, C.~W., Hainline, L.~J., Chen, B., et al.\ 2012, \apj, 756, 52 \\

\reference{} Mosquera, A.~M., Kochanek, C.~S., Chen, B., et al.\ 2013, \apj, 769, 53 \\

\reference{} M{\"u}ller, A., \& Camenzind, M.\ 2004, A\&A, 413, 861 \\

\reference{} Noble, S.~C., Krolik, J.~H., Schnittman, J.~D., \& Hawley, J.~F.\ 2011, \apj, 743, 115 \\

\reference{} Novikov, I.~D., \& Thorne, K.~S.\ 1973, Black Holes (Les Astres Occlus), 343 \\

\reference{}  Page, D.~N., \& Thorne, K.~S.\ 1974, \apj, 191, 499 \\

\reference{} Pooley, D., Blackburne, J.~A., Rappaport, S., \& Schechter, P.~L.\ 2007, \apj, 661, 19 \\

\reference{} Pooley, D., Rappaport, S., Blackburne, J.~A., Schechter, P.~L., \& Wambsganss, J.\ 2012, \apj, 744, 111 \\

\reference{} Parker, M.~L., Wilkins, D.~R., Fabian, A.~C., et al.\ 2014, \mnras, 443, 1723 \\

\reference{} Penna, R.~F., S{\c a}owski, A., \& McKinney, J.~C.\ 2012, \mnras, 420, 684 \\

\reference{}  Protassov, R., van Dyk, D.~A., Connors, A., Kashyap, V.~L., \& Siemiginowska, A.\ 2002, \apj, 571, 545 \\

\reference{} Reis, R.~C., Reynolds, M.~T., Miller, J.~M., \& Walton, D.~J.\ 2014, \nat, 507, 207 \\

\reference{} Reynolds, C. S. 2014, Space Sci. Rev., 183, 277 \\

\reference{} Reynolds, C.~S., \& Begelman, M.~C.\ 1997, \apj, 488, 109 \\

\reference{} Reynolds, C.~S., \& Fabian, A.~C.\ 2008, \apj, 675, 1048-1056 \\

\reference{} Reynolds, C.~S., \& Nowak, M.~A.\ 2003, \physrep, 377, 389 \\

\reference{} Reynolds, M.~T., Walton, D.~J., Miller, J.~M., \& Reis, R.~C.\ 2014, \apjl, 792, L19  \\

\reference{} Ross, R.~R., \& Fabian, A.~C.\ 2005, \mnras, 358, 211 \\

\reference{} S{\c a}dowski, A.\ 2016, \mnras, 459, 4397 \\

\reference{} Schneider, P., Ehlers, J., \& Falco, E.~E.\ 1992, Gravitational Lenses, XIV, 560 pp.~112 figs..~Springer-Verlag Berlin Heidelberg New York.~ Also Astronomy and Astrophysics Library, 112 \\

\reference{} Schnittman, J.~D., \& Krolik, J.~H.\ 2010, \apj, 712, 908 \\ 

\reference{} Schnittman, J.~D., \& Krolik, J.~H.\ 2009, \apj, 701, 1175 \\

\reference{} Shakura, N.~I., \& Sunyaev, R.~A.\ 1973, \aap, 24, 337 \\

\reference{} Sluse, D., Hutsem{\'e}kers, D., Courbin, F., Meylan, G., \& Wambsganss, J.\ 2012, \aap, 544, A62 \\

\reference{} Uttley, P., Cackett, E.~M., Fabian, A.~C., Kara, E., \& Wilkins, D.~R.\ 2014, \aapr, 22, 72  \\

\reference{} Vasudevan, R.~V., Fabian, A.~C., Reynolds, C.~S., et al.\ 2016, \mnras, 458, 2012 \\

\reference{} Volonteri, M., Sikora, M., Lasota, J.-P., \& Merloni, A.\ 2013, \apj, 775, 94 \\

\reference{} Walton, D.~J., Reynolds, M.~T., Miller, J.~M., et al.\ 2015, \apj, 805, 161 \\

\reference{} Wambsganss, J.\ 2006, Saas-Fee Advanced Course 33: Gravitational Lensing: Strong, Weak and Micro, 453--540, arXiv:astro-ph/0604278 \\

\reference{} Witt, H.~J., Kayser, R., \& Refsdal, S.\ 1993, \aap, 268, 501 \\

\reference{} Wyithe, J.~S.~B., Agol, E., \& Fluke, C.~J.\ 2002, \mnras, 331, 1041 \\

\reference{}  Wyithe, J.~S.~B., Webster, R.~L., \& Turner, E.~L.\ 2000, \mnras, 318, 762 \\


\end{references}
\end{document}